\documentclass[journal]{IEEEtran}
\usepackage{footnote}
\makesavenoteenv{tabular}
\usepackage{comment}

\usepackage[T1]{fontenc} 
\usepackage{color}
\usepackage{multirow}
\usepackage{array}
\usepackage{amsmath,amsthm,amsfonts,amssymb,bm,mathrsfs}
\usepackage{url}
\usepackage{soul} 
\usepackage{graphicx} 
\usepackage{stfloats}
\usepackage{enumitem}
\usepackage{algorithmic}
\usepackage{algorithm}
\usepackage{overpic}
\usepackage{cite} 
\usepackage{wasysym}
\usepackage[colorlinks,citecolor=black, filecolor=black, linkcolor=black,urlcolor=black]{hyperref}

\usepackage[table]{xcolor}
\definecolor{slg}{gray}{0.95}

\makeatletter
\newcommand{\removelatexerror}{\let\@latex@error\@gobble}
\makeatother
\usepackage{booktabs, makecell, tabularx}
\usepackage[para]{threeparttable}
\usepackage{caption}

\newcolumntype{C}{>{\centering\arraybackslash}X} 

\usepackage{siunitx}
\graphicspath{}

\def\eg{{e.g.}}
\def\ie{{i.e.}}
\def\etc{{etc}}
\def\ngram{{$n$-gram}}

\def\etal{\emph{et al.}~}

\usepackage{hyperref}
\begin{document}
\title{Semantics-Empowered Communications:~A Tutorial-cum-Survey}
\author{Zhilin~Lu,~Rongpeng~Li,~Kun~Lu,~Xianfu Chen,~Ekram Hossain,~Zhifeng~Zhao,~and~Honggang~Zhang

    \thanks{
  Zhilin~Lu,~Rongpeng Li,~Kun~Lu are with Zhejiang University, China (email: \{lu\_zhilin, lirongpeng, lukun199\}@zju.edu.cn);}\thanks{Xianfu Chen is with the VTT Technical Research Centre of Finland, Oulu, Finland (email: xianfu.chen@vtt.fi);}\thanks{ Ekram Hossain is with the University of Manitoba, Winnipeg, Canada (email: ekram.hossain@umanitoba.ca);}\thanks{ Zhifeng Zhao and Honggang Zhang are with Zhejiang Lab as well as Zhejiang University, China (email: \{zhaozf,honggangzhang\}@zhejianglab.com).}
}

\maketitle

\begin{abstract}
    Along with the springing up of the semantics-empowered communication (SemCom) research, it is now witnessing an unprecedentedly growing interest towards a wide range of aspects (\eg, theories, applications, metrics and implementations) in both academia and industry. In this work, we primarily aim to provide a comprehensive survey on both the background and research taxonomy, as well as a detailed technical tutorial. Specifically, we start by reviewing the literature and answering the ``what'' and ``why'' questions in semantic transmissions. Afterwards, we present the ecosystems of SemCom, including history, theories, metrics, datasets and toolkits, on top of which the taxonomy for research directions is presented. Furthermore, we propose to categorize the critical enabling techniques by explicit and implicit reasoning-based methods, and elaborate on how they evolve and contribute to modern content \& channel semantics-empowered communications. Besides reviewing and summarizing the latest efforts in SemCom, we discuss the relations with other communication levels (e.g., conventional communications) from a holistic and unified viewpoint. Subsequently, in order to facilitate future developments and industrial applications, we also highlight advanced practical techniques for boosting semantic accuracy, robustness, and large-scale scalability, just to mention a few. Finally, we discuss the technical challenges that shed light on future research opportunities.
\end{abstract}

\begin{IEEEkeywords}
    Semantic communications, tutorial-cum survey, semantic information theory, semantic similarity metrics, semantic transmission and reasoning, enabling techniques, challenges and opportunities
\end{IEEEkeywords}

\IEEEpeerreviewmaketitle

\section{Introduction}

\IEEEPARstart{F}{rom} 1G in the 1980s to recently commercialized 5G, modern mobile communications have experienced substantial technical revolutions in almost every decade. Meanwhile, tremendous new applications, such as massive machine-type connectivity with stringent latency and reliability constraints, spawn and exhibit significant differences from the classical throughput-oriented transmission. Correspondingly, it is highly demanded to re-design smarter and more efficient communications to satisfy the diversified quality of service (QoS) requirements whilst balancing the dilemma of finite communication resources and infrastructures. Basically, an intelligent communication system, which is a vital ingredient in both industry and academia endeavors, is envisaged to not merely memorize data flows with rigid rules, but also understand, process, and express the underlying semantics. In that regard, apart from post-processing in downstream information carriers, compression and understanding from the very beginning would be efficient \cite{strinati20216g,shi2021new}, which also emerges as a driving force to push the frontier of exploring ``meanings'' in a communication system. Building upon the classical information theory (CIT) as revealed by Shannon \cite{shannon1948mathematical}, modern transmission schemes provide cutting-edge protocols (\eg, CDMA, OFDM, MIMO, etc.) for reliable communications, securing bit-level transmission accuracy. In spite of its popularity, treating each bit with equal importance can be content- and intelligence-agnostic in real applications and inconsistent with human perception as well \cite{lu2022rethinking}. Essentially, bit-level accuracy is not sufficient for ``intelligent'' and semantic transmission - intuitively, different expressions may exactly convey a similar meaning, while an expression with a high word error rate (WER) \cite{farsad2018deep} may still convey the meaning (or meaningful representation) of the message (\ie, the semantics \cite{shannon1948mathematical}). Therefore, instead of fully transmitting every single bit, it will be desirable to develop a semantics-empowered paradigm that seeks to communicate the underlying meanings.

Transmission with semantics can date back to as early as 1949, when Weaver first introduced this idea \cite{weaver1949recent}. Unfortunately, owing to the technical constraint and a more urgent need for Shannon's CIT, related researches entered a period of silence. As a matter of fact, it took the community several decades to find concrete clues to break out the ``Shannon's trap'' \cite{yang2022semantic2}, since mastering semantics is widely beyond the scope of CIT. Benefiting from the latest advance in artificial intelligence (AI), the prerequisite techniques are becoming gradually available and ready for building a prototype model. As a piece of evidence, research on semantics-empowered communications (SemCom for short) is now experiencing unprecedented growth. According to Google Scholar, the number of papers titled ``semantic'' and ``communication'' in 2022 exceeds the summation of those in the previous two years. Notably, this surging trend is expected to continue in the near future since SemCom is a promising enabler for 6G and beyond applications (e.g., Metaverse applications) \cite{strinati20216g}.\par
Nevertheless, coming along with its popularity, researches on SemCom are relatively scattered and with varied research interests. The discussions on ``\emph{what}'', ``\emph{why}'' and ``\emph{how}'' are at their infancy. We argue that a comprehensive tutorial-cum-survey on SemCom covering basic definitions, research taxonomy, and developing trends will be very beneficial for researchers and practitioners.

\begin{table*}[ht]
    \vspace{-4mm}
    \centering
    \caption{Summary and comparison of related survey and tutorial papers on SemCom}
    \label{tab:scope}

    \begin{tabular}{m{2.5cm} m{1.5cm} m{1.0cm} m{1.0cm} m{1cm} m{1cm} m{7.4cm}} 
    \hline
    Survey    & History \&  Definitions & Taxonomy       & Coverage       & Tutorial       & Future Prospect & Brief Description      \\
    \hline
    Strinati \etal \cite{strinati20216g}     & $\CIRCLE$   & $\RIGHTcircle$ & $\RIGHTcircle$ & $\RIGHTcircle$ & $\RIGHTcircle$  & Components along with the basic mathematical theory, and 6G applications     \\
    Lan \etal \cite{lan2021semantic}   & $\RIGHTcircle$    & $\RIGHTcircle$ & $\CIRCLE$      & $\RIGHTcircle$ & $\CIRCLE$       & New insights from human-machine communications  \\
    Qin \etal \cite{qin2021semantic}   & $\CIRCLE$   & $\RIGHTcircle$ & $\RIGHTcircle$ & $\CIRCLE$      & $\RIGHTcircle$  & A brief tutorial on the definition, components, metrics and multi-modal applications       \\
    Iyer \etal \cite{iyer2022survey}   & $\RIGHTcircle$    & $\RIGHTcircle$ & $\RIGHTcircle$ & $\RIGHTcircle$ & $\RIGHTcircle$  & Basic introduction and opportunities on wireless networks   \\
    Yang \etal \cite{yang2022semantic,yang2022semantic2} & $\RIGHTcircle$    & $\CIRCLE$      & $\CIRCLE$      & $\CIRCLE$      & $\CIRCLE$       & Detailed technical taxonomy and challenges, along with future applications   \\
    Wheeler \etal \cite{wheeler2022engineering}    & $\RIGHTcircle$    & $\CIRCLE$      & $\CIRCLE$      & $\RIGHTcircle$ & $\RIGHTcircle$  & Summary of semantic theory and four enabling approaches with detailed analysis       \\
    Chaccour \etal \cite{Chaccour2022Less}   & $\RIGHTcircle$    & $\RIGHTcircle$ & $\RIGHTcircle$ & $\CIRCLE$      & $\CIRCLE$       & Design of a semantic communication network on the basis AI-enabled semantic language and knowledge reasoning \\
    This paper     & $\CIRCLE$   & $\CIRCLE$      & $\CIRCLE$      & $\CIRCLE$      & $\CIRCLE$       & Complete background, techniques and research taxonomy, balanced with adequate coverage and tutorial    \\
    \hline
    \multicolumn{7}{>{\footnotesize\itshape}r}{Notations: \rm{${\CIRCLE}$} \emph{indicates fully included;} \rm{${\RIGHTcircle}$} \emph{means partially included.}}
    \end{tabular}
\end{table*}

\subsection{Related Work and Scope}

SemCom is currently undergoing extensive discussions and the understanding of SemCom is continually evolving. Looking back at the footprints, works on SemCom can be generally grouped into three main categories (i.e., conceptual exploration, detailed technical research, and surveys \& tutorials) from the perspectives of contributions and research directions. The first category mainly concentrates on the ``\emph{what}'' and ``\emph{why}'' problems, by envisioning possible features and architectures, as well as explaining the necessity of this new communication scheme. The second category answers the ``\emph{how}'' question and provides concrete implementation details and techniques for enabling a SemCom system. Belonging to the last category, this paper focuses on the summary and tutorials for facilitating further researches.

Prior to our work, publicly-available surveys \& tutorials on SemCom have shed some light on summarizing the latest developments and future challenges. Strinati \etal \cite{strinati20216g} provide a holistic overview of the components along with the basic mathematical theory that supports a semantic level transmission, and highlight its potential in 6G scenarios. Lan \etal \cite{lan2021semantic} incorporate the semantics and efficiency problem from an inspiring human-machine modality perspective and put an emphasis on the role of knowledge graph (KG) in SemCom. On the other hand, Qin \etal \cite{qin2021semantic} concentrate more on the components, metrics, and multi-modal applications. Iyer \etal \cite{iyer2022survey} and Yang \etal \cite{yang2022semantic} focus on the potential of intelligent wireless networks and edge intelligence respectively. Later, Yang \etal \cite{yang2022semantic2} further envision possible applications and benefits along with technical challenges in future SemCom-powered 6G scenarios. Very recently, Wheeler \etal \cite{wheeler2022engineering} provide the latest survey, incorporating the history, taxonomy, techniques, and challenges of SemCom, while Chaccour \etal \cite{Chaccour2022Less} present an end-to-end vision of SemCom from the perspective of AI-enabled semantic language and knowledge reasoning.

Similar to these works \cite{strinati20216g,lan2021semantic,qin2021semantic,iyer2022survey,yang2022semantic,yang2022semantic2,wheeler2022engineering,Chaccour2022Less}, this tutorial-cum-survey is also designed to provide a clearer and more in-depth insight into SemCom. In a nutshell, we first provide a holistic overview of the historical developments, motivations, and research directions. Meanwhile, we clarify a few ambiguous concepts that are not specifically considered in previous works, and present the pioneering researches from varied directions. Outlined by the proposed taxonomy, we then present a comprehensive and thorough tutorial-cum-survey on the concrete techniques, frameworks, and most recent proposals that support the goal of semantic transmission. Moreover, from end-to-end semantic transmission to semantics-assisted networks, we identify the potential applications and future challenges, along with new possible directions to facilitate future researches. However, there exist some significant differences between this tutorial-cum-survey and the literature \cite{strinati20216g,lan2021semantic,qin2021semantic,iyer2022survey,yang2022semantic,yang2022semantic2,wheeler2022engineering,Chaccour2022Less}.
In particular, we summarize and highlight the differences from five key aspects (\ie, basic history \& definitions, research taxonomy, coverage of techniques, tutorial, and future prospect) in Table \ref{tab:scope}. Meanwhile, the major contributions of this paper relative to the recent literature \cite{strinati20216g,lan2021semantic,qin2021semantic,iyer2022survey,yang2022semantic,yang2022semantic2,wheeler2022engineering,Chaccour2022Less} are as follows:

\begin{itemize}
    \item (\emph{History \& Definitions}) Compared to other survey papers in the field, this tutorial-cum-survey provides clearer motivation \& definitions by introducing the history of key SemCom ingredients (\ie, modern mobile communications and AI) and discussing the strong incentive to converge these ingredients. We also highlight ``what is semantics'', ``how semantics is established, shared, and represented'', as well as the conditions under which a communication system can be semantic.
    \item (\emph{Taxonomy}) We provide a holistic research taxonomy of enabling techniques. Moreover, we explore the relationship between the semantic level and other levels (\eg, conventional communications). Taxonomy tables to summarize the key contributions, frameworks, metrics, datasets \& toolkits,  and limitations are also provided.
    \item (\emph{Coverage}) Besides the basic structure and working pipeline, we summarize the corresponding ecosystems like theoretical advancement, learning metrics, wireless optimizations, \etc.
    \item (\emph{Tutorial}) We provide a tutorial-cum-survey spanning from history, theories to applications of SemCom, and present key lessons learned from the survey, which will help researchers and practitioners to quickly dive into this topic.
    \item (\emph{Future Prospect}) We comprehensively survey potential applications for SemCom, and also identify future trends as well as technical and business challenges.
\end{itemize}

\subsection{Contents and Structures}

\begin{figure*}[ht]
    \centering
    \setlength{\abovecaptionskip}{-0.2cm}
    \includegraphics[width=.85\linewidth]{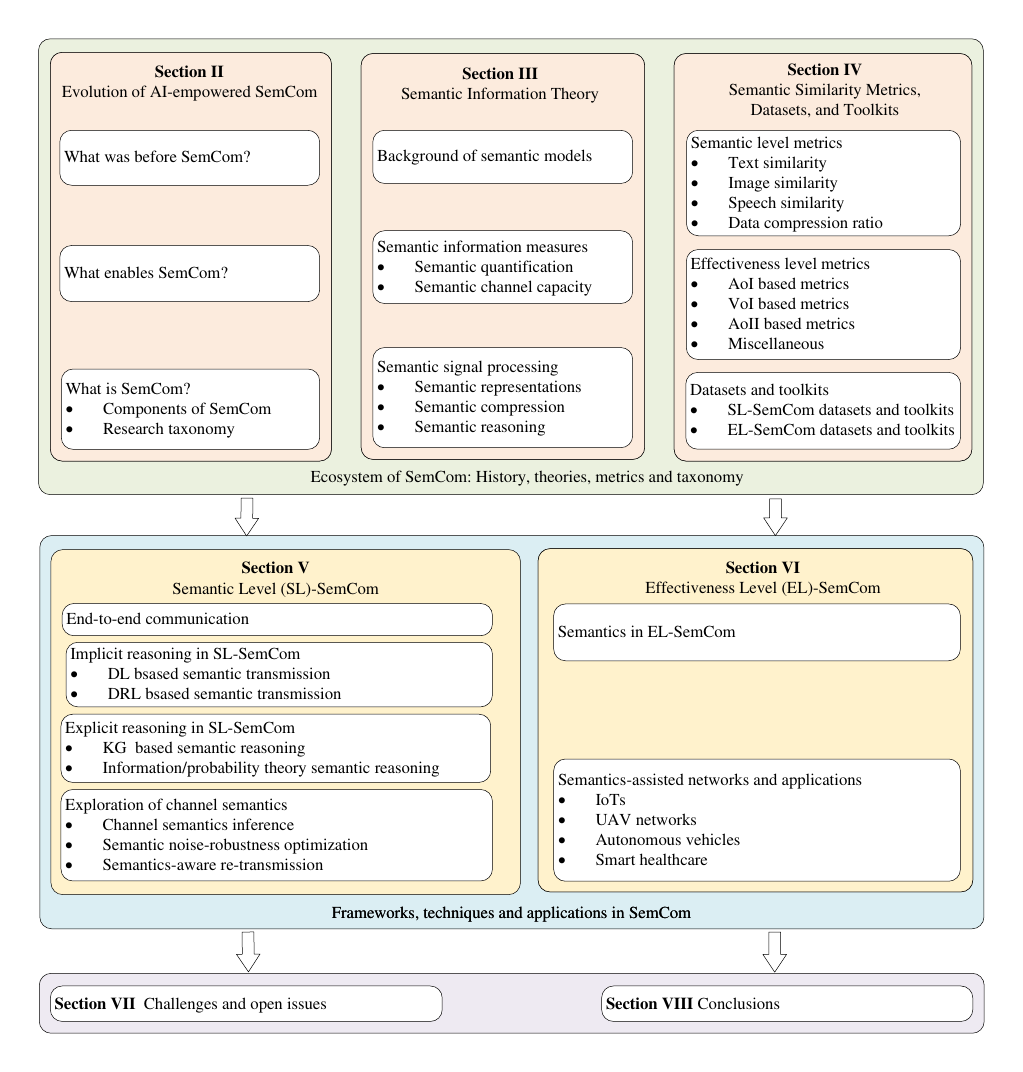}\\
    \caption{Main contents of this paper.}
    \label{totalsection}
\end{figure*}

This work is a tutorial-cum-survey covering the basic definitions, taxonomy, applications and challenges for SemCom. The organization of its contents and the relations is illustrated in Fig. \ref{totalsection}.	Our first goal is to provide a clear definition and explanation of semantic-empowered mechanism in typical transmission scenarios and illustrate how it differs from conventional communication schemes, which serves as the cornerstone for further in-depth analysis. Beforehand, in Section \ref{sec:history}, we will revisit the historical development of semantic transmission and its related driving forces for readers who are not familiar with the surveyed topic. Afterwards, following the logical order, we will discuss the components of SemCom and provide the in-depth research taxonomy. In Section \ref{sec:SemanticTheory} and Section \ref{sec:metrics}, we would clarify the definitions and elaborate on the corresponding ecosystems like theoretical guidance, metrics, toolkits, \etc. Next, we highlight each of the enabling techniques. In particular, in Section \ref{sec:sl_SemCom}, we delve into the semantic-level SemCom (SL-SemCom) by presenting the exploration of both content semantics and channel semantics. Notably, content semantics focuses on extracting semantic information from the transmitted data to improve transmission efficiency and effectiveness, while channel semantics, which encompasses both channel state information (CSI) and surrounding environment information, puts more emphasis on leveraging channel and/or environment characteristics to facilitate the design of coding schemes, thereby better adapting to the channel and increasing transmission reliability. Furthermore, we propose to classify the utilization of content semantics in SL-SemCom to two major categories (\ie, implicit reasoning and explicit reasoning), wherein the former category typically relies on structured or unstructured parameterized models like deep neural networks (DNNs), while the latter involves explicitly defined rules or entities. Moreover, targeted at practical and large-scale commercial applications, recent progress on techniques to explore and exploit both content and channel semantics is also included. On top of these end-to-end SemCom approaches, we then transfer our viewpoint to effectiveness-level SemCom (EL-SemCom) and semantics-assisted networks in Section \ref{sec:el-SemCom}, where potential applications and opportunities in 6G (and beyond) are envisioned. Finally, some future challenges and open issues are discussed in Section \ref{sec:challenge} while Section \ref{sec:conclusion} concludes the paper. 

\section{From Past to Future of AI-empowered SemCom}
\label{sec:history}
In this part, we begin with a brief history of conventional communications before SemCom, and talk about the potentially encountered obstacles and limits therein. Subsequently, inspired by the astonishing advance in AI, we provide the history of this driving force towards SemCom, and try to answer why it contributes to breaking out ``Shannon's trap \cite{yang2022semantic2}''. Afterwards, we formally discuss the differences between SemCom and conventional communications, and present our viewpoint of SemCom. Besides, the research taxonomy is summarized to provide a succinct overview of SemCom.

\subsection{What Was Before SemCom?}
\begin{figure*}[t]
    \centering
    \setlength{\abovecaptionskip}{-0.7cm}
    \includegraphics[width=\linewidth]{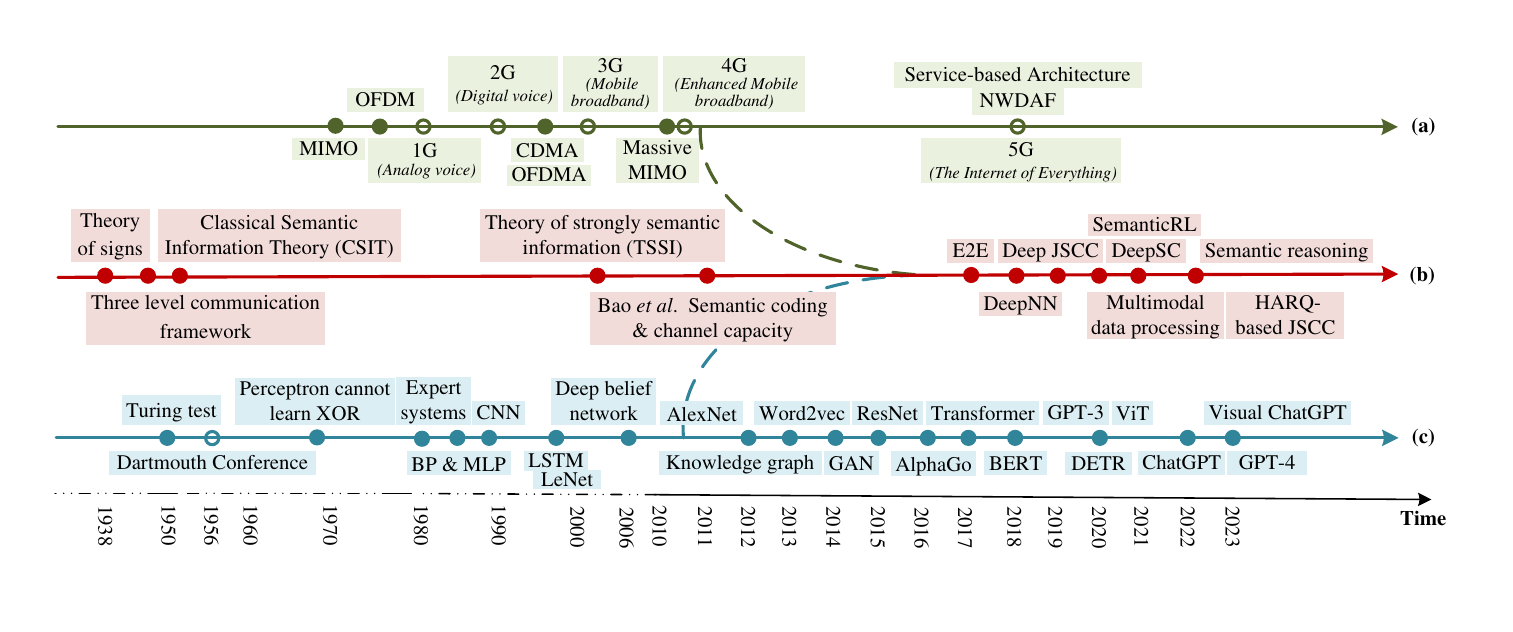}\\
    \caption{Evolution of SemCom and its enabling ingredients: (a) Evolution of modern mobile communications (green dotted line), (b) Evolution of SemCom (red dotted line), and (c) Evolution of AI techniques (blue dotted line).}
    \label{Evolution}
\end{figure*}

Before delving into the SemCom, it is meaningful to recall the evolution of mobile communications, which can be traced back to 1897~\cite{baker2013history,qiao2021survey}.
Since then, we have witnessed tremendous progress in communications. In particular, 5G capably connects both humans and things, and provisions diversified services to vertical industries \cite{jiang2021road}. For example, 5G has begun to support three different service classes including enhanced mobile broadband (eMBB), ultra-reliable and low-latency communications (URLLC), and massive machine type communications (mMTC), and built up a service-based architecture (SBA) with network data analytics function (NWDAF) to make the implementation of AI feasible. Meanwhile, as summarized in Fig. \ref{Evolution}(a), the drastic evolution of communications is also accompanied by cutting-edge techniques such as FDMA, TDMA, W-CDMA, and OFDMA \cite{patwary2020potential}. 

In order to realize AI-based ultra-sensitive tactile and haptic communications in 6G \cite{qiao2021survey}, it is natural to explore more high-frequency bands and fully leverage powerful technologies like multi-layer precoding/beamforming and ultra-massive MIMO \cite{mchangama2020mmwave}. However, such an intuitive evolution often implies more intense resource demand in terms of both power and bandwidth, and gives rise to new challenging issues. For instance, though spreading signals at higher frequencies could partially resolve the over-exploitation issue of sub-6GHz spectra, it faces disadvantages such as disruptive attenuation and cost-ineffective deployment. Meanwhile, according to Shannon's CIT \cite{shannon1948mathematical}, the existing communication technologies have nearly approached the physical-layer capacity limit \cite{luo2022semantic}. Therefore, we will need to develop new and more efficient technologies (e.g., SemCom) for future emerging applications.

\subsection{What Enables SemCom?}
\label{sec:what_enable_SemCom}
Considering many successful implementations of SemCom (\eg, semantic transmission \& reasoning \cite{qin2021semantic,yang2022semantic2,Chaccour2022Less}), it can be observed that the development of SemCom is often accompanied by the flourishing of AI. Specifically, as shown in Fig. \ref{Evolution}(c), albeit its tortuous progress at the early stages, AI, specifically deep learning (DL) based on DNNs, has made remarkable achievements recently. Inspired by this interesting observation, we go through the most commonly used AI technologies and try to explain the importance of breakthroughs in AI to SemCom\footnote{Notably, besides AI, there are alternative theoretical means (\eg, \cite{choi2022unified}) to implement SemCom.}.

One major category of modern AI comes from discriminative models. For vision tasks, convolutional neural networks (CNNs) take advantage of the correlation of neighboring pixels, and play a key role in both high-level (\eg, detecting, tracking, classification) and low-level processing (\eg, inpainting, enhancing, denoising) \cite{long2015fully,girshick2015fast,he2016deep,carion2020end}. To model the sequence data, recurrent neural networks (RNNs) update their inner hidden states at each time step. In prominent research on semantics, sequential data (typically, text messages) is the most widely studied media. Benefiting from advanced AI models like long short-term memory (LSTM) \cite{hochreiter1997long} and attention-based transformer \cite{vaswani2017attention}, both short-time and long-time semantic dependencies can be represented and extracted for downstream processing \cite{mikolov2013efficient,devlin2018bert,radford2018improving,liu2017improved}. Besides structured stream data, edge-node-based graph models handle the dynamics of data topology by aggregating information from their neighborhoods, which shows great success on a variety of graph media such as social networks and knowledge graphs \cite{hamilton2017inductive,kipf2017semisupervised,velickovic2017graph}. Besides, deep transfer learning (DTL) \cite{li_tact_2014} emerges as a useful technique for effectively using previously learned knowledge to solve novel tasks with minimum training. Meanwhile, theoretical researches such as information bottleneck (IB) theory \cite{alemi2017deep,tucker2022towards,yuan2021graphcomm} and probabilistic models \cite{choi2022unified, seo2021semantics} contribute to easing concerns on the interpretability of DNN-based schemes.

Another category of AI models concentrates on the generative process, where the most representative ones are variational auto-encoder (VAE) \cite{kingma2014auto}, generative adversarial network (GAN) \cite{NIPS2014_5ca3e9b1}, and diffusion models \cite{ho2020denoising}. These models are designed to mimic the source data distribution, which can be further used to augment finite data sets, support intelligent generation \& editing, and even provide surrogate functions for non-differentiable processes like random communication channels \cite{lu2022contrastive,hua2019gan,ye2020deep}. Moreover, ChatGPT\footnote{Interesting readers could refer to \url{https://openai.com/blog/chatgpt/} for further details.}, one representative large language model (LLM) fine-tuned from the generative pre-trained transformer (GPT)-3.5, has attracted intense interest due to its remarkable performance to handle contexts (e.g., the chain of thoughts). 
Last but not least, deep reinforcement learning (DRL) promises to learn by maximizing the total expected \emph{reward} instead of relying on conventional loss functions. Learning by repetitive interactions with the environment, RL is believed to be an advanced form of AI, and shows its unprecedented superiority in decision-making tasks such as gaming, driving, and scheduling \cite{silver2016mastering,li2018deep,kiran2021deep}. It is also commonly used to provide another alternative for non-differentiable functions since the reward can take any scalar value.

Notably, as data-driven DL remarkably boosts learning performance and raises increasing attention, conventional communications also benefit from this cutting-edge technique and evolve towards enhancing the awareness and cognition capability. In this new area as illustrated in Fig. \ref{Evolution}(b), SemCom witnesses its new spring on the tacit intersection of both modern mobile communications and AI developments. Typically, AI/DL-based physical layer schemes \cite{gao2018comnet,wen2018deep} and end-to-end communications \cite{o2017introduction,o2017deep,erpek2018learning} are now serving as important precursors of SemCom, demonstrating superior performance in a variety of tasks \cite{xie2021deep,lu2021reinforcement,wang2021performance}. In other words, AI is becoming an indispensable ingredient for communications \cite{li_collective_2020,li_intelligent_2017}. When it emerges an urgent incentive to design a new communication paradigm (\ie, SemCom), motivated by the gradual maturity of current AI and DL technologies, it is natural for researchers to seek effective and promising AI-based strategies.

\subsection{What is SemCom?}
\label{sec:whatisSemCom}
In a classical communication model, for a transmitted message $x$ encoded from the source signal $s$, the receiver tries to accurately decode a signal $\hat{s}$ from the received message $y$ in terms of the bit-level accuracy. In particular, the received signal could be approximately derived from the channel information $H$ as $y=Hx+\varepsilon$, where $H$ is the channel gain, and $\varepsilon$ denotes the channel noise. In order to pursue the bit-level accuracy, various channel-oriented \cite{neumann2018learning,he2020model,ye2020deep} and source-oriented methods \cite{o2017deep,erpek2018learning} have been proposed. The former category mainly focuses on signal estimation, channel detection, and resource allocation; while the latter puts emphasis on modulation and coding technologies, as well as constellation diagram design. Building on the accuracy principle, significant progress on reliable and efficient transmission has also been made, approaching the Cramer-Rao lower bound (CRLB) \cite{kay1993fundamentals} and Shannon limit. However, this conventional design of communication systems faces a performance dilemma with the surging demand for data traffic and limited resources. It is expected and becomes imperative to revolutionize and explore alternative dimensions to solve the limitation and bottleneck of conventional communications \cite{shi2021new}.

\begin{figure*}[htb]
    \centering
    \setlength{\abovecaptionskip}{-0.2cm}
    \includegraphics[width=.9\linewidth]{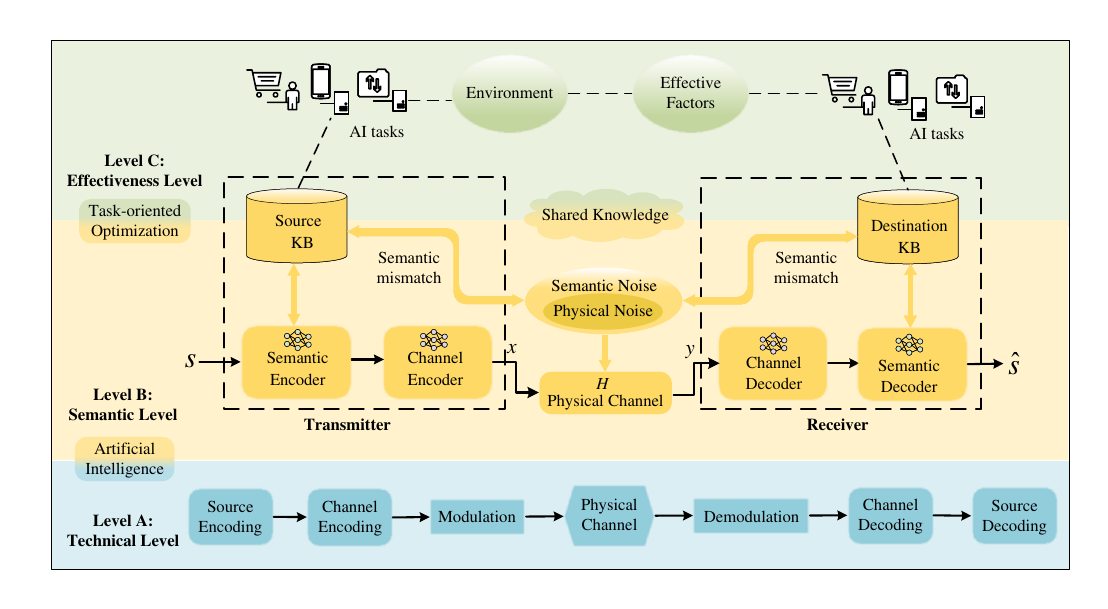}\\
    \caption{The three levels of communications.}
    \label{three}
\end{figure*}

Different from securing bit-level accuracy, Weaver proposes an alternative communication model that is designed following a completely different philosophy \cite{weaver1949recent}. In Weaver's vision, communications can take place on three levels. The first level is designed for \emph{reliably} transmitting symbols (\ie, bits) of data from the sender to the receiver, and has been extensively investigated. The second level, which is called the \emph{semantic} level, aims to extract the \emph{semantic} information and transmits it via a channel with both semantic and physical noise; while the third level focuses on \emph{effectively} performing tasks instantiated by the sender. Following Weaver's path, Carnap \cite{carnap1952outline} envisions that efficient SemCom can be realized by quantifying semantic information and considering semantic compression errors. The proposal of semantic transmission is attractive, as it naturally ``compresses'' the original message and only transmits the core meaning beneath any complicated symbolic representation. Due to the remarkable reduction in information overhead, this philosophy allows SemCom to stand out from high-rate \& reliability-based designs and better touches the essence of communications \cite{strinati20216g,lu2022rethinking}.
In a word, SemCom focuses on accurately transmitting the semantic meanings rather than bits.

Despite its promising features, SemCom did not gain sufficient attention for quite a long period. The reasons could be bi-folded. On one hand, Shannon believed that ``the semantic aspects of communication are irrelevant to the engineering aspects'' since the meaning of messages can be related to specific physical and conceptual entities \cite{weaver1949recent}. As a result, semantic transmission inevitably affects the generality of mathematical model \cite{yang2022semantic}. Moreover, owing to the much more urgent need for high-rate reliable communications at that time, researchers mostly follow the path of the CIT established by Shannon \cite{shannon2001mathematical}. It has been pointed out in recent work that merely securing a symbolic recurrence, however, can be regarded as a means of preserving the syntactic information while lacking the capability of semantic awareness \cite{jiang2015error,zhong2017theory,lu2022rethinking}. Analogous to the CIT, Carnap \etal \cite{carnap1952outline} put forward the semantic information theory (SIT), which is henceforth referred to as classical semantic information theory (CSIT). Later, Floridi \etal \cite{floridi2004outline} extend CSIT to explore the measurements of information, i.e., \emph{theory of strongly semantic information} (TSSI\footnote{The details of CSIT and TSSI shall be given in Section \ref{sec:semantic_uncertainty}.}).
Recently, Bao \etal \cite{bao2011towards} propose a theoretical framework to measure the semantic information and channel capacity.
These works clearly demonstrate the positive impact of semantics in communications, but are unable to provide theoretical analysis and feasible implementation methods for practical large-scale communication networks. Instead, benefiting from the impressive advance in data-driven AI or machine learning (ML) technologies, AI-powered SemCom is ushering into a new era and has achieved some preliminary results, on which will be elaborated later.


Fig. \ref{Evolution}(b) highlights some early theoretical foundations including CSIT \cite{carnap1952outline}, TSSI \cite{floridi2004outline} and semantic channel capacity \cite{bao2011towards}, as well as representative achievements from the perspective of AI-powered SemCom. Generally, these SemCom works have significantly advanced Weaver's vision. Beforehand, in order to summarize and keep our presentations succinct and easy to follow, we feel it essential to first outline the key components and prominent research directions of SemCom.

\subsubsection{Components of SemCom}
\label{sec:component_semcom}

From top-layer design to bottom-layer implementation, we illustrate a mainstream structure in recently popular SemCom system in Fig. \ref{three}. Serving as a cornerstone for the second and third levels of Weaver's model, SemCom focuses on ``how'' to transmit and ``what'' to transmit in a more intelligent and semantically-compact way. Ideally in SemCom, the sender transmits the most useful information in terms of semantics but can ignore the redundant information. Meanwhile, the receiver in SemCom could understand and infer the expected semantic meanings correctly. This means that SemCom is only interested in semantic meanings and can tolerate the loss of certain semantically-irrelevant contents, wherein a pair of sender and receiver may not need to exchange all bits or pursue the bit-level accuracy. Equivalently, the bandwidth required for data transmission in SemCom can be saved significantly.

\begin{figure}[t]
    \centering
    \setlength{\abovecaptionskip}{-0.2cm}
    \includegraphics[width=8.6cm]{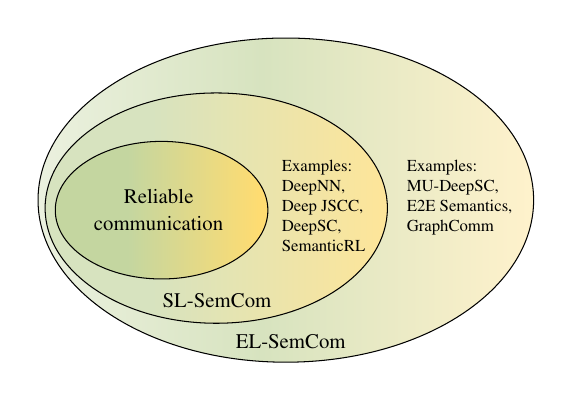}\\
    \caption{The relationship among the three levels of communications.}
    \label{venn}
\end{figure}

Meanwhile, to be consistent with Weaver's vision \cite{weaver1949recent}, SemCom can be further divided into SL-SemCom and EL-SemCom. In particular, the SL-SemCom systems mainly focus on semantic transmission for data reduction and put more emphasis on the delivery of content meaning rather than bit-level accuracy from a sender to a receiver, while the EL-SemCom systems consider more how to effectively utilize the semantic information at an appropriate time, thus facilitating successful task execution. Furthermore, Weaver believes that ``the purpose of all communication is to influence the conduct of receiver'', and ``it is clear that communication either affects conduct or is without any discernible and probable effect at all''. Therefore, we hold the view that all communications can fall into the scope of EL-SemCom from an engineering viewpoint. However, it is worth noting that EL-SemCom can achieve a goal on top of either extracted semantics or simple bits. Typical works in the former case include MU-DeepSC \cite{xie2021task}, E2E Semantics \cite{kountouris2021semantics}, and GraphComm \cite{yuan2021graphcomm}, while \cite{pappas2021goal, zhang2022goal} belong to the latter ones. In this way, when the goal of communications turns to accurate delivery of semantics only, it degenerates to SL-SemCom, such as \emph{joint source-and-channel coding} (JSCC) on text or image transmission (\eg, DeepNN \cite{farsad2018deep}, deep JSCC \cite{bourtsoulatze2019deep}), DeepSC \cite{xie2021deep} and SemanticRL \cite{lu2021reinforcement}. It is worth noting that the terminology ``joint'' in JSCC implies the training of the source and channel DNNs together, while the adopted DNN modules therein accomplish the functionalities like the source and channel codec in a separately designed yet end-to-end training manner. Hence, in Fig. \ref{three}, we still decompose it into a semantic encoder and channel encoder, so as to cover as many implementation means as possible. SemCom further falls back to the more fundamental and classical \emph{reliable communications} when the data transmission focuses on the bit level accuracy. Intuitively, Fig. \ref{venn} depicts the relationship between the three-level communications.

Technically, SL-SemCom usually introduces an extra semantic encoder and decoder to process the information, while EL-SemCom adds some goal-oriented or task-oriented\footnote{In this paper, we use the terminologies ``task-oriented'' and ``goal-oriented'' interchangeably.} factors to enhance the communication efficiency. As shown in technical and semantic levels in Fig. \ref{three}, SemCom performs various functions similar to conventional communications, and a SemCom architecture can encompass components such as \emph{semantic encoder}, \emph{channel encoder}, \emph{channel decoder} and \emph{semantic decoder}. Nevertheless, in contrast to independent modules in conventional communications to tackle statistical properties of symbols, SemCom is commonly contingent on \emph{jointly} (i.e., end-to-end) trained DNNs on top of an additional, essential component (i.e., the KB). Specially, the semantic encoder extracts semantic features based on its KB from the input sequence $s$, during which the redundant information is filtered. Meanwhile, the channel encoder is responsible for generating the symbol (embedding representation) $x$ to facilitate the subsequent transmission. Then, the extracted useful and relevant information will go through the physical channel. Similarly, the receiver is composed of a channel decoder for detecting the received symbol $y$ from the received semantically noisy signals, and a semantic decoder for semantic-level or effectiveness-level source estimation and recovery based on the KB \cite{zhou2021semantic}. Apparently, the most distinctive component of such a typical SemCom system attributes to a shared and continually updated KB, which can learn from the perceived environment and continuously evolve by repeated training or sharing via communications \cite{luo2022semantic}. Another noticeable difference lies in the \emph{semantic noise}, which extends the concept of physical noise in conventional communications, and can be defined as any noise that possibly leads to semantic infidelity \cite{carnap1952outline,bao2011towards}. This extended definition implies that semantic noise could take into account the fact that even when a semantic concept is perfectly described and encounters no physical distortion during transmission, the receiver may not find a precise symbolic or semantic interpretation that matches the original meaning \cite{carnap1952outline,bao2011towards} (typically when the KB of two communicating participants are not perfectly aligned). Therefore, besides physical noise and turbulence (e.g., Gaussian noise and multi-path fading), semantic noise also involves semantic mismatch, ambiguity and interpretation errors. Consequently, beyond simple \emph{data compression}, SemCom strives to effectively combat the semantic noise and transmit the semantic meaning (\eg, maximizing the semantic similarity between the source data $s$ and reconstructed data $\hat s$) with adequate encoding and decoding schemes. Moreover, the performance metrics such as semantic similarity and the compression ratio of the transmitted data constitute an important pillar of SemCom as well.


\subsubsection{Research Taxonomy}
In this paper, we take a coarse-to-fine routine where a macro research taxonomy is first provided for the community followed by detailed micro technical taxonomy based on the preliminary semantic information theory and the semantic similarity metrics in Section \ref{sec:SemanticTheory} and Section \ref{sec:metrics}, respectively.



\begin{figure*}[htb]
    \centering
    \setlength{\abovecaptionskip}{-0.2cm}
    \includegraphics[width=\linewidth]{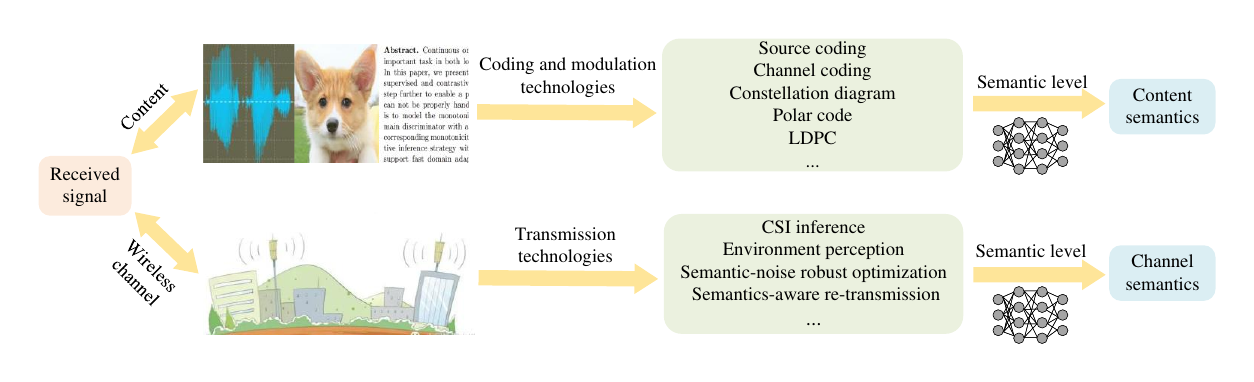}\\
    \caption{A potential direction of communications evolving toward SemCom.}
    \label{signal}
\end{figure*}

As mentioned earlier, SemCom can be classified as SL-SemCom and EL-SemCom. As for the research of SL-SemCom, the exploration of semantics includes content semantics and channel semantics learned by DNNs shown in Fig. \ref{signal}, wherein the former is realized over coding and modulation techniques, and the latter is contingent on transmission techniques. Specifically, the learning manner of content semantics can be categorized into implicit reasoning and explicit reasoning. In particular, explicit reasoning characterizes contextual probabilities with real entities or reasoning rules explicitly, wherein these real entities can be updated via pre-defined rules. We will mainly discuss mathematical probabilities and graph-modeled probabilities for this category \cite{choi2022unified,seo2021semantics,xiao2022reasoning}. On the contrary, the implicit reasoning methods, which are typically DL or DRL-based, learn how to express the semantics without a real mathematical probability distribution and infer with a set of learnable parameters supervised by human-crafted semantic similarity metrics as in Section \ref{sec:semantic-level_metric}, but in a discriminative manner. In particular, DL-based methods like JSCC-based frameworks \cite{farsad2018deep, bourtsoulatze2019deep, xie2021deep, zhou2022adaptive} assume the existence of differentiable objectives and adopt pre-trained DNNs as encoders and decoders.  On the other hand, regardless of the differentiability, DRL-based methods \cite{lu2021reinforcement,narvekar2020curriculum} model the SemCom as a Markov decision process (MDP) and dynamically learn the most appropriate transmission methods by trial-and-error. Notably, without explicitly modeling the semantic expression and inference principles at the receiver side, these models embed the contextual probabilities in their parameters, but are more difficult to adapt to new and varying semantic environments (\eg, channels).

In addition to the aforementioned content-related semantics, some channel-related semantics are explored as well, so as to better adapt to the varying physical channel and further optimize the entire SemCom system. As a comparison, the content semantics of the transmitted information can be mined through semantic coding (SC), while by means of environmental perception, the channel semantics can be further inferred for semantics-aware transmission techniques and  optimizing semantic noise-robustness.

In terms of EL-SemCom, we discuss the effectiveness of goal-oriented communications and some semantics-assisted networks. In this part, the SemCom pays more attention to the impact of semantic information on task execution, which can be evaluated in the corresponding effectiveness-level metrics in Section \ref{sec:task-metrics}, such as energy efficiency \cite{chen2022performance, chen2022neuromorphic}, \emph{age of information} (AoI) \cite{yates2021age}, \emph{value of information} (VoI) \cite{ayan2019age, molin2019scheduling} and the success or accuracy of task execution \cite{kountouris2021semantics, farshbafan2022common, yuan2021graphcomm}, rather than merely semantic transmission.

\section{Semantic Information Theory}
\label{sec:SemanticTheory}
\subsection{Background of Semantic Models}

Before going through SIT, we would first recall the basic analytic structure of SemCom and how semantics are extracted and processed. Basically, in a SemCom system, KB, which consists of all semantically related ``common knowledge'' between the transmitter and the receiver, is deployed on both sides as a consensus. For example, the common knowledge for English speakers could be the lexical and grammatical English rules, while that for any two educated persons (even with no shared languages) comes from socialization. We assume that common knowledge is shared (but possibly not completely aligned) on both sides \emph{before} any communications really takes place, \ie, an intrinsic property in SemCom. To start communicating, for any entity in the real world, a semantic concept would be first derived from the related observation $s$ and KB \cite{ogden1923meaning}, after which this semantic concept is transformed into a symbolic representation or semantic message $x$. Next, the message is encoded and transmitted through a physical channel, as illustrated in Fig. \ref{three}. Upon receiving the signal, a receiver could then decode and infer the corresponding semantic concept or symbol on the basis of the KB. Notably, a message $x$ can represent different types of media such as images, text (\eg, sentences) and audio, in which multiple semantic concepts can be described via simple or complex logical relations.

The SIT, hence, aims to quantitatively measure the amount of semantics, and explain how semantics flows in the communication process. Different from CIT which leverages the statistical probability of the message to measure the amount of information, in CSIT, Carnap \etal point out that the \emph{information} carried by a sentence within a given language system is treated as synonymous with its content normalized in a certain way \cite{carnap1952outline}. In this sense, the semantic information can be explicated by various measures of content based on logical probability, which is known as epistemic probability and refers to the degree of rational belief or confidence in a proposition within a given background knowledge and logical inference  \cite{ognjanovic2009probability,ellerman2013introduction}. Notably, in line with the consensus that the logical probability is defined within the context of a specific set \cite{ellerman2021new}, the logical probability discussed in CSIT and its extensions \cite{carnap1952outline, floridi2004outline, bao2011towards} is based on propositional logic within \emph{specific} KB (\ie, sets), and thus significantly different from the independently defined statistical probability (and its induced Shannon theory), which is applicable to \emph{any} given sets. Hence, in order to derive the proposition-oriented logical probability, the corresponding semantic universe and their induced messages (\eg, sentences) shall be firstly clarified. Specifically, taking the example of a language system $L_m^\pi$ with $m$ different individuals (\eg, things, events, or positions) and $\pi$ designating primitive properties for the individuals \cite{carnap1952outline,bar1953semantic}. Any message can be formed by atomic sub-messages with the help of five customary connectives (\ie, $\sim$ (\texttt{Not}), $\vee$ (\texttt{Or}), $\wedge$ (\texttt{And}), $\supset$ (\texttt{If} $\cdots$ \texttt{then}), and $\equiv$ (\texttt{If and only if})). Analogously, complicated molecular predictors can be built by a combination of primitive properties and connectives. In other words, the ``logical probability'' of a message is measured by the likelihood of the atomic sub-messages inferred from the message using the logical connectives being true \cite{carnap1952outline}.  Mathematically, to stand in consistent with \cite{carnap1952outline,bao2011towards}, we will use $m(\cdot)$ to indicate the logical probability. For example, suppose a language system $L_2^1$ consisting of two independent atomic sub-messages ``A'' and ``B'' (\ie, $m=2$), which mean ``true'' or ``false'' with equal probability (\ie, 0.5). 
Hence, the logical probability for the message ``A and B'' is true is 0.25, while that for ``A or B'' is 0.75. Mathematically, $m(A \wedge B) = 0.25$ and $m(A \vee B) = 0.75$ \cite{bao2011towards, carnap1952outline}.

Furthermore, Bao \etal \cite{bao2011towards} extend this theory into semantic universe and introduce the interpretations from semantic source $s$ to message $x$. Specifically, a random \emph{observation} $S$ takes a value from a \text{semantic world model} $\mathcal{S}$ (\ie, a virtual world of all possible semantics), wherein the world model can be modeled as \emph{interpretations} with a statistical probability distribution $\mu(\cdot)$, which can be regarded as a process of ``semantic coding''. Specifically, for a semantic message $x$,
\begin{equation}
    \label{eq:logical_prob}
    m(x) = \frac{{\mu({\mathcal{S}_x})}}{{\mu(\mathcal{S})}} = \frac{{\sum\nolimits_{s \in \mathcal{S},s\models x} {\mu(s)} }}{{\sum\nolimits_{s \in S} {\mu(s)} }},
\end{equation}
where ${\mathcal{S}_x}$ denotes a set that $s \in \mathcal{S}_x$ semantically entails $x$, i.e., $\mathcal{S}_x = \{s \in \mathcal{S},s\models x\} $, $x$ is ``true'' in semantic world model $\mathcal{S}$ \cite{bao2011towards}.

Based on the above basic background, some researchers have further studied and expanded the semantic information theory. Next, we will discuss in detail from the aspects of semantic information measures and semantic signal processing.

\subsection{Semantic Information Measures}
\label{sec:semantic_uncertainty}

\subsubsection{Semantic Quantification} In CSIT, Carnap \etal \cite{carnap1952outline} measure the information by the logical possibility defined above, rather than statistical distributions in CIT. In particular, for a message $x$ with the logical probability $m(x)$, the \emph{content measure} of a message is defined negatively related to the logical probability as $\text{cont}(x) = 1 - m(x)$.  Subsequently, the \emph{information measure}  $\text{inf} (x)$ of one message $x$ is further defined as
\begin{equation}
    \label{eq:information_measure}
    \text{inf} (x) = {\log}\frac{1}{1 - \text{cont}(x)}.
\end{equation}
\eqref{eq:information_measure} reveals the relationship between the amount of information and the semantic probability. Carnap's theory \cite{carnap1952outline,bar1953semantic} can be viewed as an inspiring and classical extension to Shannon's work.

However, CSIT measures the information for a statement based on the logical probability only, and ignores the influence of the truth on the amount of information, which will lead to a dilemma that the same logical probability of a false message and a true message imply the same amount of information. Meanwhile, it encounters the Bar-Hillel-Carnap (BHC) Paradox, that is, the ironical assignment of maximum information to the contradictions. For instance, for the message ``A and not A'', the corresponding logical probability equals $0$, thus the information measure going to infinity. To address these issues, Floridi \etal \cite{floridi2004outline} propose TSSI to model the \emph{informativeness} from both the polarity of a message $x$ (\ie, true or false) and the degree of discrepancy between $x$ and a given observation. Intuitively, for a received true message $x$ perfectly conforming to a given world model $\mathcal{S}$, the expected semantic discrepancy is zero. Otherwise, depending on the truth value of $x$, a function $f(\cdot)$ is further introduced as
\begin{equation}
    f(x) = \left\{
    \begin{aligned}
    & \frac{e}{l},  	    \quad\quad		\textrm{if} \, x \, \textrm{is false;} \\
  - & \frac{n}{n_l},   	\quad\quad	\textrm{if} \, x \, \textrm{is true},
    \end{aligned}
    \right.
    \label{equ:Floridi_DegreeCombined}
\end{equation}
where ${e}$ is the number of false atomic sub-messages in the message $x$, and \emph{l} is the total number of atomic sub-messages. \emph{n} denotes the cardinality of possible semantics consistent with message ${x}$, and ${n_l}$ is the number of all possible semantics that can be expressed in an $l$-length message.  \eqref{equ:Floridi_DegreeCombined} can be interpreted as that for a true message, it is the most indiscriminate, and can be portrayed by \emph{degree of vacuity}; while for a false message, $f(\cdot)$ depicts the \emph{degree of inaccuracy}. Based on \eqref{equ:Floridi_DegreeCombined}, \cite{floridi2004outline} further defines the \emph{degree of informativeness} as
\begin{equation}
    \label{eq4}
    \tau (x) = 1 - f^2(x).
\end{equation}

In addition, as the degree of vacuity and inaccuracy is largely dependent on the specific model of messages, it is not easy to quantify these two metrics for more complicated sentences rather than merely simple statements. As such, D'Alfonso \cite{d2011quantifying} measures the semantic information by introducing the ``truthlikeness'', which can be further quantified by numerous distance metrics, such as Tichie-Oddie approach and Niiniluoto approach \cite{d2011quantifying}. To some extent, these efforts \cite{floridi2004outline, d2011quantifying} partially solve the BHC paradox by effectively quantifying the semantic value of messages.

On the other hand, from the perspective of CIT \cite{shannon1948mathematical}, entropy, which is defined by the statistical probability of a random variable, could indicate the average level of information or ``uncertainty'' inherent to the random variable's possible outcomes. For example, when a random variable takes a deterministic value, we say it has no uncertainty (a low entropy). Consistent with CIT, Bao \etal \cite{bao2011towards} employ Shannon entropy to quantify information of semantic source (\ie, source semantic entropy), which can be formulated as
\begin{equation}
    \label{equ:SeEntropy_Source}
    H(S) = - \sum\nolimits_{s\in \mathcal{S}} \mu(s) \log \mu(s),
\end{equation}
where $\mu(s)$ is the statistical probability of $s$ defined above.
Analogized to the CIT, for each given semantic message $x$ encoded from $s$, Bao \etal \cite{bao2011towards} defines the semantic entropy of a message $x$ from the perspective of logical probability as
\begin{equation}
    \label{equ:SeEntropy_Message}
    {H_s}(x) =  - \log m(x).
\end{equation}
We use the subscript $s$ in \eqref{equ:SeEntropy_Message} to distinguish a semantic entropy based on the logical probability in \eqref{eq:logical_prob} from the statistical probablity-based classical entropy in CIT. Besides, it is worthwhile to note here that in terms of a logarithmic measure of a message, Bao's definition of the semantic entropy puts more emphasis on the informativeness of the message semantically encoded from the source, and is slightly abused from the standard definition of entropy in CIT, which takes the statistically average value.

In order to reduce semantic source entropy, D'Alfonso \cite{d2011quantifying} exemplifies the positive contribution of the KB. Specifically, with the help of KB, even when the receiver is unable to directly decode a semantic message, the message could be correctly inferred given a set of logical relations, thus leading to possibilities of lossless semantic compression with fewer encoded bits. In that regard, Choi \etal \cite{choi2022unified} further investigate the uncertainty and quantification of KB (denoted as $\mathcal{K}$) by using practical logic programming language, ProbLog, in which $\mathcal{K}$ consists of some logical clauses and each logical clause is annotated with a probability that indicates the degree of belief in the clause. Specifically, they define the \emph {knowledge entropy} of a $\mathcal{K}$ as the uncertainty of answers it computes. As for a query $q$, a probability $p_q$ is computed with respect to $\mathcal{K}$ following the semantics of ProbLog (\ie, $\mathcal{K} \models q$). Therefore, the knowledge entropy is expressed as the average semantic entropy of query (message) $q$ computable from the KB $\mathcal{K}$ \cite{choi2022unified}. Mathematically,
\begin{equation}
    \label{equ:SeEntropy_Knowledge}
    H(\mathcal{K}) = \frac{1}{\vert \mathcal{H_K}\vert} \sum\nolimits_{q\in \mathcal{H_K}} - H_\mathcal{K}(q),
\end{equation}
where $H_\mathcal{K}(q) = [{p_q}\log {p_q} + (1 - {p_q})\log (1 - {p_q})]$, and $\mathcal{H}_\mathcal{K}$ denotes the set of the terms which are the heads of all clauses in $\mathcal{K}$. For instance, suppose a knowledge base $\mathcal{K}$ with $\mathcal{H}_{\mathcal{K}} = \{a,b\}$. Accordingly, the possible queries that $\mathcal{K}$ can answer are $a$ and $b$. Besides, assume that $p_a = 0.2$, $p_b = 0.3$, and $p(b \to a) = 0.5$ (where $\to $ denotes ``implies''). Therefore, the probability can be deduced as $p(\mathcal{K} \models  a) = 1-(1-p_a)(1-p_b\times p(b \to a) ) = 0.32$ and $p(\mathcal{K} \models b) = 0.3$. Correspondingly, $H(\mathcal{K}) = \frac{1}{2}(H_\mathcal{K}(a) + H_\mathcal{K}(b))  \approx 0.839 $. It can be observed that different from the logical probability that can be deduced from implicit reasoning rules, the probability in ProbLog is determined by the explicit rules defined in the knowledge base. Finally, (\ref{equ:SeEntropy_Knowledge})  implies that in order to reduce the uncertainty of KB, a timely exchange and updating of the KB plays a critical role, and eventually contributes to less decoding uncertainty and more precise semantic transmission.

\subsubsection{Semantic Channel Capacity}
\label{sec:SemanticCoding}

In this subsection, we present the related results on semantic channel capacity, and discuss the theoretical insights on semantic source coding and channel coding. As discussed in Section \ref{sec:component_semcom}, the received message in SemCom may simultaneously suffer from physical noise and semantic mismatch \& ambiguity, which leads to semantic infidelity. In other words, conventional reliable communications are concerned about the syntactic preservation of the message, whilst SemCom focuses on the semantic similarity between the input and output messages. In this sense, not all syntactic errors will lead to semantic errors due to the presence of semantic redundancy \cite{bao2011towards}, while considering the possible incorrect inferring rules and unaligned KBs, a syntactic error-free process does not always guarantee perfect semantic decoding. 
To quantify the amount of information flow like in classical communication scheme, the semantic channel capacity $C$ for discrete memoryless channels can be developed in \cite{bao2011towards} as
\begin{equation}
    \label{eq7}
    C = \mathop {\sup }\limits_{P(X|S)} \{ I(X;Y) - H(S|X) + \overline {{H_s}(Y)}\},
\end{equation}
where $I(X;Y)$ is the mutual information between the set of transmitted messages $X$ and received messages $Y$. $H(S|X)$ measures the semantic ambiguity as mentioned above; $\overline {{H_s}(Y)}$ denotes the average semantic entropy of received symbol $Y$, i.e., $\overline {{H_s}(Y)} =  - \sum\nolimits_y {p(y){H_s}(y)}$, where $p(y)$is the distribution of received $y$. Compared with Shannon theory, which defines the channel capacity as the supremum of mutual information between the channel input $X$ and channel output $Y$, i.e., $C = \sup \{ I(X,Y)\} $, \eqref{eq7} states that there is a channel coding strategy through which the maximal probability of semantic error can be arbitrarily reduced within the semantic channel capacity. Notably, in SemCom, since both the encoding process and decoding process are dependent on inference rules and related local knowledge, the physical channel is no longer the only factor in shaping the semantic channel capacity. Instead, the relationship between ``semantic channel capacity'' and the standard ``channel capacity'' ($C=\sup{I(X,Y)}$) becomes intricate, as the exact value of $H(S|X)$ or $\overline {H_s(Y)}$ in \eqref{eq7} is governed by the practically implemented semantic inference rules and the shared knowledge base. It is also shown in \cite{d2011quantifying} that a semantic encoder with low semantic ambiguity and a semantic decoder with strong inference capability, together with the help of a large shared KB, can jointly contribute to a better and precise SemCom system.

\emph{Semantic source coding}, which is the process of mapping the observation $s$ into a symbolic representation (\ie, $x \in{\mathcal{X}}$), can be essentially formulated as a conditional distribution $P(x|s)$. Correspondingly, $P(x) = \sum\nolimits_{s \in {\mathcal{S}}} {\mu (s)P(x|s)}$. \cite{bao2011towards} points out that for a given semantic source, the source semantic entropy $H(S)$ and message entropy $H(X)$ satisfy
\begin{equation}\label{eq6}
    \begin{array}{l}
    H(X) = H(S) + H(X|S) - H(S|X),
    \end{array}
\end{equation}
where $H(X|S)$ and $H(S|X)$ measure the semantic coding redundancy and semantic ambiguity respectively. \eqref{eq6} provides a general view of semantic source coding. It can be first concluded that depending on the coding redundancy and semantic ambiguity, message entropy can be larger or smaller than the source semantic entropy. Further, 
for a given semantic source $S$ with fixed coding redundancy $H(X|S)$, as we use more language symbols to describe, the message entropy tends to increase while the coding ambiguity decreases accordingly.

\begin{figure}[t]
    \centering
    \setlength{\abovecaptionskip}{-0.2cm}
    \includegraphics[width=8.6cm]{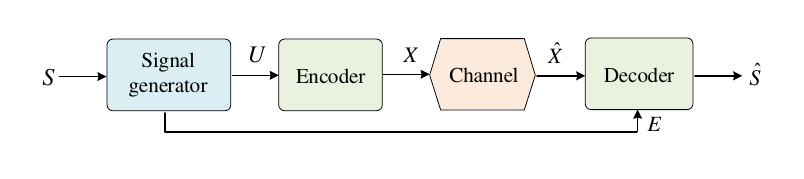}\\
    \caption{A SemCom model consists of a signal generator of semantic information source $S$, an encoder transforming $U$ to $X$, a channel $P(\hat X|X)$, a decoder to obtain $\hat S$ from the received $\hat X$ as well as the side information $E$.}
    \label{side_info}
    \vspace{0cm}
\end{figure}

In order to quantify the beneficial side information (\ie, background knowledge and/or information obtained from previous communications), Xiao \etal \cite{xiao2022rate} propose to extend \eqref{eq6} to characterize the positive impact of side information on decoding. Beforehand, the authors \cite{xiao2022rate} give a showcase of side information by referring to a smart factory or smart city scenario, wherein a simple instruction sent by the controller may involve complex interactions between system components, and the side information corresponds to the underlying system structure and possibly related conditions to perform certain functions for individual components. Furthermore, as shown in Fig. \ref{side_info}, the authors \cite{xiao2022rate} assume that the semantic information source $S$ includes some intrinsic states and features unable to be directly obtained by the encoder. Instead, the encoder can only leverage part of intrinsic observations, termed as the input signal $U$; while the decoder can have the privilege to access side information $E$ to assist the decoding from received signals. Therefore, in \cite{xiao2022rate}, \eqref{eq6} is reformulated as
\begin{equation}\label{eq6-ext}
    I(S, U; Z|E)= I(S,U;Z,E) - I(S,U;E),
\end{equation}
where $Z$ is an auxiliary $E$-independent random variable introduced in the Wyner-Ziv \cite{merhav2003joint} coding, denoting the output of a ``test channel''. \eqref{eq6-ext} implies the benefit of side information source under which the uncertainty of semantic source $S$ can be reduced at the semantic decoder side.

Taking into account the unique property of semantics that one semantic concept can correspond to multiple symbolic descriptions, it remains challenging for semantic source coding to determine an appropriate language system (typically those with low $H(X)$) to represent the semantic world. In that regard, \cite{bao2011towards} shows that due to the redundancy of semantic representation $X$, one can always build another language system $X^{\prime}$ that is \emph{equivalent} to the semantic representations but yields lower or equal message entropy. To experimentally approximate the ideal semantic source coding system, \cite{farshbafan2022common} provides a plausible solution by selecting the smallest subset of (learnable) semantic tokens via RL, which exhibits a remarkable encoding efficiency.

Besides semantic source coding, there also emerges significant research interest towards \emph{semantic channel coding} to combat the semantic noise. A desired semantic channel coding strategy aims to eliminate the impact of semantic noise introduced by a physical channel and KB discrepancy. Practical semantic channel coding solutions often borrow ideas from classical channel coding schemes like introducing semantic redundancy, exploiting semantic noise-robust adversarial coding schemes, re-transmissions, semantic-correcting techniques, \etc \cite{hu2022robust,jiang2022deep,zhou2022adaptive,lu2021reinforcement,liang2022life}. It is noteworthy that due to the difficulty of theoretical semantic modeling for practical high-dimensional media, most of the existing semantic coding schemes follow a (typical) JSCC methodology. Meanwhile, it remains difficult to implement modular designs similar to the classical Shannon paradigm, which may be addressed as intensely new interest continues emerging.

\subsection{Semantic Signal Processing}
Based on the aforementioned semantic information measures, which provide valuable insight into understanding why SemCom is feasible, we further discuss the design of semantic signal processing techniques in typical scenarios (\eg, semantic representations, compression, and reasoning).

\subsubsection{Semantic Representations} Information theory-based semantic methods provide rigorous and convincing means of measuring the absolute amount of semantic information. In spite of its promising features, decomposing a general message (\ie, complex text, audio, images) into atomic semantic representations and modeling the complex semantic relations is rather challenging in reality. Till now, most semantic information theories are limited to addressing simple semantic concepts in a low-dimensional scenario. For real semantic representations, measuring the relative semantic information (\ie, semantic similarity) provides another solution. It is also experimented in \cite{lu2021reinforcement} that SemCom can be established from either semantic information rules or maximizing mutual semantic similarity.

Classical semantic similarity measurements are generally based on structured rules. One of the representative approaches is tree or graph-structured semantic representations like WordNet \cite{miller1995wordnet}. The distance of two semantic symbols can be measured by counting the path, depth or edit distance \cite{rada1989development}. Besides building an explicit structured representation, statistical contextual metrics like \emph{term frequency inverse document frequency} (TF-IDF) and the \emph{information content} \cite{jones1972statistical,resnik1995using} counts the similarity from semantic probabilities. In addition to the token's semantic similarity, more advanced approaches address phrase or sentence similarity by considering the contexts since similar semantic meaning often emerges from frequently used words. Some of the representative methods include $n$-gram contexts, customizing strategies for different  syntactic and lexical structures (\eg, verb, noun, synonym), and machine learning-based decomposition like latent semantic analysis (LSA) and latent Dirichlet allocation (LDA) \cite{amir2017sentence,le2018acv,landauer1998introduction,blei2003latent}.

While structured representations provide a plausible solution for simple semantic concepts, it is hard to generalize for complex relations and varied meanings. To address this issue, DL approaches build an automatic representation by exploiting the underlying semantic meanings from a unified semantic latent space \cite{mikolov2013efficient,sarzynska2021detecting,devlin2018bert}. Specifically, by comparing the distance of two semantic tokens or clauses in that latent space, DL-based representative learning methods, which can be interpreted as implicit probability modeling and automatic semantic atom discovery, can easily handle task-specific scenarios with flexible similarity measurements.

\subsubsection{Semantic Compression} The aforementioned semantic information measures pave the way for lossless transmission. While in practical scenarios, semantic compression, which aims to transmit adequate information with certain tolerance of distortion at or below an acceptable level, is also of great importance. \emph{Rate distortion function} belongs to one of the possible solutions to explore semantic compression based on information theory. In particular, for a simplified communication model under certain distortions, the optimization objective is to minimize the transmitted information volume between the semantic source $S$ and the semantic encoder output $X$ (denoted by mutual information $I(S;X)$), while reducing distortion function $d(s,x)$ as much as possible. To enable a reliable transmission under constrained expected distortion, the maximum bit rate can be solved by introducing a Lagrange multiplier $\beta$ and minimizing the following function:
\begin{equation}\label{equ:RD_Optimize}
    \mathcal L[p(s,x)] = I(S;X)) + \beta d(s,x).
\end{equation}
\eqref{equ:RD_Optimize} indicates that a smaller $\beta$ implies a higher compression ratio, and can be typically solved by Blahut–Arimoto algorithm \cite{tishby2000information}.

However, as illustrated in (\ref{equ:RD_Optimize}), the rate distortion theory mainly focuses on the distance between source information $S$ and the encoded $X$, which
ignores the information recovery process at the receiver side. In this case, IB theory \cite{tishby2000information} extends rate distortion and provides theoretical guidance on balancing the distortion rate and compression rate. Typically in \cite{tishby2000information}, IB aims to minimize the mutual information between $S$ and $X$ (which leads to a compression), whilst maximizing the mutual information between $X$ and $\hat S$ (which leads to fidelity). Mathematically,

\begin{equation}\label{equ:IB_Optimize}
    \mathcal L[p(s,x)] = I(S; X) - \beta I(X; \hat S),
\end{equation}
whose optimal solution satisfies \cite{tishby2000information}
\begin{equation}\label{equ:IB_Optimize_1}
    p(x|s) = \frac{p(x)}{Z(s, \beta)} \exp \left[ -\beta \sum\nolimits_{\hat s} \textrm{KL}(p(\hat s|s) || p(\hat s|x) \right],
\end{equation}
where $Z(s,\beta )=\sum\nolimits_x {p(x)\exp ( - \beta \textrm{KL}[p(\hat s|s)||p(\hat s|x)])}$ is the normalized function, and $\textrm{KL}$ denotes the Kullback–Leibler divergence. Actually, \eqref{equ:IB_Optimize} implies the means to combine semantic similarity metrics with the theoretical output. With the tunable parameter $\beta$, one can explore the tradeoff between the preserved semantic information and data compression at different resolutions. Hence, by defining proper distortion metrics, rate-distortion and IB-based loss can be mathematically converted to mean squared error (MSE) and cross-entropy (CE) loss and implemented with or without extra classifiers that model $p(\hat s|s)$ \cite{liu2022task,xie2021deep}.

\subsubsection{Semantic Reasoning} Semantic reasoning is a concept that depicts the process of converting existing semantic cues to logical answers, which is mostly studied on the decoder side. A successful semantic decoder is expected to infer the true meaning from its local knowledge and received messages.

Seo \etal \cite{seo2021semantics} investigate the progressive relationship among \emph{entities}, \emph{observations}, \emph{concepts}, and \emph{semantic symbols}, wherein they propose to model each of the semantic transmission processes (in the semantic flow) with contextual probabilities. 
Under their semantics-native framework \cite{seo2021semantics}, it is proved that the expected semantic coding length is bounded by the frequency of semantic concepts and decoding state; while semantic reasoning is equivalent to establishing a close communication context (termed as the \emph{focal point} in their paper) for semantic agents on both encoder and decoder side. Iterative solutions and optimal transitions towards a semantics-native reasoning system are also involved.

Besides explicit mathematical probability-based representations, \cite{liang2022life,wang2021performance,wang2023knowledge} model the reasoning process as completing a knowledge graph with entities and linked edges. Their experiments demonstrate that semantic contextual information can be successfully inferred with semantic concepts and logical rules, and verify that semantic reasoning is able to fill a few missing entities that can not be directly observed from the sent messages. In parallel, the authors in \cite{choi2022unified} additionally investigate the dynamic update of the reasoning KB, whose results demonstrate that benefiting from a self-updating KB, semantic inference contributes to the uncertainty reduction at the receiver side and even decreases the necessity for information delivery. These approaches can also be viewed as probability-based reasoning like \cite{seo2021semantics,xiao2022rate}, wherein the task-specific probability distributions are now developed from large-scale training data.

\subsection{Discussions}

As the aforementioned semantic information theories contribute to a preliminary yet meaningful exploration of SemCom, they are still evolving and not fully ready to become a complete theory. For example, most of the existing works succeed in modeling the semantic information with a pre-defined, proposition-oriented (and usually assumed to be easy-to-compute) logical probability, under which the basic semantic relationships can be theoretically measured but is hard to get generalized for more complex scenarios in real-life applications. Meanwhile, as a less-investigated yet critical component in the reasoning process, researches on the KB are scattered. A few key questions related to semantic KB, for instance, to what extent a shared KB influences the communication process; how to quantitatively model the semantic flow in partially-shared KBs, remain elusive and deserve more in-depth exploration. Meanwhile, theories towards a general semantic information theory, such as semantic security \& robustness, semantic efficiency \& generalization trade-off, and semantic computability are also expected to become important research topics.

Besides theoretical advances, experimental contributions serve as another strong and prominent driving force in SemCom. Among all techniques, there are encouraging signs that DNN-powered techniques are currently a promising direction and dominating methodology. These approaches take advantage of representative learning and provide a practical solution to measure semantic probability and semantic similarity. Opportunities can also be found in other techniques like JSCC, advanced AI pipelines (RL, adversarial learning, federated learning (FL), \etc.), and game theory, and have demonstrated superior capability in handling a variety of challenging tasks including varying channels, KG based semantic reasoning, semantic compression \cite{xie2021deep, zhou2022adaptive, lu2021reinforcement}, \etc.

It is worth mentioning that advances in theoretical and experimental studies are complementary. Exploration of the experimental methodology provides plausible ways for evaluating and verifying the semantic theory. On the other hand, theoretical advances pave the way for discovering new opportunities for new SemCom systems. Unfortunately, both methodologies progress in a quasi-independent manner in the literature, where theoretical insight is generally a ``later starter'' due to the difficulties in semantic representation. In that regard, despite the initial success of SemCom, the semantic information theory \cite{carnap1952outline, bao2011towards, bar1953semantic, floridi2004outline} is often limited to the measurement of semantic information or uncertainty, yet fails to directly gauge the semantic similarity of messages, with a few exceptions like the adoption of the IB theory \cite{tishby2000information} and its variants into the loss functions to optimize the DNN-enabled SemCom systems \cite{xie2021deep,sana2022learning}. Therefore, on one hand, it implies the necessity to continually explore the theoretical output of SemCom faced with semantic noise, given the apparent benefits of theoretical advances. For example, the integration of ProbLog \cite{nugues2006introduction} and practical SemCom systems bring a substantial performance boost in explicit reasoning models \cite{choi2022unified}. Meanwhile, the analysis of semantic coding and channel capacity also provides additional interpretability for DNN-based coding schemes. On the other hand, in order to avoid the inconsistency between theoretical results and implementation means, it becomes inevitable to incorporate semantic similarity metrics with ease of computation and adoption, which shall be given later in Section \ref{sec:metrics}.

To summarize, SIT tries to quantify how semantic information flows and how it is processed, along with some related features for the deployment purpose. These powerful theoretical tools will guide the design of future communications systems.

\section{Semantic Similarity Metrics, Datasets, and Toolkits}
\label{sec:metrics}
During the establishment of either SL-SemCom or EL-SemCom, the semantic similarity metrics, datasets, and toolkits belong to one of the most important pillars. Basically, attributed to a part of SIT, semantic similarity metrics contribute to evaluating the differences of meaning rather than bits between the transmitted and recovered content, and are largely involved as the optimization objective in the design and measurement of SemCom systems. As for semantic similarity metrics, considering the differences among different types of media (\ie, text, image, and speech), metrics could be rather different, and could be classified dependent on the practical scenarios, such as error-based transmission metrics, data compression metrics (\ie, to measure the reduction ratio of data volume), and goal-oriented effectiveness metrics (\eg, age \& value of information). Similarly, datasets and toolkits can be categorized as SL-SemCom-oriented (\eg, text, images, speech, and videos) and EL-SemCom-oriented. Table \ref{metricsT} summarizes the typical metrics in SemCom. Meanwhile, the datasets and toolkits frequently used in existing SemCom systems are provided in Table \ref{datasets} as well.

\subsection{Semantic-Level Metrics}
\label{sec:semantic-level_metric}
\subsubsection{Text Transmission Similarity}
Besides MSE or CE for training the DNN, some  text similarity metrics can be used to quantify the difference between the transmitted $s$ and received $\hat s$. As a widely-accepted metric for text similarity, the \emph{average semantic distortion} measures the distortion between the transmitted word and the recovered one \cite{guler2014semantic}. Meanwhile, inheriting from measuring the speech-to-text accuracy of automatic speech recognition systems, \emph{word error rate (WER)} \cite{farsad2018deep} is another common metric in this field. Unfortunately, both of them lack the understanding capability of word ambiguity and pay more attention to the distance between individual words.

In order to avoid this word ambiguity issue and further measure the sentence similarity, the frequency of $n$-gram co-occurrence between the transmitted source $s$ and the recovered $\hat s$ is often leveraged. Specifically, the terminology $n$-gram refers to a group of $n$ consecutive words in a sentence. Essentially, a higher \ngram~can reflect more contextual information, while as a special case of \ngram, $1$-gram score totally ignores the contextual semantics and is equivalent to a word-level measurement. Typically, high-order $n$-gram similarities are often adopted in the literature \cite{liu2017improved,lu2021reinforcement} as they reflect more on the semantic contexts.

On top of \ngram, several sentence similarity metrics like bilingual evaluation understudy (BLEU) and consensus-based image description evaluation (CIDEr) are proposed to measure the degree of consistency between the ``candidate messages'' and ground-truth references \cite{xie2021deep,sana2022learning,lu2021reinforcement}. Benefiting from these similarity metrics, the authors in \cite{lu2021reinforcement} further put forward a SemanticRL framework that directly optimizes the semantic distances rather than via word-level supervision.

In particular, BLEU score \cite{papineni2002bleu} is initially proposed for machine translation and has quickly become a dominating method against previous state-of-the-arts at that time. Without loss of generality, let ${l_s}$ and ${l_{\hat s}}$ denote the length of $s$ and $\hat s$. The BLEU score $\Theta _{\text{BLEU}}$ in the logarithmic domain is defined as

\begin{equation}\label{eq8}
    \begin{array}{l}
  {\Theta _{\text{BLEU}}} = \text{BP} \cdot \exp \left(\sum\limits_{n = 1}^4 {{u_n}\ln {p_n}} \right),
    \end{array}
\end{equation}
where ${u_n}$ and ${p_n}$ denote the weight and the frequency of co-occurrence for $n$ consecutive words, respectively. Notably, it is commonly recommended that $n$ is up to $4$ and $u_n$ takes the average value \cite{papineni2002bleu}, \ie, ${u_n}=1/4, \forall n \in \{1,\cdots, 4\}$. 
Besides, in order to avoid the misleading high matching degree under cases where a short candidate sentence is part of the long reference one, \emph{brevity penalty} (BP) is introduced to punish short sentences and can be formulated as
\begin{equation}
    \text{BP} = \begin{cases}
  1,\qquad \  \; \text{ if} \  {l_{\hat s}} > {l_s}; \\
  e^{1 - {l_s}/{l_{\hat s}}},  \text{ if} \ {l_{\hat s}} \le  {l_s},
    \end{cases}
\end{equation}
The BLEU score $\Theta _{\text{BLEU}}$ ranges from 0 and 1, and a higher score indicates larger sentence similarity. Taking an example of the reference sentence ``Going to play basketball in this afternoon?'' and the received sentence ``Going to play basketball this afternoon?'', it is apparent that ${l_s} = 8$ and ${l_{\hat s}} = 7$, thus $\text{BP} = e^{1-\frac{8}{7}} = e^{-1/7}$. As for the calculation of \ngram, for instance, the co-occurrence of 2-gram in these two sentences include ``Going to'', ``to play'', ``play basketball'', ``this afternoon'', thus ${p_2} = 4/6$. Similarly, ${p_1} = 6/7$, ${p_3} = 2/5$, and ${p_4} = 1/4$. Hence, $\text{BLEU} = \text{BP} \cdot \exp \left((\ln {p_1} + \ln {p_2} + \ln {p_3} + \ln {p_4})/4\right) \approx  0.4238$.

On the other hand, the CIDEr score \cite{vedantam2015cider}, which has also proven effective for image captioning task \cite{liu2017improved}, is adopted in \cite{lu2021reinforcement} as a semantic metric for texts. Specifically, in order to calculate the accuracy and diversity of decoded \ngram ~phrases, CIDEr score $\Theta _{\text{CIDEr}}$ measures the cosine similarity of the TF-IDF weights and can be formulated as

\begin{equation}\label{eq10}
    \begin{array}{l}
  {\Theta _{\text{CIDEr}}} = \sum\limits_{n = 1}^4 {{u_n}\frac{{\Vert{g_n}(s) \circ {g_n}(\hat s)\Vert}}{{\Vert {g_n}(s)\Vert  \Vert {g_n}(s)\Vert}}},
    \end{array}
\end{equation}
where $\Vert \cdot \Vert$ denotes an $l_2$-norm and $( \circ )$ denotes the Hadamard product. Besides, the TF-IDF vector $g_n$, corresponding to all $n$-grams of length $n$, indicates the frequency of $n$-gram.

Nevertheless, both BLEU and CIDEr scores are still largely affected by polysemy. In that regard, \cite{xie2021deep} proposes a BERT (bidirectional encoder representation from transformers) \cite{devlin2018bert}-based similarity metric (\ie, BERT-SIM \cite{xie2021deep}) as
\begin{equation}\label{eq11}
    {\Theta _{\text{BERT - SIM}}} = \frac{\Vert {B_\phi }(s) \circ {B_\phi }(\hat{s}) \Vert}{\Vert {B_\phi }(s)\Vert \Vert B_\phi (\hat{s}) \Vert },
\end{equation}
where ${B_\phi }$ is a pre-trained BERT model \cite{devlin2018bert}. Apparently, ${\Theta _{\text{BERT - SIM}}} \in [0,1]$, and an increase in the BERT-SIM implies a more accurate semantic recovery. However, the BERT model consumes significant resources for training and might be difficult to be extended to different tasks. As a remedy, the authors in \cite{jiang2022deep} introduce a transformer-based DNN called \emph{Sim32} to detect the meaning errors in the received sentences, which can tolerate some recovery error of several words as long as the meanings of the context remain unchanged. Sim32 could be effectively adopted in saving transmitting resources, since quite a large number of bit-lossy yet semantics-reserving sentences no longer need to be re-transmitted.

\subsubsection{Image Transmission Similarity}
In the context of image processing, \emph{peak signal-to-noise ratio} (PSNR) \cite{bourtsoulatze2019deep} and \emph{structural similarity index method} (SSIM) \cite{sara2019image} are the most popular metrics for evaluating the image similarity. In particular, PSNR is the most widely adopted metric in early SemCom systems \cite{bourtsoulatze2019deep}. However, due to the existence of estimation absolute errors, PSNR is error-sensitive and sometimes yields perception-inconsistent evaluation results \cite{zhang2015semantic}. As a result, it can not sufficiently incorporate the perception characteristics from the human vision perspective.

On the other hand, belonging to a perception-based model, the SSIM is established on the idea that the pixels have strong inter-dependencies especially when they are spatially close. By leveraging these dependencies that carry information about the structure of the objects in the visual scene, the SSIM score $\Theta_\text{SSIM}(s,\hat s)$ compares \emph{s} and $\hat s$ in terms of the \emph{luminance}, \emph{contrast} and \emph{structure} \cite{sara2019image} at the pixel level, and can be computed as
\begin{equation}\label{eq12}
    \Theta_\text{SSIM}(s,\hat s) = {(\rho_\text{l}(s,\hat s))^{\lambda_1} } \cdot {(\rho_\text{c}(s,\hat s))^{\lambda_2} } \cdot {(\rho_\text{s}(s,\hat s))^{\lambda_3} },
\end{equation}
where the functions $\rho_\text{l}$, $\rho_\text{c}$ and $\rho_\text{s}$ reflect the perceived changes in terms of luminance, contrast, and structural patterns between two images. Besides, $\lambda_1$, $\lambda_2$ and $\lambda_3$ denote the exponential coefficients. However, the performance of SSIM becomes less effective when it is used to rate blurred and noisy images. Henceforth some variants of SSIM, such as 3-SSIM \cite{li2009three} and FSIM \cite{zhang2011fsim} are also proposed.

In order to evaluate the human perception consistency between generated images and natural images, some \emph{distribution divergence}-based metrics like FID (Fr\'echet inception distance) \cite{heusel2017gans} and KID (Kernel inception distance) \cite{binkowski2018demystifying} are proposed by computing the Wasserstein distance \cite{vaserstein1969markov} instead of the MSE (and its variants), as the Wasserstein distance measures the distance between probability distributions on a given metric space, and could capture more information than the sole comparison of specific realizations between $s$ and $\hat s$ in MSE \cite{Soloveitchik2021conditional}. Meanwhile, FID and KID compare the distribution of generated images with the distribution of a set of real images (``ground truth''). More specifically, lower FID proves to correlate well with higher-quality images, while KID measures the maximum mean discrepancy (MMD) in the feature space of a classifier, and a lower KID score implies better sampling quality \cite{huang2021deep}. Both of them are adopted to evaluate the efficiency of semantic exchange. However, based on feature extraction, FID and KID can not fully utilize the spatial relationship between features.

\subsubsection{Speech Transmission Similarity}

Given the similarity between the audio data and the text data, similar metrics have also been applied to measure the accuracy between speech signals. Quite similar to WER in text similarity evaluation, the \emph{character error rate} (CER) metric operates on characters instead of words, so as to evaluate the accuracy of speech recognition \cite{weng2022deep}. On the other hand, in order to measure the semantic discrepancy, the authors in \cite{weng2021semantic} adopt \emph{signal to distortion ratio}  (SDR) \cite{vincent2006performance} as the semantic metric, which is defined on top of MSE and can be formulated as
\begin{equation}\label{eq13}
    \Theta_{\text{SDR}} = 10{\log _{10}}\left(\frac{{\Vert s\Vert^2}}{{\Vert s - \hat s\Vert^2}}\right).
\end{equation}
Notably, as implied in \eqref{eq13}, SDR and MSE are correlated and can be easily interpreted from one metric to the other. Specifically, a lower MSE often leads to higher SDR. Besides, as SDR produces clear performance differences, it can be conveniently used to optimize the DNNs.

\begin{figure}[t]
    \centering
    \includegraphics[width=8.8cm]{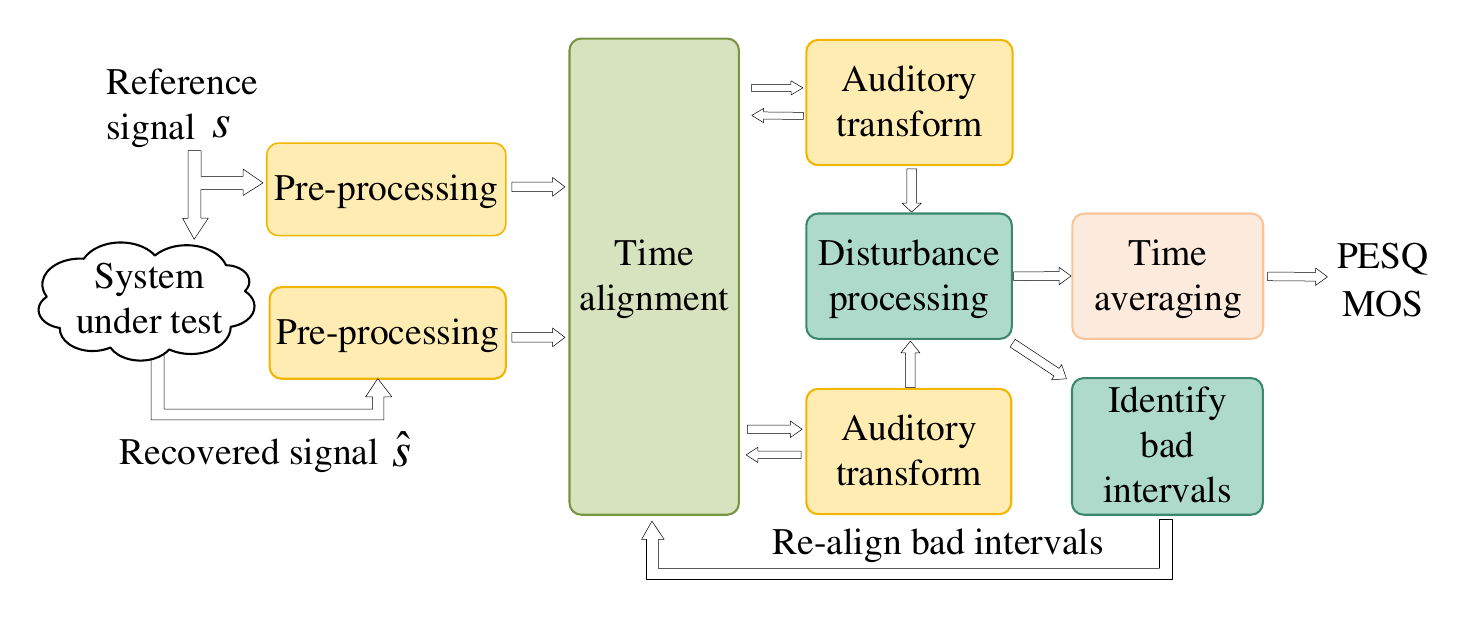}\\
    \centering
    \caption{The procedures of PESQ \cite{loizou2011speech}.}
    \label{PESQ}
\end{figure}

Highlighted by International Telecommunication Union for Telecommunication Standardization Sector (ITU-T) recommendation P.862 \cite{loizou2011speech}, the \emph{perceptual evaluation of speech quality} (PESQ) \cite{rix2001perceptual} takes into account the short memory in human perception and provides a popular tool for assessing speech quality \cite{weng2021semantic}. As shown in Fig. \ref{PESQ}, PESQ first preprocesses the reference source \emph{s} and the reconstructed one $\hat s$, after which the two signals are aligned in time to correct the time delays, and undergoes the auditory transform into a representation of perceived loudness in both time and frequency domain. Then the disturbance in the frequency and time domain is aggregated by the ``disturbance processing'' module, and the output is further mapped to compute the PESQ subjective mean opinion score (MOS) score \cite{rix2001perceptual}. In addition, if the ``time alignment'' module fails to correctly identify a delay change, the related bad intervals will be realigned and the PESQ MOS score will be recalculated. Clearly, in spite of its accuracy, PESQ retains high computational complexity, which on the other hand limits its light-weight applications.

Besides these carefully-designed empirical metrics, the distributions of two speech signals can also be quantified, providing another statistics-related interpolation. For instance, Fr\'echet deep speech distance (FDSD) and  kernel deep speech distance (KDSD) \cite{binkowski2019high} are adopted in the SemCom system \cite{weng2022deep} to assess the speech similarity and evaluate the quality of recovered speech signals. Specifically, a lower FDSD or KDSD score indicates higher similarity between real and synthesized speech sample sequences.

\subsubsection{Ratio of Data Compression}

Considering that SemCom aims to maximize the semantic similarity with as minimal amount of data as possible, the above error-based semantic similarity metrics focus more on the measurement of the semantic similarity, but neglect to measure the data volume before and after semantic transmission. As implied in \eqref{equ:RD_Optimize}, a higher compression ratio means lower distortion, but requires corresponding more bandwidth. Therefore, it entails a tradeoff to minimize distortions with the smallest possible bit rate. Hence, besides error-based semantic similarity metrics such as BLEU, PSNR and SDR, subject to a particular semantic loss, it is still essential to design some semantic similarity metrics to measure the reduction of data volume compared to conventional communications.

In the context of text compression, Wang \etal \cite{wang2022performance} propose a metric named \emph{reduction ratio of the number of bits} to compute the ratio of data size between the semantic information and the classical encoded text. Specifically, assuming that each English letter is encoded by $8$ bits and each token includes $10$ letters. Therefore, a token can conventionally be encoded with $80$ bits by classical $16$-dimensional encoding vectors, while the SemCom method can represent the semantics of the original token by a DNN, wherein a decimal is encoded into $32$ bits. Compared with the classical methods, the simulation results demonstrate that their proposed SemCom schemes can actually reduce the data size by up to $41.3\%$. Moreover, Zhou \etal \cite{zhou2022cognitive} adopt a similar metric \emph{total number of bits} to evaluate the compression ratio. Jiang \etal \cite{jiang2022deep} leverage the metric of \emph{average bits} consumed per sentence to measure the efficiency of the SemCom. On the other hand, as for the image compression, Sun \etal \cite{sun2022semantic} adopt the \emph{semantic mutual information} (SMI) and \emph{accuracy} (ACC) simultaneously to evaluate the semantic-level distortion, wherein the SMI is estimated by utilizing CLUB \cite{cheng2020club} due to its excellent accuracy.

\begin{table*}
    \caption{A summary of semantic similarity metrics for SemCom}
    \vspace{-1mm}
    \label{metricsT}
    \rowcolors{1}{white}{slg}

    \begin{tabular}{m{2cm} m{3cm} m{4.5cm} m{4.5cm} m{1.5cm}}
    \toprule
    \textbf {Metric type}  & \textbf {Semantic Similarity Metrics }   & \textbf {Advantages}  & \textbf {Drawbacks}       & \textbf {Ref}   \\
    \midrule
    \cellcolor{white}      & Average Semantic Distortion      & It utilizes the semantic distances based on lexical taxonomies as a distortion measure.   & It only computes the semantic similarity between individual words, and is difficult to calculate for large data sets. & \cite{guler2014semantic}    \\
    ~    & Word Error Rate (WER)       & It can reflect the semantic similarity to a certain extent and is easy to calculate.      & It does not capture the effects of synonyms or semantic similarity.  & \cite{farsad2018deep}       \\
    \cellcolor{white}      & Bilingual Evaluation Understudy (BLEU)   & It considers the linguistic law that semantically similar sentences are consistent in the semantic space.  & It can only compare the differences between words in two sentences rather than their semantic meaning in sentences.   & \cite{xie2021deep}    \\
    ~    & Consensus-based Image Description Evaluation (CIDEr)       & Compared to BLEU, it does not evaluate semantic similarity on the basis of a reference sentence, but a group of sentences with the same meaning. & It concentrates more on the middle part of a sentence, and thus the middle part possesses more $n$-gram weight.       & \cite{lu2021reinforcement}  \\
    \cellcolor{white}      & BERT-based Similarity (BERT-SIM)       & It can explain semantics at the sentence level due to BERT's sensitivity to polysemy.     & It is not easy to generalize the pre-trained model on others.  & \cite{xie2021deep}    \\
    ~    & Peak Signal-to-Noise Ratio (PSNR)      & It is easy to calculate and understand, and can roughly reflect the image similarity.     & It is not always consistent with human perception.       & \cite{zhang2015semantic}    \\
    \cellcolor{white}      & Structural Similarity Index Method (SSIM)      & Compared with PSNR, SSIM is more consistent with human perception in image quality evaluation.  & It reflects a higher evaluation than the actual.   & \cite{sara2019image}  \\
    ~    & Fr\'echet Inception Distance (FID) and Kernel Inception Distance (KID) & It exhibits distinctive robustness to noise.   & Based on feature extraction, the spatial relationship between features cannot be fully utilized.    & \cite{huang2021deep}  \\
    ~ \cellcolor{white}    & Signal to Distortion Ration (SDR)      & It is easy to calculate and can reflect the quality of voice to a certain extent.   & Its evaluation results are sensitive to the volume of audios.  & \cite{weng2021semantic}     \\
    ~    & Perceptual Evaluation of Speech Quality (PESQ)       & Its evaluation is objective and close to human perception.      & It has high computational complexity.      & \cite{weng2021semantic}     \\
    \cellcolor{white}  \multirow{-18}{2cm}{\textbf{Semantic-Level Metrics}} & Semantic Mutual Information (SMI)      & It can reflect the semantic-level distortion.  & SMI needs to be estimated by extra module.   & \cite{sun2022semantic}      \\
    ~    & Reduction ratio of the number of bits  & It can intuitively reflect the reduction of data volume.  & High computational complexity  & \cite{zhou2022cognitive}    \\ \cellcolor{white} &  Total number of bits & It can intuitively reflect the amount of data volume. & High computational complexity  &\cite{wang2022performance}\\
    \midrule
    ~    & Age of information (AoI) and peak age of information (PAoI)      & It can reflect the freshness of information.   & It may misjudge the value of information.  & \cite{yates2021age}   \\
    \cellcolor{white}      & VoI based approach    & It can capture the value of information.       & It is not easy to design the VoI function for some complicated systems.    & \cite{ayan2019age, molin2019scheduling} \\
    \multirow{-5}{2cm}{\textbf{Effectiveness-Level Metrics}}    & Age of Incorrect Information (AoII)    & It combines the age and value of information to reflect the significance of updates.      & The optimal estimation of the penalty function needs to be further investigated.      & \cite{maatouk2022age}       \\
    \bottomrule
    \end{tabular}
\end{table*}

\subsection{Effectiveness-Level Metrics}
\label{sec:task-metrics}
On top of the semantic-level metrics, effectiveness-level metrics are primarily leveraged to evaluate the contribution of information to the task accomplishment.

\subsubsection{Age of Information-based Metrics}
Different from the traditional metrics which ignore the temporal impact of one message, AoI is put forward by describing the time lag between the current time and timestamp of packets \cite{yates2021age}. In this end, the receiver is more interested in the freshness of the received information, so as to avoid repeated transmission of useless information and bandwidth wastage \cite{maatouk2022age}. AoI has been widely used in cyber-physical systems, such as unmanned aerial vehicle (UAV) \cite{abedin2020data,yi2020deep,abd2019deep} and sensor networks~\cite{zheng2019age,ning2020mobile}. Notably, the definition of AoI can be slightly different across tasks. For example, \cite{yates2021age} introduces AoI as the average age of packets during a period, while the \emph{peak age of information} (PAoI) \cite{costa2014age} represents the peak age of packets during a period.

\begin{figure}[t]
    \centering
    \includegraphics[width=8.8cm]{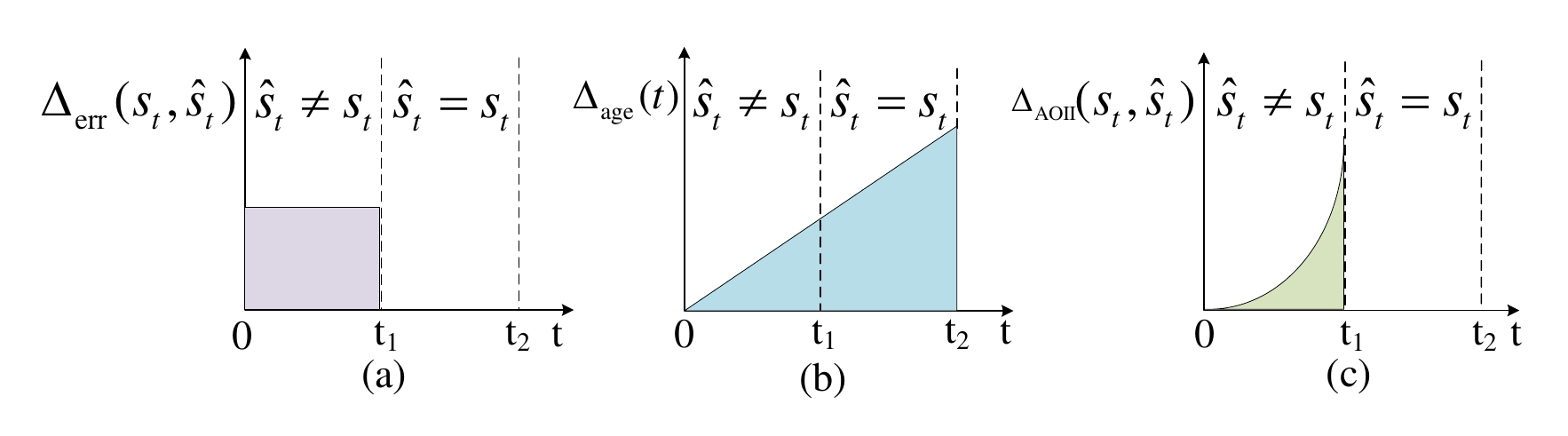}\\
    \centering
    \caption{Example of the different penalty functions: (a) Error-based penalty function, (b) AoI-based penalty function, (c) AoII-based penalty function \cite{maatouk2022age}.}
    \label{AoIVoI}
\end{figure}

Although AoI can indicate the freshness of data in various applications, it also encounters some critical flaws. Taking the example of sensor networks involving temperature monitoring and control \cite{maatouk2022age}, communications could act as an anchor to ensure the controller responds swiftly to any abnormal temperatures. As shown in Fig. \ref{AoIVoI}, intuitively, the error-based metrics are penalized only when $s \ne \hat s$. It can also observed from Fig. \ref{AoIVoI}(b) that although ${s_t} = {\hat s_t}$ in the time interval $t\in [{t_1},{t_2}]$ and the monitor perfectly estimates the actual process, the system still penalizes the AoI metric since messages become less valuable as time goes by \cite{popovski2021internet}. In other words, the AoI-based metrics focus more on the freshness of data and are more suitable for applications with stringent time requirements.

\subsubsection{Value of Information-based Metrics}
Similar to AoI, the exact computation method of \emph{Value of information} (VoI) is task-dependent as well \cite{molin2019scheduling}. However, different from the AoI which is concerned about the freshness but ignores the content itself, VoI measures the contribution of content towards achieving a specific goal, and becomes especially popular in the networked control systems. For example, in \cite{ayan2019age}, VoI is defined to minimize the quadratic norm of estimation errors, and can be regarded as a measure of the uncertainty reduction from the information at the receiver after a successful transmission. Moreover, VoI can be utilized as a part of a prioritizing scheduler \cite{molin2019scheduling} in which the essential information with a larger VoI will be given higher priority.

\subsubsection{Age of Incorrect Information-based Metrics}
In addition to the aforementioned metrics, different SemCom systems also design specific measurements according to different tasks. The \emph{age of incorrect information} (AoII) \cite{maatouk2020age} deals with the shortcomings of the AoI and conventional error-based metrics, as it neatly extends the notions of information updates to ``informative'' updates \cite{maatouk2020age} and focuses on the usefulness to the communication's goal. Afterwards, \cite{maatouk2022age} takes a step forward by further considering the objective of communications towards enabling the SemCom. In this sense, as semantics can reflect the significance and usefulness of information, in order to reflect the effectiveness to this task, AoII is introduced to incorporate the semantics of data more meaningfully. More specifically, AoII measures the captured information by two aspects (\ie, information-penalty and time-penalty function), which can be formulated as ${\Theta _{\text{AoII}}}(t) = \kappa(t) \times \iota({s_t},{\hat s_t})$, where $\kappa(t)$ is a monotonically increasing time penalty function, while $\iota({s_t},{\hat s_t})$ denotes the information penalty function to reflect the difference between the actual data ${s_t}$ and estimate data ${\hat s_t}$ at time $t$. Meanwhile, in different scenarios, the functions $\kappa ( \cdot )$ and $\iota ( \cdot )$ need to fit the interest of applications. As illustrated in Fig. \ref{AoIVoI}(c), in the time interval $[{t_1},{t_2}]$, compared to the error and AoI-based metrics, AoII-based metrics allow to capture a more meaningful purpose of the transmitted data. The authors also show that the metrics adopted in many goal-oriented communications can be regarded as variants of the AoII after tweaking specific parameters \cite{maatouk2022age}.

\subsubsection{Miscellaneous}

In addition, some goal-oriented metrics are also proposed to achieve the prescribed purpose. For instance, in order to evaluate the sample updates, the authors in \cite{kountouris2021semantics} adopt the metrics like real-time reconstruction error and the cost of actuation error. In particular, the reconstruction error reflects the discrepancy in real-time data exchange, while the cost of actuation error captures the significance of the error at the actuation point considering that there may have a mismatch between the original source $s$ and the  reconstructed one $\hat s$. For example, assume that there exist three states for the original source. If $s$ belongs to State $1$, while $\hat s$ is wrongly projected to other states (\eg, State $2$ or State $3$), an actuation error occurs. Besides, compared to State $2$, State $3$ deviates further from State $1$. Therefore, the cost of actuation error for $\hat s = \text{State } 3$ will be larger than that for $\hat s = \text{State } 2$. When the $s$ and $\hat s$ belong to the same state, there is no actuation error \cite{kountouris2021semantics}.

\subsection{Datasets and Toolkits}

\begin{table*}
    \caption{A summary of datasets and toolkits for SemCom.}
    \vspace{-1mm}
    \label{datasets}
    \begin{threeparttable}
    \begin{tabular}{m{1.5cm} m{2cm} m{8cm} m{1.5cm} m{2.5cm}}
        \toprule
        \textbf {Applications}   & \textbf {Datasets}    & \textbf {Descriptions}  & \textbf {Toolkits} & \textbf {Reference Samples}     \\
        \midrule
        \rowcolor{white}   & \multirow{2}{2cm}{Proc. European Parliament\tnote{a}} & \multirow{2}{8cm}{The English version has $2$ million sentences, each with $4$ to $30$ words (or equivalently $53$ million words).}    & TensorFlow   & \cite{farsad2018deep}      \\
        \multirow{7}{1.5cm}{\textbf{Text}}    &      &       & PyTorch      & \cite{xie2021deep,lu2021reinforcement,zhou2022adaptive,jiang2022deep} \\
        \rowcolor{slg} \cellcolor{white}      & Tatoeba Project\tnote{b}      & A large database of sentences and translations (e.g., from English to French).  & $\backslash$       & \cite{sana2022learning}    \\
        & WebNLG English dataset\tnote{c}     & It comprises of sets of triplets describing facts with around $600$ entities and $20$ relations.  & $\backslash$       & \cite{zhou2022cognitive}   \\
        \rowcolor{slg} \cellcolor{white}      & WordNet-WN18\tnote{d}   & The subset of WordNet with $18$ scraped relations for roughly $41,000$ synsets.       & $\backslash$       & \cite{liang2022life}       \\
        & WMT 2018 Chinese-English news track\tnote{e}    & It contains $202,221$ Chinese-English pairs for training while $50,556$ pairs for testing.  & $\backslash$       & \cite{xie2022task}   \\
        \midrule
        \rowcolor{slg} \cellcolor{white} \multirow{3}{1.5cm}{\textbf{Speech}}      & LJSpeech\tnote{f}     & A corpus of English speech at the sampling rate of $22,050$ Hz.   & TensorFlow   & \cite{weng2022deep}  \\
        & Librispeech\tnote{g}  & A speech-to-text library based on public-domain audiobooks containing $960$ hours of speech for training and $2,703$ utterances for testing. & $\backslash$       & \cite{han2022semantic}     \\
        \midrule
        \rowcolor{slg} \cellcolor{white}      &      &       & Tensorflow   & \cite{bourtsoulatze2019deep,kurka2020deep}    \\
        \rowcolor{slg} \cellcolor{white}      &      &       & $\backslash$       & \cite{hu2022robust,yoo2022real,zhang2022deep}       \\
        \rowcolor{slg} \cellcolor{white}      & \multirow{-3}{2cm}{CIFAR-10\tnote{j}}     & \multirow{-3}{*}{It consists of $60,000$ $32\times32$ color images in $10$ classes.}  & TensorFlow, Keras  & \cite{xu2021wireless}      \\
        & MSCOCO\tnote{k}       & A large-scale object detection, segmentation, key-point detection, and captioning dataset $123,287$ images.  & PyTorch      & \cite{zhang2022wireless}   \\
        \rowcolor{slg} \cellcolor{white}      & ADE20K\tnote{l}       & Scene parsing benchmarking dataset with $27,574$ images of $150$ semantic labels.     & PyTorch      & \cite{zhang2022wireless}   \\
        & MNIST\tnote{m}  & A handwritten digit dataset consists of $60,000$ $28\times28$ images.    & $\backslash$       & \cite{zhang2022deep}       \\
        \rowcolor{slg} \cellcolor{white}      & PASCAL-VOC2012\tnote{n}       & A object class recognition dataset  consists of $2,913$ RGB images
        with sizes $513\times513$ in $20$ categories.     & $\backslash$    & \cite{zhang2022deep}     \\
        & CUHK03\tnote{o}       & A pedestrian re-identification dataset including roughly $14,000$ images of over $1400$ pedestrians.       & $\backslash$       & \cite{jankowski2020deep}   \\
        \rowcolor{slg} \cellcolor{white}      & Stanford Online Products\tnote{p}   & It consists of $120,053$ online product images in $22,634$ categories.    & $\backslash$       & \cite{xie2022task}   \\
        \multirow{-15}{1.5cm}{\textbf{Image}}       & CLEVR\tnote{q}  & A diagnostic dataset with $70,000$ images and $699,989$ questions that capably tests a range of visual reasoning abilities.      & $\backslash$       & \cite{xie2022task}   \\
        \midrule
        \rowcolor{slg} \cellcolor{white} \multirow{3}{1.5cm}{\textbf{Video}}  & UCF101\tnote{h}       & It contains $13,000$ clips ($27$ hours) of realistic action videos from YouTube in $101$ action categories.  & PyTorch      & \cite{tung2022deepwive}    \\
        & Vimeo-90k\tnote{i}    & It consists of $89,800$ video clips with a large variety of scenes and actions.       & $\backslash$       & \cite{wang2022wireless}    \\
        \midrule
        \rowcolor{slg} \cellcolor{white} \multirow{4}{1.3cm}{\textbf{UAVs}}  & $\backslash$    & A self-defined 2D $3, 750 \textrm{m} \times 3, 750 \textrm{m}$ grid with $4$ UAV, a server and several users.  & Self-defined       & \cite{yun2021attention}    \\
        & Drone detection dataset\tnote{r}    & It consists of $7$ different types of UAVs, and includes three flight modes (\ie, switched on, hovering, and flying).      & $\backslash$       & \cite{liang2022few}  \\
        \midrule
        \rowcolor{slg} \cellcolor{white} \multirow{3.5}{1.5cm}{\textbf{IoTs}}      & $\backslash$    & The CartPole consists of a cart that can be moved
        to the left or right and a pole positioned vertically above it; while Acrobot is a two-link pendulum that only activates at the second joint. & Gym (\eg, Cartpole and Acrobot)\tnote{s}  & \cite{lotfi2022semantic}       \\
        & $\backslash$    & The IC3Net toolkits include traffic junction and predator prey Environments.    & IC3Net\tnote{t}    & \cite{yuan2021graphcomm}   \\
        \bottomrule
    \end{tabular}
    \begin{tablenotes}\footnotesize
        \item[a] \url{https://www.statmt.org/europarl/}
        \item[b] \url{http://www.manythings.org/anki/}
        \item[c] \url{https://paperswithcode.com/dataset/webnlg}
        \item[d] \url{https://wordnet.princeton.edu/}
        \item[e] \url{https://statmt.org/wmt18/translation-task.html}
        \item[f] \url{https://keithito.com/LJ-Speech-Dataset/}
        \item[g] \url{https://librivox.org/}
        \item[h] \url{https://www.crcv.ucf.edu/data/UCF101.php}
        \item[i] \url{http://toflow.csail.mit.edu/}
        \item[j] \url{http://www.cs.toronto.edu/~kriz/cifar.html}
        \item[k] \url{https://cocodataset.org}
        \item[l]\url{http://groups.csail.mit.edu/vision/datasets/ADE20K/}
        \item[m]\url{http://yann.lecun.com/exdb/mnist/}
        \item[n]\url{http://host.robots.ox.ac.uk/pascal/VOC/voc2012/index.html}
        \item[o]\url{http://www.ee.cuhk.edu.hk/~xgwang/CUHK_identification.html}
        \item[p] \url{https://github.com/rksltnl/Deep-Metric-Learning-CVPR16}
        \item[q]\url{http://cs.stanford.edu/people/jcjohns/clevr/}
        \item[r]\url{https://ieee-dataport.org/open-access/dronedetect-dataset-radio-frequency-dataset-unmanned-aerial-system-uas-signals-machine}
        \item[s]\url{https://github.com/openai/gym}
        \item[t]\url{https://github.com/IC3Net/IC3Net}
        \normalsize
    \end{tablenotes}
    \end{threeparttable}
\end{table*}

On the basis of semantic similarity metrics above, datasets and toolkits, which are typically originated from authoritative institutions and organizations like Google, OpenAI and Facebook, manifest their significance in the development of SemCom as well.

\subsubsection{SL-SemCom Datasets and Toolkits}
As the primary case to demonstrate the potential of SemCom, text-oriented SemCom has rather abundant datasets, which can be categorized into varying-length sentence datasets and KG-based entity-relationship-entity triplets. In particular, the former category includes \emph{Proceedings of the European Parliament} and \emph{WMT 2018 Chinese-English news track} \cite{xie2021deep,lu2021reinforcement,zhou2022adaptive,jiang2022deep,xie2022task}, while the latter encompasses \emph{WebNLG English dataset} and \emph{WordNet-WN18} \cite{zhou2022cognitive,liang2022life}. Furthermore, since text has a pivotal role in providing training and evaluation labels for multimedia transmission, some cross-modal datasets are applied in SemCom as well. For instance, LJSpeech belongs to a popular text transcription dataset, and is particularly suitable for testifying voice speech recognition and synthesis results \cite{weng2022deep}. Meanwhile, Librispeech, which is derived from read audiobooks and provides a corpus of approximately $1,000$ hours of read English speech, has been leveraged to examine the efficacy of speech transmission. On the other hand, the availability of image datasets like CIFAR-10 and MNIST greatly promotes the development of AI and facilitates the verification of semantic image transmission  \cite{bourtsoulatze2019deep,kurka2020deep,hu2022robust,yoo2022real,zhang2022deep,xu2021wireless}. Furthermore, some researchers, who apply the semantics from 2D discrete images to enhance video transmission accuracy, leverage UCF101 and Vimeo-90k video datasets to investigate the performance  \cite{tung2022deepwive, wang2022wireless}. Finally, SL-SemCom commonly experiments on Python-based toolkits like PyTorch or TensorFlow.

\subsubsection{EL-SemCom Datasets and Toolkits}
EL-SemCom apparently enjoys the applicability of aforementioned text and image datasets and toolkits \cite{xie2022task,liew2022economics,johnson2017clevr}. For instance, on top of the CLEVR dataset \cite{johnson2017clevr}, \cite{xie2022task} designs a visual question answering (VQA) task to investigate visual reasoning abilities in response to a text-based question. In addition, EL-SemCom leverages some well-designed UAV or Internet of things (IoT) experimental toolkits as well, and typical examples include the OpenAI Gym toolkit and the IC3Net toolkit. Specifically, the former toolkit, which provides a standard API to communicate between learning algorithms and environments, as well as a standard set of environments (\eg, CartPole and Acrobot) compliant with that API, is becoming the de-facto scenario for evaluating RL algorithms \cite{lotfi2022semantic}. Meanwhile, the latter toolkit includes traffic-junction and predator-prey environments \cite{yuan2021graphcomm}, and benefits the study on learning when to communicate at scale in multi-agent cooperative and competitive tasks. Besides, some propriety datasets on UAVs and IoTs start to emerge, and the drone detection dataset \cite{liang2022few} can be categorized into this scope. In particular, it consists of $7$ different types of UAVs with three flight modes (\ie, switched on, hovering, and flying) captured.

To sum up, there is no doubt that datasets play an essential role in boosting the development of AI and DL, and act as an important pillar to support the validation of techniques in SemCom. Additionally, toolkits significantly facilitate the fostering and reproduction of interesting ideas. Together with semantic similarity metrics, the datasets and toolkits lay the foundation of SemCom.

\section{Toward Semantic Level SemCom}
\label{sec:sl_SemCom}
The design of models in SL-SemCom is mainly explored from two aspects (i.e., content semantics and channel semantics). For the sake of clarity, we will begin with some preliminary works, i.e., DL-based end-to-end communications, which treat the entire communication system as an end-to-end reconstruction task \cite{qin2019deep}. Furthermore, most of the existing SL-SemCom frameworks explore and extract the semantic information in an implicit reasoning process by a pre-trained model within a given KB, while some schemes explicitly reason part of hidden semantics that cannot be directly observed in the data by introducing the reasoning mechanisms, such as KG-based inference rules and probability theory-based semantic reasoning rules. As such, we will then introduce these frameworks from the perspectives of implicit reasoning and explicit reasoning. Finally, we provide an overview of channel semantics inference and exploitation, while optimization techniques around the channel semantics are also discussed.

\begin{table*}
    \centering
    \caption{A summary of techniques and applications for SemCom}
    \label{techniques}
    \rowcolors{1}{white}{slg}
    \begin{tabularx}{\linewidth}{m{2cm} m{3cm} m{3cm} m{3cm} m{1.5cm} m{1.5cm} m{1.201cm}}
    \toprule
    
    \textbf {Communication Level}       & \textbf {Techniques}   & \textbf {Applications}      & \textbf {Contributions}      & \textbf {Objective Function}       & \textbf {Semantic Similarity Metrics}     & \textbf{References}    \\
    \midrule
    \cellcolor{white}     & DL, BLSTM, JSCC  & Text transmission     & First implementation of the JSCC for text; Lower WER; Preservation of semantic information.      & $\backslash$   & WER       & \cite{farsad2018deep}$^{\square}$  \\
    ~    & DL, Transformer, DTL       & Text transmission     & Optimization of the JSCC at the semantic level; Robustness to the varying channel.       & CE and MI      & BLEU, BERT-SIM    & \cite{xie2021deep }$^{\square}$    \\
    \cellcolor{white}     & DL, Transformer, NN compression    & IoTs       & A lite DeepSC; Robustness especially under low SNR.   & CE  & MSE, BLEU   & \cite{xie2020lite}$^{\square}$     \\
    ~    & DL, Universal Transformer, JSCC    & Text transmission     & Enhanced flexibility for semantic coding; Better adaptation to varying channels.   & CE  & SER, BLEU   & \cite{ zhou2021semantic}$^{\square}$     \\
    \cellcolor{white}     & DL, Transformer, JSCC, HARQ  & Text transmission     & Robustness with adaptive bit rate control; Lower communication cost.    & CE  & Average bits for each word, BLEU    & \cite{ zhou2022adaptive}$^{\square}$     \\
    ~~   & DL, Transformer, JSCC  & Text transmission     & Effective SemCom       & CE, KL    & BLEU      & \cite{sana2022learning}$^{\square}$      \\
    \cellcolor{white}     & DRL, JSCC      & Text and image transmission   & Suitability for non-differentiable metrics and channels.    & CE, rewards    & WER, BLEU, CIDEr  & \cite{lu2021reinforcement}$^{\square}$   \\
    ~    & DRL, JSNC      & Text transmission     & Semantic confidence based semantic distillation mechanism.  & MSE, CE and rewards  & WER, BLEU   & \cite{lu2022rethinking}$^{\square}$      \\
    \cellcolor{white}     & DRL, GAN, VGG  & Semantic segmentation and object detection      & A remarkable reduction of data volume with high semantic similarity; Naturalness.  & Rate-semantic-perceptual     & FID, KID, PSNR, SSIM, mIoU, Semantic loss & \cite{huang2022towards}$^{\square}$      \\
    ~    & DL, JSCC, CNN  & Image transmission    & Initial introduction of JSCC to image transmission.   & MSE       & PSNR      & \cite{bourtsoulatze2019deep}$^{\square}$ \\
    \cellcolor{white}     & DL, JSCC, CNN, GDN/IGDN\cite{balle2016density} & Image transmission    & Practical implementation of JSCC to fully exploit channel feedback; Bandwidth saving.    & MSE       & PSNR, Average bandwidth ratio       & \cite{ kurka2020deep}$^{\square}$  \\
    ~    & DL, CNN, ResNet-152, LSTM, JSCC    & Image transmission    & Multi-level semantic extraction.  & MSE       & PSNR, SSIM  & \cite{zhang2022wireless}$^{\square}$     \\
    \cellcolor{white}     & DL, CNN, JSCC  & Image transmission, Digit recognition, Image classification & Adaptive to the observable datasets with high performance.  & MSE, CE, KL divergence       & PSNR, Accuracy    & \cite{zhang2022deep}$^{\square}$   \\
    ~    & DL, CNN, ViT, Residual block       & Image transmission    & Real-time SemCom.      & MSE       & SSIM      & \cite{yoo2022real}$^{\square}$     \\
    \cellcolor{white}     & DL, CNN, Attention mechanism, ResNet     & Telephone systems and multimedia transmission systems       & Higher accuracy of signal recovery with more weights given to essential speech information.      & MSE       & MSE, SDR, PESQ    & \cite{weng2021semantic}$^{\square}$      \\~
    
       & FL, CNN  & Audio      & Application of FL to the edge devices and server to improve audio semantic accuracy.     & Normalized root mean squared error (NRMSE)     & MSE       & \cite{tong2021federated}$^{\square}$     \\
    \cellcolor{white}     & DL, CNN, GRUs  & Speech recognition and synthesis    & Reduction of the volume of transmitted data without performance degradation.  & The connectionist temporal classification (CTC) loss & CER, WER, FDSD, KDSD    & \cite{weng2022deep}$^{\square}$    \\
    \multirow{-55}*{\textbf{SL-SemCom}} \cellcolor{white} & DL, LSTM, VGG, Transformer, GAN    & Speech-to-text and speech-to-speech transmission      & Speech signal recovery by combining text-related features and additional speech-related features with a pre-trained GAN. & CTC loss, CE, MSE    & Bert-SIM, WER     & \cite{han2022semantic}$^{\square}$       \\
    \bottomrule
    \multicolumn{7}{>{\footnotesize\itshape}r}{To be continued on the next page.}
    \end{tabularx}
\end{table*}%
\begin{table*}
    \ContinuedFloat
    \centering
    \caption{A summary of techniques and applications for SemCom (\emph{cont.})}
    \rowcolors{1}{slg}{white}
    \begin{tabularx}{\linewidth}{m{2cm} m{3cm} m{3cm} m{3cm} m{1.5cm} m{1.5cm} m{1.2cm}}
    \toprule
    \textbf {Communication Level}    & \textbf {Techniques}       & \textbf {Applications}    & \textbf {Contributions}     & \textbf {Objective Function}   & \textbf {Semantic Similarity Metrics}  & \textbf{References}       \\
    \midrule
    ~       & DL, Transformer, HARQ      & Text transmission   & Significant reduction in the amount of transmitted data; Decreased sentence error rate.  & MSE, CE      & BLEU, WER, average bits    & \cite{jiang2022deep}$^{\square}$      \\
    \cellcolor{slg}    & DL, CNN, Attention mechanism, IB, DIB    & IoT, Image classification, Multi-view object recognition  & A Flexible control of the communication overhead.   & $\backslash$       & Accuracy, Relevance  & \cite{shao2022task}$^{\square}$       \\
    ~       & DL, JSCC, GDN, IGDN, CNN   & Image transmission  & A channel-wise soft attention mechanism to scaling features according to SNR       & MSE  & PSNR, Storage      & \cite{xu2021wireless}$^{\square}$     \\
    \cellcolor{slg}    & DL, ViT, Codebook    & Image classification      & Higher robustness to semantic noise.    & CE   & Classification accuracy    & \cite{hu2022robust}$^{\square}$       \\
    ~       & DRL, KG   & Text transmission   & A graphical representation of semantic meaning.     & Semantic distance-based rewards      & Loss values, Accuracy      & \cite{xiao2022reasoning}$^{\blacksquare}$   \\
    \cellcolor{slg}    & DL, KG    & Text transmission   & Utilization of the life-long learning to automatically update the reasoning rules.       & Semantic distance-based rewards      & Error rate, Loss values, Accuracy      & \cite{liang2022life}$^{\blacksquare}$       \\
    ~       & KG, DL    & Text transmission   & A simple and general solution with higher compression ratio and reliability of communications. & $\backslash$       & Number of bits, Bert-SIM, BLEU   & \cite{zhou2022cognitive}$^{\blacksquare}$   \\
    \multirow{-25}{*}{\textbf{SL-SemCom}} \cellcolor{slg}    & KB, ProbLog     & Message choice problem    & A probabilistic logic approach to improve the KB.   & Entropy      & $\backslash$       & \cite{choi2022unified}$^{\blacksquare}$     \\
    \midrule
    ~       & DL, JSCC, ResNet-50  & Image retrieval, Edge devices     & Robust JSCC retrieval-oriented image compression.   & CE   & Accuracy, CE loss  & \cite{jankowski2020deep}$^{\square}$  \\
    \cellcolor{slg}    & DL, Transformer      & VQA, Image retrieval, Machine translation     & Multi-user SemCom systems for transmitting both single-modal data and
    multi-modal data.  & CE, MSE   & BLEU, answer accuracy, recall@1   & \cite{xie2021task}$^{\square}$     \\
    ~       & DL, Transformer, Codebook  & VQA, sentiment analysis, Image classification       & A unified multi-task SemCom for multimodal data.    & MSE  & BLEU, Accuracy, PSNR       & \cite{zhang2022unified}$^{\square}$   \\
    \cellcolor{slg}    & DRL, Curriculum learning   & IoTs, Control tasks or automotive production in factories & Lower task execution time; Reduced transmission cost.    & Transmission cost and execution time & Transmission cost and task execution time    & \cite{farshbafan2022common}$^{\square}$     \\
    ~       & DRL, KG, VD based DQN      & IoTs, traffic control     & A significant reduction of transmission delay and energy consumption.  & Rewards      & Energy, Similarities       & \cite{chen2022performance}$^{\blacksquare}$ \\
    \cellcolor{slg}    & DRL, KG   & IoTs, heterogeneous agents  & Semantics-aware collaborative DRL for wireless networks.       & Rewards      & Average return, Maximum average return       & \cite{lotfi2022semantic}$^{\square}$  \\
    ~       & DRL, IB, GIB, GCN    & IoTs, Predator-prey game  & Semantics-empowered multi-agent cooperation to relive the IB.  & Rewards and IB     & Success rate, entropy, reward    & \cite{yuan2021graphcomm}$^{\square}$  \\
    \cellcolor{slg}    & DL  & IoTs,  edge devices       & A trading mechanism with higher utilities for semantic models.   & Rewards      & Average utility of winning sellers, BLEU, Bert-SIM & \cite{liew2022economics}$^{\square}$  \\
    ~       & DRL, QMIX\cite{ rashid2018qmix }, RNN, Attention mechanism & UAVs     & Lower latency; Inter-UAV collision avoidance.       & Temporal difference(TD) error  & Latency, Error rate  & \cite{yun2021attention}$^{\square}$   \\
    \multirow{-35}{*}{\textbf{EL-SemCom}}    \cellcolor{slg} & DL, ResNet-18, SVM, Few-shot learning    & UAVs     & A higher recognition accuracy than other few-shot learning schemes.    & CE and Euclidean distance      & Accuracy     & \cite{liang2022few}$^{\square}$       \\
    \bottomrule
    \multicolumn{7}{>{\footnotesize\itshape}r}{Notations: The superscript ${\blacksquare}$ indicates the learning process is ``explicit''; while the superscript ${\square}$ denotes an ``implicit'' learning process.}
    \end{tabularx}
\end{table*}

\subsection{End-to-End Communications}
As stated in Section II-B, conventional communication systems are divided into multiple independent blocks, each of which is independently optimized. Despite its simplicity of implementation in engineering, such a design is known to be sub-optimal \cite{aoudia2018end}. Benefiting from the rapid development of DL, data-driven methods pave the way to completely interpret the communication system as an end-to-end one, and promise performance improvement in complex communications scenarios that are conventionally difficult to describe with tractable mathematical models.

\begin{figure}[t]
    \centering
    \includegraphics[width=8.7cm]{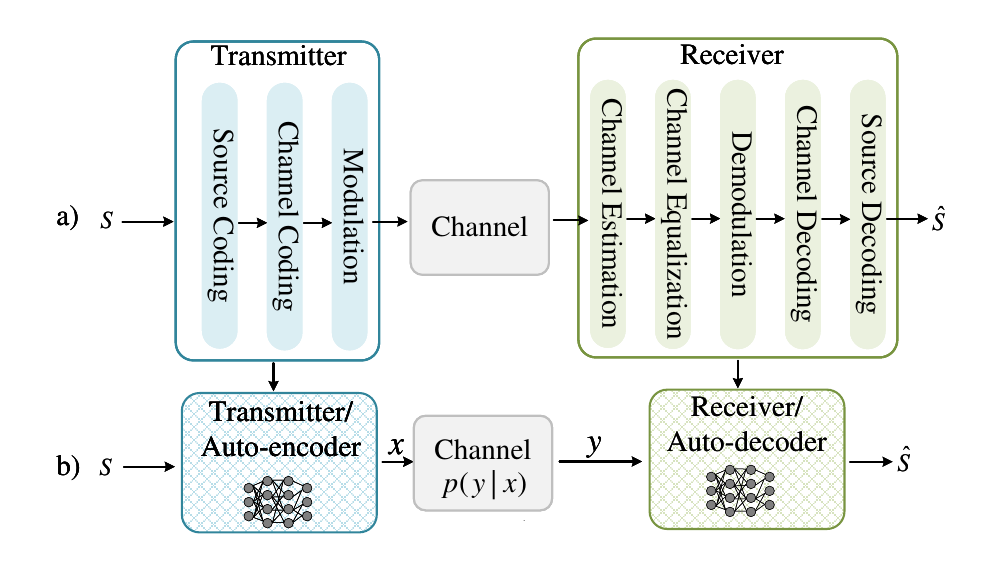}\\
    \centering
    \vspace{-0.5cm}
    \caption{a) The conventional communication system; b) End-to-end communication system \cite{o2017introduction}.}
    \label{endtoend}
\end{figure}

End-to-end communication systems with a given CSI usually assume the AWGN channel model. On top of such an assumption, O'Shea \etal \cite{o2017introduction} first propose to use the autoencoder to implement an end-to-end communication system by replacing the transmitter and receiver with DNNs. Fig. \ref{endtoend} depicts the comparison of conventional and end-to-end communication systems \cite{o2017introduction}, wherein the transmitter and receiver can be implemented by DNNs (\ie, the auto-encoder and auto-decoder respectively). Specifically, the DNNs can be pre-trained from both data and expert knowledge (\eg, the KB) by optimizing the end-to-end loss function (\eg, CE and MSE) in a supervised manner. Afterwards, the auto-encoder-based transmitter removes the redundancy of the source information and converts them into encoded symbols, which are further transmitted through the channel. Based on the received symbols with channel noise, the auto-decoder-based receiver attempts to accurately recover the information. The result \cite{o2017introduction} shows that in terms of the block error rate (BLER), i.e., $ P (s \ne \hat s)$, the end-to-end system can achieve competitive performance in comparison with the conventional communication system employing binary phase-shift keying (BPSK) modulation and a Hamming $(7,4)$ code, and also outperforms the combination of a Hamming $(7,4)$ code with either binary hard-decision decoding or maximum likelihood decoding.

Based on the pioneering work of \cite{o2017introduction}, some variants \cite{o2017deep,erpek2018learning} are proposed to extend and optimize the end-to-end communication system. For example, the authors in \cite{o2017deep} focus on communications under Rayleigh fading channels and extend the use of autoencoders from SISO systems to MIMO systems by designing multi-antenna-oriented DNNs. Erpek \etal \cite{erpek2018learning} study the physical layer scheme for MIMO systems in the presence of annoying interference, and build two autoencoder-based communication systems, which share the same channel and can be jointly optimized to minimize their symbol error rate (SER). These works also verify that based on auto-encoders, the end-to-end systems can obtain significantly superior performance than conventional modulation and coding schemes, which lay the foundation for SemCom. Notably, despite the performance improvement, this end-to-end system is still dedicated to accurately recovering data at the bit level, and there is no significant improvement in data compression and semantic transmission.
\subsection{Implicit Reasoning in SL-SemCom}
\subsubsection{DL-Based Semantic Transmission}
In some sense, the aforementioned DL-based end-to-end communications lay the basic foundation for the development of SemCom, as DL proves itself as an effective tool to improve the reliability of data transmission. Therefore, by shifting from the bit-level accuracy to a semantic-level one, the DL-based semantic transmission also promises a new communication paradigm. In other words, DL can be leveraged in SemCom to extract and reproduce the semantics of information, so as to implicitly understand the semantics rather than decode the received bits. Therefore, basically, the joint optimization of the DL-based encoder and decoder could achieve the end-to-end semantic delivery and reconstruction of the source information. This section will present the latest work in this aspect.

\paragraph{Text Processing}
The rapid advance of natural language processing (NLP) which lays the very foundation for understanding the semantics behind the text, inspires researchers to redesign the transceiver to achieve SemCom. With the aid of a KB and some preliminary information from the CSI, SemCom primarily leverages a semantic encoder to understand and extract feature information from the source information, while the semantic decoder attempts to reconstruct the received semantic symbols into the semantic information as consistent with that from the transmitter as possible. For instance, the words ``three'' and ``3'' share the same meaning at the semantic level, but they are completely different in terms of the encoded bit sequences. Meanwhile, channel coding is also complemented to combat channel impairment. Since the semantic encoder and decoder are commonly implemented in DNNs, an essential part of content-related SemCom becomes to design an appropriate loss function to minimize the semantic error and distortion caused by semantic noise, so as to achieve end-to-end optimization.

LSTM enabled JSCC for text transmission \cite{farsad2018deep}, i.e., DeepNN, belongs to one of the pioneering works, which is capable of preserving semantic information by embedding semantically similar sentences located in a closer semantic space, and recovering the transmitted sentence in terms of the WER rather than the bit error rate (BER). Compared with the traditional end-to-end optimization, \cite{farsad2018deep} shows the great potential of JSCC-based SemCom, especially in scenarios with sentences encoded by a small number of bits. Although the concept of SemCom is not mentioned in \cite{farsad2018deep}, the idea of semantic extraction and preservation has been put forward and greatly inspires later researchers.

\begin{figure}[t]
    \centering
    \includegraphics[width=8.6cm]{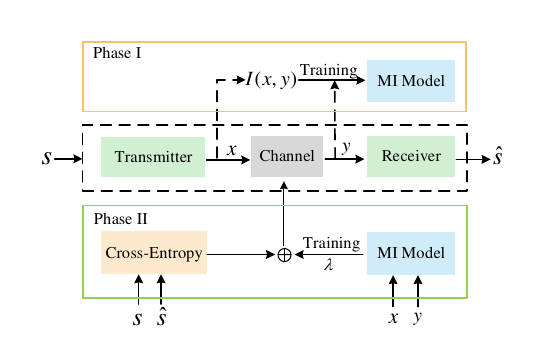}\\
    \caption{The training framework of the DeepSC~\cite{xie2021deep}.}
    \label{DeepSC}
\end{figure}

In order to measure the semantic similarity at the sentence level, Xie \etal \cite{xie2021deep} propose a DL-based SemCom system (\ie, DeepSC), the framework of which is similar to the semantic level of Fig. \ref{three}. Similar to end-to-end communications, the physical-layer blocks in the conventional communication system are merged together. The DNN structure of DeepSC consists of a transformer \cite{vaswani2017attention}-based source encoder to extract the embedding information and dense layers to generate semantic symbols, thus facilitating subsequent transmission. As for the channel environment, the AWGN channel is interpreted as one layer of the pre-trained model. Likewise, the receiver is composed of a channel decoder for symbol detection and a transformer decoder for text estimation. It should be noted that a loss function ${L_{{\rm{total}}}} = {{\rm{CE}}}(s,\hat s) - \gamma I (x,y)$ is first developed in \cite{xie2021deep} to train the whole process, where the first term takes the CE loss between $ s $ and $\hat s $ for minimizing semantic differences. Besides, $\gamma$ is a parameter to balance the relative importance of CE and MI. As depicted in Fig. \ref{DeepSC}, the training framework for DeepSC by stochastic gradient descent (SGD) consists of two phases due to the introduction of the two-term loss function for jointly training the whole DNN. Particularly, Phase I trains the estimation model for MI, while Phase II trains the whole DNN in terms of CE and MI. Moreover, DTL can also be adopted to better adapt to different channel environments. By carefully calibrating the layers of semantic encoders and channels, DeepSC can effectively extract semantic information and ensure reliable transmission. Finally, the simulation results \cite{xie2021deep} show that DeepSC outperforms other conventional communication systems, especially for low signal-to-noise ratio (SNR) scenarios. The BLEU score of DeepSC also converges to about $90\%$, which implies that DeepSC can capture most of the relationship of words and features of the syntax to facilitate the understanding of the texts.

Furthermore, on the basis of DeepSC, some variants are developed. For example, in \cite{xie2020lite}, a lite version ``L-DeepSC'' is studied, in which the authors consider an affordable structure for context transmission in IoT devices by pruning the model redundancy and decreasing the resolution of DNN weights. In order to capture the effects of end-to-end semantic distortion and minimize the errors during reconstructing the sequences, Sana \etal \cite{sana2022learning} propose a novel loss function, which incorporates the MI and Kullback-Leibler divergence and is defined as ${L_{\text{total}}} = I(x,s) - (1 + \alpha )I(x,y) + \beta \text{KL}(y,\hat s)$. Besides, $\alpha$ and $\beta$ are the parameters to reflect the relative importance. The mutual information $I(x,s)$ and $I(x,y)$ is leveraged to ensure compactness and informativeness, respectively. Meanwhile, $\text{KL}(s,\hat s)$ minimizes the semantic distortion between the intended $s$ and the decoded $\hat s$. Another contribution in \cite{sana2022learning} lies in a semantics-adaptive mechanism, which can dynamically adjust the number of symbols per word to better balance the tradeoff between accuracy and complexity. By experiments, for the same accuracy, the adaptive method in \cite{sana2022learning} could dynamically use fewer symbols for each word, thus outperforming the fixed one.

Many existing methods \cite{xie2021deep, xie2020lite} utilize the transformer-based encoder to extract semantic features from the source. However, these methods utilize a fixed transformer and ignore the impact of the difference in length and semantic complexity for each sentence. Therefore, it becomes essential to design a flexible architecture for the semantic encoder to better adapt to different inputs and channel variations. Zhou \etal \cite{zhou2021semantic} adopt a universal transformer (UT) \cite{dehghani2018universal}-based encoder to dynamically process sentences, wherein UT introduces a circulation mechanism to adjust the DNN structure of the transformer and an adaptive computation time (ACT) \cite{graves2016adaptive} to adaptively allocate the number of computational resources to process each symbol. In this way, the semantic system can be more flexible to deal with different input and channel conditions.

Similarly, considering complicated content and channel conditions, the encoder side in \cite{zhou2022adaptive} adopts an adaptive bit rate control mechanism for the transmission of semantic information. Specifically, this bit rate control mechanism consists of two parts, that is, an encoder capable of transmitting sentences at different rates and a policy network to choose the suitable rate for the encoder. In \cite{jiang2022deep}, the transformer-based semantic coding adjusts the code length according to the sentence length, so as to enable semantic coding to outperform other coding methods with fixed length and reduce SER under different channel conditions.

Most of the aforementioned SemCom systems directly use semantic error as the performance metrics but ignore the semantic loss in the process of signal detection. Hence, in order to compensate the semantic loss, Guo \etal \cite{guo2022signal} propose a signal shaping method for SemCom system, which measures the semantic loss during detecting signals by the BERT model \cite{devlin2018bert}. Afterwards, an efficient projected gradient descent method is adopted to solve the signal optimization problem subject to the power constraint.

\paragraph{Image Processing}

Nowadays, the image and video data accounts for 75$\%$ of IP traffic and is more informative than the text \cite{forecast2019cisco}. Hence, it becomes imperative to build an image SemCom (ISC) system that can significantly reduce the amount of data without sacrificing the fidelity of the image. In this part, we focus on the DL-based image compression and transmission for ISC.

\emph{DL-based image compression}: The conventional image compression methods such as JPEG \cite{wallace1991jpeg} and JPEG2000 \cite{rabbani2002overview} rely on its quantization and entropy coder to compress images \cite{cheng2018deep}. However, these methods might be inflexible to encode all types of image content and image formats. By contrast, the DL-based end-to-end compression methods \cite{cheng2018deep,prakash2017semantic,theis2017lossy} exhibit an excellent learning ability wherein their deep autoencoder encodes the images into low dimensional vectors to realize efficient compression. Nevertheless, the legacy image compression approaches mostly focus on pixel-level consistency but neglect the signal distortion caused by the channel transmission on downstream tasks. The latest image compression technology transforms traditional pixel-level reconstruction into a DL-based semantic-level one, which becomes capable to maintain semantic consistency and stably guarantee the performance for downstream tasks.

Many DL solutions have been proposed for image compression. Dependent on the type of utilized DNNs, those solutions can be mainly classified as CNN-based \cite{prakash2017semantic,kong2020spectral}, RNN-based \cite{toderici2017full,punnappurath2019learning}, and GAN-based \cite{rippel2017real,agustsson2019generative,kudo2021gan}.
\begin{itemize}
    \item \emph{CNN-based methods}: CNNs have prime importance in feature extraction and image compression. In \cite{prakash2017semantic}, the authors present a CNN-based method, in which semantically-salient regions are highlighted by an intentional map and encoded at a higher quality than those background regions. The authors in \cite{kong2020spectral} propose a CNN-based end-to-end multi-spectral image compression method for feature partitioned extraction, which consists of two parallel parts (\ie, one for spectral features and the other for spatial features). Then, a rate-distortion optimizer is adopted to make the representation data more compact, and yields superior performance in terms of PSNR than JPEG2000 \cite{kong2020spectral}.
    \item \emph{RNN-based methods}: Commonly, the RNN-based methods \cite{toderici2017full, punnappurath2019learning} also involve some convolutional layers. For instance, Toderici \etal \cite{toderici2017full} present a set of full-resolution lossy image compression methods consisting of an RNN-based encoder and decoder, a binarizer, and LSTM modules for entropy coding. The RNN encoder encodes the residuals between the previous reconstruction and the uncompressed layer, and combines the extracted key features with the output of the previously hidden layer at each recurrent step. The performance of typical RNNs (\eg, LSTM, associative LSTM, a new hybrid DNN of gate recurrent unit (GRU) and residual network (ResNet)) has been validated in \cite{toderici2017full} and indicates fine image compression quality.
    \item \emph{GAN-based methods}: GAN encompasses one generator and one discriminator, the game of which contributes to generating new, synthetic data samples that seem authentic and exhibit many realistic characteristics \cite{jamil2022learning}. Rippel \etal \cite{rippel2017real} introduce GAN to realize a real-time adaptive compression. By adopting the pyramidal decomposition and regularization for feature extraction and processing with adaptive codelength, the codec in \cite{rippel2017real} efficiently produces files $2.5$ times smaller than JPEG and JPEG 2000. Agustsson \etal \cite{agustsson2019generative} also present a full-resolution image compression framework based on GAN operating under low bit rates, and successfully build an extremely-effective image compression system by constraining the application domain to street images. In particular, on the basis of available semantic labels, this system can fully synthesize unimportant regions in decoded images with preserved ingredients such as streets and trees.
\end{itemize}
In addition, researchers also resort to signal processing methods to boost the semantic preserving performance of DNNs \cite{zhou2020multi, li2021learning}. In summary, all these efforts increase the interpretability of image semantics, by learning the natural distribution of images instead of merely local information construction.

\begin{figure*}[t]
    \centering
    \includegraphics[width=\linewidth]{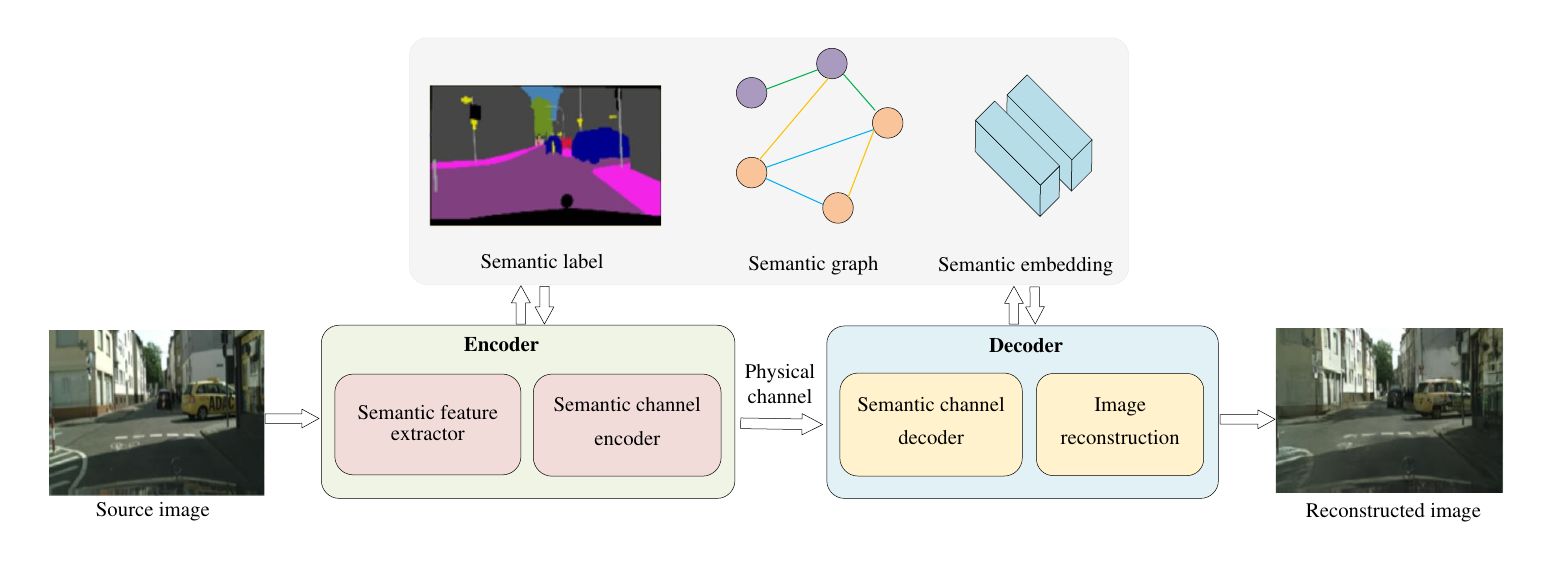}\\
    \centering
    \caption{A general SemCom framework for image transmission.}
    \label{general}
\end{figure*}

\begin{figure}[t]
    \centering
    \includegraphics[width=8.6cm]{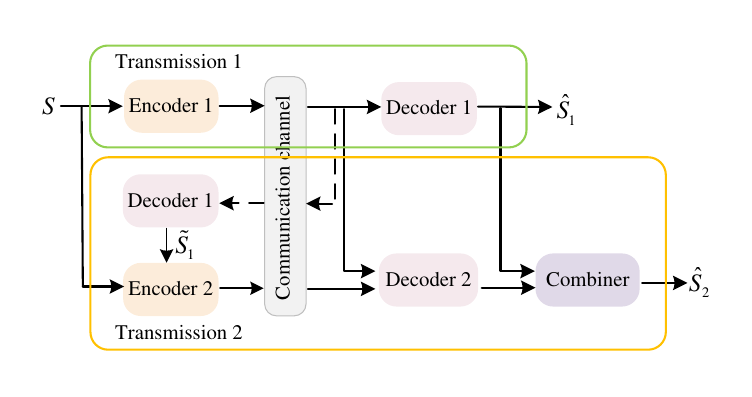}\\
    \caption{Architecture of DeepJSCC-f for the $L$ = 2 case \cite{kurka2020deep}.}
    \label{DeepJSCC-f}
\end{figure}

\emph{SemCom for image transmission}:
Classical methods such as JSCC for image transmission aim to directly map pixel-level values into complex-valued symbols, and jointly learn the encoder and decoder with feedback from the channel. Fig. \ref{general} depicts a general framework for semantic image transmission \cite{zhang2015semantic,sara2019image}, wherein the DL-based semantic encoder first extracts original low-dimensional information as a form of semantic representations (e.g., semantic labels, graph, and embedding) by a joint source-channel optimization. Based on the received semantic signals, the semantic decoder recovers the images according to some semantic similarity metrics such as PSNR and SSIM. Assuming the consistency between the data distribution at the transmitter and the shared KB, Bourtsoulatze \etal \cite{bourtsoulatze2019deep} propose a JSCC technique for wireless image transmission (i.e., deep JSCC) by modeling the encoder and decoder as two CNNs and incorporating the channel in the DNNs as a non-trainable layer, which can significantly outperform the separation-based transmission scheme under both AWGN and Rayleigh fading channels. Kurka \etal \cite{kurka2020deep} introduce an autoencoder-based JSCC with flexibly varying-length coding scheme by utilizing channel feedback, called \emph{DeepJSCC-f}, and adopt CNN-based layered autoencoders encompassing an encoder, a decoder and a combiner. The transmission of each image $s$ is divided into $L$ layers, in which each layer tries to improve the quality of recovered images by additional information of the residual error from the previous layer. Especially, as depicted in Fig. \ref{DeepJSCC-f}, for a $2$-layer DeepJSCC-f, the estimation ${{\tilde s}_1}$ in the previous Layer 1 works as feedback to Layer 2 for obtaining the final recovered image ${\hat s_2}$, so as to reduce construction errors. Furthermore, Kurka \etal re-design a DeepJSCC-l framework \cite{kurka2021bandwidth} for adaptive transmission, in which images are transmitted over multiple channels, allowing a flexible and bandwidth-adaptive transmission. By simulations, DeepJSCC-l leads to superior performance over state-of-the-art progressive transmission schemes in challenging scenarios with low SNRs and limited bandwidth.

Furthermore, Zhang \etal \cite{zhang2022wireless} propose a multi-level semantic feature extractor for image transmission, named MLSC-image, which consists of three feature extraction modules (\ie, a semantic feature module composed of ResNet-152 and LSTM \cite{he2016deep} for text-form semantic features, a pre-trained SPNet-based module \cite{hou2020strip} for high-level segmentation feature, as well as a ``Concat'' module for low-level features such as local details of images). These extracted features are further aggregated and encoded by the joint source and channel encoder into semantic symbols. In addition, the adaptive rate control can also be used for ISC. In that regard, Yang \etal \cite{yang2022deep} propose an SNR-adaptive module and a policy DNN to dynamically adjust the number of active intermediate features over various SNRs. Meanwhile, the authors demonstrate that this scheme can maintain a comparable visual quality compared to a specially trained model for a particular rate.

In order to address the issue of distribution discrepancy between the actually observed data by the transmitter and the shared KB, the authors in \cite{zhang2022deep} develop a semantic coding DNN to learn how to extract and transmit the semantic information using a receiver-leading training process. In particular, on the basis of a data adaptation (DA) DNN, the semantic coding DNN can directly leverage the converted quasi-empirical distribution from the observed data rather than complete retraining.

Instead of sticking to pixel-based metrics (\eg, PSNR and SSIM), Huang \etal present a GAN-based coarse-to-fine ISC for multimedia \cite{huang2021deep} by adopting a base layer and an enhancement layer. Specifically, the base layer of an image is first generated to reconstruct the semantic information while the residuals coded by better portable graphics (BPG) \cite{Bellard2017} further refine details as the enhancement layer. The proposed model can output images visually more similar to the source and significantly outperforms the baselines (\ie, BPG, JPEG2000) in terms of perception metrics such as FID and KID.

There emerge exciting prototype results for ISC as well. For instance, in a field-programmable gate array (FPGA) prototype, Yoo \etal \cite{yoo2022real} demonstrate the feasibility of ISC in a real-time wireless channel with promising results compared to the traditional $256$-QAM in actual wireless environments, especially at low SNRs. Meanwhile, Zhang \etal \cite{zhang2023model} propose a model division multiple access (MDMA) scheme to capably excavate the shared and personalized information for different users. On this basis, shared semantic information is transmitted within the same time-frequency resources, while personalized semantic information is delivered separately. Experimental results show a $5$-dB improvement over the conventional non-orthogonal multiple access (NOMA) scheme \cite{ding2015application} under low SNRs.

\paragraph{Speech Processing}

Unlike text that only contains characters, the speech signal is more complex and difficult to be semantically understood \cite{luo2022semantic}, since the speech content not only involves signal values and frequency, but also its loudness and tone. For example, the text ``come on'' can express a kind of encouragement for someone. Meanwhile, it can also be a modest and negative emotion. In this case, the speech possibly produces more confusing meanings. Besides, dialect also becomes an obstacle to speech recognize \& transmission. Therefore, semantic speech processing puts more emphasis on  technical enhancement to semantic speech transmission, and also sheds light on cross-modal transmission involving both text and speech.

In order to enhance the recovery accuracy at the semantic level, Weng \etal \cite{weng2021semantic} propose a DL-enabled SemCom system for speech, named DeepSC-S. Both the speech encoder and speech decoder in DeepSC-S are based on an attention mechanism employing squeeze-and-excitation (SE) modules, named SE-ResNet, in which one or multiple SE-ResNet modules are sequentially connected to learn attention weights of features by capturing inter-dependencies. \cite{weng2021semantic} shows that in terms of SDR and PESQ, DeepSC-S outperforms the conventional communication systems, especially at the low SNRs. Moreover, FL can also be utilized in designing encoder/decoder \cite{yang2022federated}, so as to mitigate the problem of privacy issues to share training data among devices. For example, Tong \etal \cite{tong2021federated} develop a FL trained model for audio SemCom to reduce overhead between edge devices and the server and improve the accuracy of semantic information extraction. Specifically, in order to collaboratively train the autoencoder over multiple devices and the server, a wave-to-vector framework is devised to extract the semantic information of audio signal and update CNN models locally. Afterwards, the local parameter updates are transmitted to the server at each time step. Finally, the server aggregates the collected local updates into a global model and broadcasts it to all participated devices. With the help of FL, this SemCom framework can not only improve the accuracy of semantic transmission, but also further alleviate the privacy issues associated with the network. Shi \etal \cite{shi2021new} design a first-understand-then-optimization methodology for semantic speech transmission. Besides the typical modules for SemCom, an additional symbol recognition module is added to further reduce errors, so as to largely alleviate the data traffic burden and improve semantic fidelity.


\begin{table}
    \centering
    \caption{Recognized sentences in different systems over Rayleigh channel (for SNR = 4 dB) \cite{weng2022deep}.}
    \label{recognized}
    \begin{tabular}{m{2.3cm} m{5.5cm}}
    \toprule   
    Original Sentence       & ``he concluded that school had nothing to offer him''       \\
    \midrule
    DeepSC-ST \cite{weng2022deep} & ``he concluded that school had noghing to offer him''       \\
    \hline
    Speech Transceiver      & ``it cood sili ite bebou a pims t lup ar of ig mote terigytit w'' \\
    \hline
    Text Transceiver  & ``h ea aahesourhhtntchchen ehaoeitdcofo offer him''   \\
    \bottomrule
    \end{tabular}
\end{table}

DeepSC-ST \cite{weng2022deep} jointly optimizes the tasks of speech recognition and speech synthesis, on top of text transcription and speech synthesis modules. As shown in Table \ref{recognized}, in terms of both speech recognition metrics (\eg, CER and WER) and speech synthesis metrics (\eg, FDSD and KDSD), DeepSC-ST \cite{weng2022deep} yields superior performance than conventional speech transceivers and text transceivers in a real-time human speech input prototype, and also significantly reduces the amount of transmitted data, since it can comprise useful characters in the long speech samples. Similarly, Han \etal \cite{han2022semantic} propose a semantics-oriented communication system for speech-to-text transmission and speech-to-speech transmission, which exploits an attention-based soft module and a redundancy removal module to extract semantics-related features. In particular, a language model-based semantic corrector computes the most matching text transcription while a connectionist temporal classification (CTC)-based speech information extractor obtains the additional semantics-irrelevant but speech-related information like duration, power and pitch.

\subsubsection{DRL-Based Semantic Transmission}

Despite the aforementioned works can successfully extract essential features of data at the semantic level, there also exist some problems that need to be addressed. First and foremost, most of the popular works commonly focus on the DNN design of autoencoder for semantic extraction while using bit-level as objective functions, such as semantics-blind CE or MSE, which introduces an extra ``semantic gap''. On the other hand, most works have now studied the optimization of differentiable objectives, there is little light towards a universal framework that allows the optimization of both differentiable and non-differentiable objectives, which is instead commonly seen in real wireless scenarios. As DRL is an efficient and stable learning for any user-defined rewards and capable to address the non-differentiablity issue, it is attracting increasing interest in SemCom.

To the best of our knowledge, the authors in \cite{ lu2021reinforcement} first introduce the DRL to the SemCom system to preserve the semantic-level accuracy instead of bit-level optimization, called SemanticRL. First, a semantic similarity-oriented JSCC solution (i.e., SemanticRL-JSCC) is designed by calibrating the semantic-level metrics like BLEU and CIDEr, so as to close the ``semantic gap''. Afterwards, the learning process is formulated as an MDP and a self-critic algorithm is leveraged to solve the non-differentiablity issue, wherein the \emph{state} is defined as the recurrent state of decoder combined with historical actions taken so far, and the \emph{action} is the decoding of words from the dictionary. Besides, the \emph{policy} is the probabilistic preference to take an action based on the current state, and the \emph{reward} is the similarity metrics between input sequence $s$ and output sequence $\hat s$. In this way, SemanticRL models the output likelihood as a probabilistic multinomial distribution and continually optimizes the semantic similarity by maximizing the total rewards. Moreover, in order to cope with non-differentiable channels, SemanticRL further regards the encoder and decoder as two independent agents and introduces the DRL into the receiver side. Therefore, in the SemanticRL-based JSCC, the whole learning system turns into a collaborative semantic transceiver. Compared with the CE baseline, SemanticRL yields competitive results in terms of BLEU, BERT-SIM, and CIDEr scores. Meanwhile, as SemanticRL performs better in BLEU ($2$-gram), BLEU ($3$-gram) and BLEU ($4$-gram), but acts slightly poorly in BLEU ($1$-gram), it strongly proves that it pays more attention to the completeness of context and has the ability to preserve semantic information. As for noisy channels, SemanticRL also exhibits strong robustness under varying AWGN and fading channels with SNR randomly fluctuating from $0$ to $20$ dB, especially in low SNRs from $2.5$ dB to $7.5$ dB. The authors \cite{ lu2021reinforcement} also extend this framework to other SemCom tasks like image transmission, which also performs well on the MNIST dataset.

In order to further tackle the impact of time-varying channels on semantic transmission, \cite{lu2022rethinking} also develops a joint DRL-based semantics-noise coding (JSNC) mechanism to adapt to the varying channel and preserve semantics. Specifically, \cite{lu2022rethinking} adopts a confidence-based distillation mechanism at both the encoder and the decoder, which automatically adjusts the depth of semantic representations according to the underlying channel state and sentence structure. As such, only when the semantic confidence reaches a pre-defined threshold, the encoder and decoder could leverage the extracted semantic information for further processing. The simulation results \cite{lu2022rethinking} prove that JSNC outperforms the baselines without a distillation mechanism, especially under low SNRs, and the distillation time also increases along with the sentence's length.

Moreover, Huang \etal \cite{huang2022towards} propose an RL-based adaptive semantic coding (RL-ASC) approach that encodes images beyond the pixel level, by optimizing a triple factor consisting of semantic loss (\ie, mean intersection over union, mIoU), perceptual loss and transmitted bit rate. In particular, the RL-ASC encompasses a semantic encoder, an RL-based semantic bit allocation for adaptive quantization, and a GAN-based semantic decoder. Specifically, the RL-based semantic bit allocation module assigns different quantization levels for different semantic concepts with varying degrees of importance in downstream tasks. In other words, the crucial semantic concepts should be encoded with higher precision so as to incur as little loss as possible, while the task-irrelevant concepts can be relaxed. Finally, the GAN-based semantic decoder can fuse both local and global semantic features via an attention-based module to reconstruct images. In this way, the data volume can be significantly reduced while maintaining a high semantic similarity.

Furthermore, since wireless resources are limited, the base station (BS) may not be able to disseminate the entire semantic information to all user equipment (UE) under stringent latency constraints. Therefore, Wang \etal \cite{wang2022performance} propose a semantics-driven wireless network in which the BS extracts the semantic information from the textual data by a KG \cite{ ji2021survey} and transmits it to attached UEs, while the users recover the original text using a graph-to-text generation model \cite{ koncel2019text}. In order to optimize the \emph{metric of semantic similarity}, which considers the semantic accuracy and semantic completeness, an attention proximal policy optimization (APPO) algorithm is leveraged by evaluating the importance of each triple in the semantic information. On top of that, RL tries to find effective policies for resource allocation and semantics transmission. The experiments show that this approach can reduce the required volume of data transmission by $41.3\%$ meanwhile significantly boosting higher semantic similarity than conventional communications.

In summary, benefiting from some advanced NLP and computer vision (CV) models to extract feature embeddings and model the context, the DRL-based works successfully improve the effectiveness of SemCom. Besides, some DRL-based works start to address semantic and channel non-differentiable issues.

\subsection{Explicit Reasoning in SL-SemCom}

The aforementioned SemCom frameworks implicitly learn from the context based on pre-trained models, and may lack inference and error correction functionalities. However, it is unwise to preclude the context, which usually includes rich information that is hard to be represented and recovered at first glance. As such, explicit reasoning can contribute to the semantic transmission by inferring incomplete information using inference rules at the receiver, and thus effectively guide the task execution. In terms of the underlying methodology, explicit reasoning can be classified as KG-based, information theory-based, and probability theory-based.

\subsubsection{Knowledge Graph-based Semantic Reasoning}
In order to find an appropriate representation of semantic meaning, a graph-based structure emerges as an intuitive way. Consistent with common settings in a graph, the KG consists of some nodes and undirected edges as well, wherein the node and edge can be formally termed as the entity and relationship, respectively. Moreover, the relationship connects distinguishable and independent entities together, and the basic unit of KG is an entity-relationship-entity triplet. Interesting readers could refer to \cite{chen2020review,ji2021survey,yu2022survey} for more aspects (\eg, knowledge graph completion) of the KG as well.

Typically, the KBs derived from real-world facts and grammar can be built in the form of a KG and represented as well-design DNNs. For example, WordNet \cite{miller1995wordnet}, which is one of the most popular lexical KBs, consists of $118,000$ English words and the corresponding relationships. The entities and relationships in a specific KB facilitate the semantic reasoning of implicit meanings. Taking the example of a sentence ``\emph{I want to buy an apple}'', the entity ``\emph{apple}'' may link to some hidden entities such as ``a smartphone brand'' or ``a kind of fruit''. If the context has explained that ``\emph{I want to buy a phone}'', the meaning of ``apple'' can be referred to as the former case. In SemCom, the semantic KBs, which possibly contain some private knowledge for specific individuals contingent on the particular environment or backgrounds, can be shared and updated among transmitters and receivers. Therefore, the receiver in SemCom can better infer the meaning of received messages based on the semantic KBs and other learned inference rules.

Recently, several works \cite{xiao2022reasoning,liang2022life,zhou2022cognitive} leverage KG to extract the essential information and realize the semantic reasoning for SemCom. For example, \cite{xiao2022reasoning} proposes an implicit SemCom to optimize the semantic representation and delivery of messages. To this end, the source signals are first mapped into a semantic KG by an entity detector, and a commonly used graph embedding solution (\ie, TransE) \cite{bordes2013translating} is adopted to obtain the semantic embeddings of entities and relations. Afterwards, a generative imitation learning-based reasoning mechanism is employed at the decoder side to learn and minimize the semantic discrepancy between the possible reasoning path and the expert path. Therefore, the decoder is capable of generating a reasoning path as similar as possible to the expert one. Liang \etal \cite{liang2022life} also utilize a KG-based approach to infer the implicit information in incomplete entities that cannot be directly observed in the received message. In particular, a life-long learning-based updating process is taken into account to automatically update the reasoning rules, so as to improve the interpretation accuracy. Due to the over-simplicity of TransE, more thoughtful graph embedding models, such as TransG \cite{xiao2016transg} and TranSparse \cite{ji2016knowledge}, are leveraged as well.

Different from mapping triples to embeddings \cite{xiao2022reasoning,liang2022life}, Zhou \etal \cite{zhou2022cognitive} propose a cognitive SemCom system, in which the ``cognitive'' characteristic is enabled by the KG. Specifically, the triples are regarded as semantic symbols while the entities or relationships are uniquely mapped as integers. \cite{zhou2022cognitive} demonstrates that this cognitive SemCom framework is capable of capturing and recovering semantic information with reduced errors.

\subsubsection{Information Theory and Probability Theory-based Semantic Reasoning}

Information theory and probability theory are also adopted to implement semantic reasoning in communication systems, in which the entropy and probability are leveraged to express the uncertainty of logical clauses (e.g., rules or facts) in the KB. Choi \etal \cite{choi2022unified} provide a unified and technical approach for SemCom through using a practical ProbLog \cite{nugues2006introduction}. In ProLog, each logical clause is annotated with a probability (by a programmer) that indicates the degree of (the programmer's) belief in the clause. Specifically, the ProLog can learn from the facts to determine the relationship between objects, and then automatically match the target that needs to be queried. Besides, a SemCom layer is additionally introduced on the basis of the reliable communication layer to exchange logically meaningful clauses in the KBs, wherein both layers interact with each other to improve the efficiency of communications.

Furthermore, Seo \etal \cite{seo2021semantics} introduce the probability distribution-based contextual reasoning to semantics-native communications, which can significantly reduce the bit-length of semantic representation with high reliability. In order to improve the interpretability of the data, Thomas \etal \cite{thomas2022neuro} propose a neuro-symbolic AI for learning the causal structure behind the observed data. In particular, a symbolic component characterizes the high-level semantic representations while a neural network (NN) component is responsible for logical deduction and learning. Meanwhile, \cite{thomas2022neuro} shows that it yields significant gain in bandwidth saving for reliable data transmission compared to conventional communications.

In addition, different from most existing works that the encoder and decoder are jointly optimized to maximize the semantic information, Xiao \etal \cite{xiao2022rate} consider a case where the transmitter and the receiver can have different distortion measures, and make rational decisions about their encoding and decoding strategies independently. On basis of this assumption, \cite{xiao2022rate} studies the impact of strategic decision-making on semantic communications and focuses particularly on the case of a transmitter committing to an encoding strategy based on rate-distortion theory. In this regard, recalling the formulation in \eqref{eq6-ext}, a transmitter can have a personal understanding of semantic messages based on its indirect source observations $U$, and the receiver can recover the full semantic information based on its KB and the assistance of side information $E$. To figure out the impact of strategic communications by utilizing the side information $E$, they study three types of equilibrium solutions including optimal Stackelberg equilibrium (OSE) \cite{bacsar1998dynamic}, robust Stackelberg equilibrium (RSE) \cite{chen1972stackelberg} and Nash equilibrium (NE) \cite{crawford1982strategic}. Finally, though different decoding strategies may have different distortion rates, Xiao \etal \cite{xiao2022rate} verify a sufficient condition, committing to which an encoding strategy can always improve distortion performance.

\subsection{Exploitation of Channel Semantics}

Given the wide adoption of JSCC-oriented DNNs in SemCom, it is challenging to independently model the channel layer. Nevertheless, channel semantics, which refers to CSI and critical surrounding environment information, still makes significant senses in SemCom. In this part, we first discuss the inference and exploitation of channel semantics, so as to facilitate the design of end-to-end DNNs in SemCom. Afterwards, given the possible existence of semantic mismatch and ambiguity, which adds to the difficulty to circumvent the semantic noise, we provide some interesting optimization techniques for noise-robustness. Finally, we show how the shift from bit-level accuracy to semantic-level accuracy could motivate the development of semantics-aware transmission/re-transmission techniques as one of the effective means to ensure transmission reliability.

\subsubsection{Channel Semantics Inference}
\label{sec:csi_inference}
In a real communication system, it is often difficult to obtain the distribution of the real channel as the channel impairments such as channel noise and annoying channel-varying properties are usually hard to be modeled mathematically and expressed analytically. However, apart from the widely assumed wireless channels (\eg, Rayleigh fading channel, AWGN), DeepSC \cite{xie2021deep} and MLSC-image \cite{zhang2022wireless} still impose an assumption that the CSI must be accurately given before the training phase to minimize the reconstruction error, though the assumptions might not always hold and possibly lead to biased DNN weights. Hence, researchers shed some attention on learning channel semantics from varying channels, so as to make the SemCom capable of agnostically adapting to varying channel conditions, and the rapid progress in AI greatly accelerates such a process \cite{qin2023generalized,yang2023environment}.

\begin{figure}[t]
    \centering
    \includegraphics[width=8.8cm]{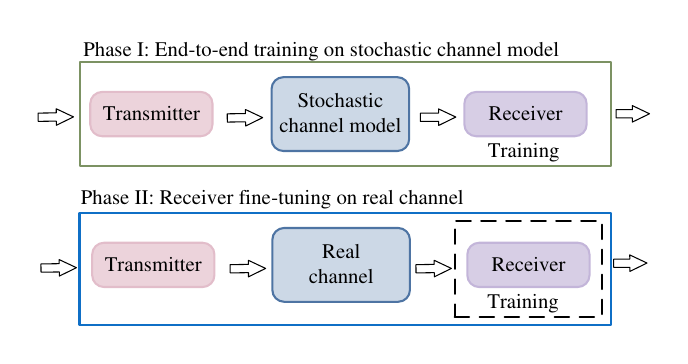}\\
    \caption{ Two-phase training strategy~\cite{dorner2017deep}.}
    \label{twophase}
\end{figure}

\paragraph{Inference from Pilot-Assisted Information}
In order to learn diverse channel effects, Ye \etal propose to approximate the distribution of channels (\ie, $p(Y|X)$) by using a conditional GAN (cGAN) from historical data \cite{ye2018channel}. In particular, \cite{ye2018channel} first adds some pilot information as conditional information for the agnostic channel, and tries to learn a surrogate channel from corresponding outputs \cite{o2019approximating,ye2020deep}. Afterwards, \cite{ye2018channel} takes advantage of an end-to-end learning method to jointly optimize the loss function. Regardless of the inaccuracy of expert knowledge about the channel, the simulation results \cite{ye2018channel} confirm the effectiveness of this method, and show competitive performance compared to those with known channel models under AWGN and Rayleigh channel configurations. On the other hand, inspired by the concept of model-agnostic meta-learning \cite{finn2017model}, \cite{park2020end} proposes to first send multiple pilot packets over varying channel conditions to learn a set of channel models, and then carry out online meta-training on any SGD-based SemCom frameworks with minimal modifications. Similarly, as shown in Fig. \ref{twophase}, \cite{dorner2017deep} introduces a two-stage method \cite{lecun2015deep}, within which DNN-based transmitter and receiver are first trained on available datasets collected from a stochastic channel that resembles the behavior of the expected channel as closely as possible. Afterwards, these DNNs get partly fine-tuned under the real channel, so as to accelerate the training process \cite{dorner2017deep}. Meanwhile, the authors extend this idea towards continuous data transmission entailing the synchronization issue by introducing a DNN-based frame synchronization module. However, due to the existence of differences between the practical channel and the stochastic channel for training, substantial efforts are still required before conducting practical over-the-air transmissions \cite{dorner2017deep}.

Additionally, some researchers \cite{aoudia2019model,raj2018backpropagating} propose to estimate the noisy channel by RL instead of supervised learning, since these RL-based systems cannot depend on any prior knowledge. Taking \cite{aoudia2019model} as an example, the transmitter is considered as an agent while both the channel and the receiver can be regarded as the environment. Furthermore, the key idea turns to approximate the gradient of the loss function with respect to the DNN parameters of the transmitter by regarding the channel input as a random variable. Specifically, in order to circumvent the issue due to the missing channel gradient, the transformer adopts RL to gradually learn how to take actions to minimize the loss function provided by the environment. Besides, the training of the receiver is formulated as a supervised learning task, under the assumption that the receiver has the privilege to access the send messages (\eg, preambles). Hence, by using the policy gradient algorithm, the transmitter can learn without requiring the gradients of the channel. However, some prior information about the channel is still required to achieve a competitive performance \cite{ye2018channel}.

\paragraph{Inference through Environment Perception}
Strong interests are emerging towards exploiting surrounding environment information for channel semantics inference, and preliminary results validate their effectiveness. Specifically, critical scatterers, which could significantly affect signal propagation in the wireless channel, can be located after filtering environment information redundancies, so as to save inference time and communication cost   \cite{wen2023vision, qin2023generalized, yang2023environment}. For example, Wen \etal \cite{wen2023vision} propose a vision-aided detection technique to build keypoint heatmaps. Accordingly, the optimal mmWave beam pair can be conveniently learned by another DNN, thus significantly saving beam selection overhead while boosting the accuracy. Analogously, \cite{yang2023environment} verifies that, the environment information can contribute to predicting beam and blockage situations in an extremely efficient and timely manner even without pilot training or costly beam scanning.

\subsubsection{Semantic Noise-Robustness Optimization}     
Semantic noise can be mainly categorized into semantic mismatch \& ambiguity and physical noise (\eg, interference-induced symbol or bit errors). In order to cope with the former category of semantic noise, it mainly requires a timely update of shared KBs to reduce the mismatch or periodic retraining of the DNNs with data containing semantic noise. For example, Hu \etal \cite{hu2022robust} propose an adversarial training method with weight perturbation and train the DNNs by samples with semantic noise \cite{goodfellow2015explaining}. For the latter, Lu \etal \cite{lu2022rethinking} establish a confidence-based distillation mechanism to continually refine the embedding in the encoder and decoder, which endows the transceiver with a proper semantic extraction means to adapt to the changing sentences and channel condition. Xu \etal \cite{xu2021wireless} propose an attention-based JSCC method for ISC under different SNRs, in which a channel-wise soft attention network is adopted to dynamically adjust the bit allocation ratio between the source coding and the channel coding. More specifically, when the channel turns worse, more bits will be allocated for the channel coding to compensate for channel changes. Otherwise, extra bits will be allocated for the source coding to improve image quality. The results \cite{xu2021wireless} also confirm that the semantic-level features are more robust against noise than pixel-level features.

\subsubsection{Semantics-Aware Re-transmission}
\begin{figure}[t]
    \centering
    \includegraphics[width=\linewidth]{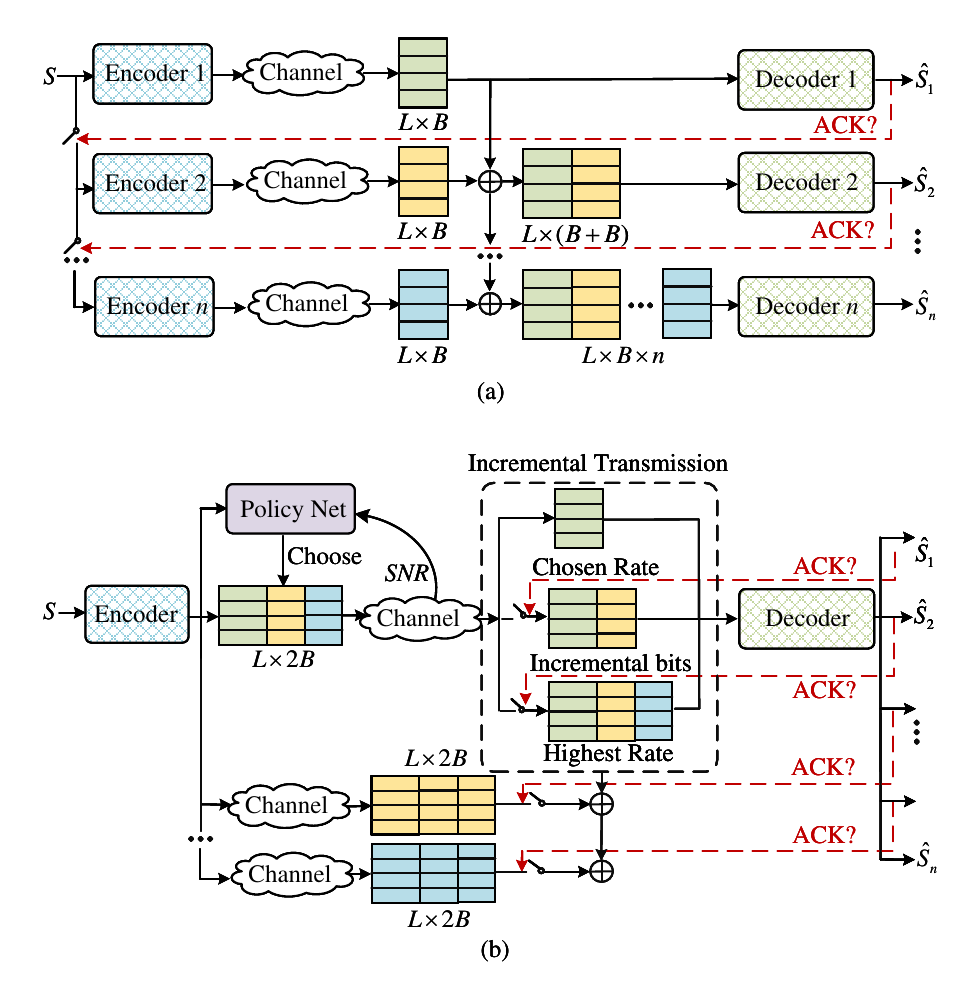}\\
    \caption{The illustration of HARQ enhancements for SemCom: (a) DNN structure of SC-HARQ with the maximum re-transmission number $n$ and a fixed bit rate $B$ \cite{jiang2022deep}, (b) DNN structure of IK-HARQ \cite{zhou2022adaptive} with the peak bit rate $2B$.}
    \label{HARQ}
\end{figure}

Semantics-Aware re-transmission (\eg, Hybrid automatic repeat request [HARQ]), which plays a vital role in guaranteeing reliability in conventional communications, is also re-designed in SemCom to reduce semantic transmission errors. For example, considering the inflexibility of directly adopting fixed-length codeword-based SemCom frameworks to transmit varying-length sentences, Jiang \etal \cite{jiang2022deep} propose an SC-RS-HARQ framework by combining semantic coding with \emph{Reed Solomon} (RS) channel coding and HARQ, and demonstrate the superiority of semantic coding and reliability of conventional methods. In particular, as shown in Fig. \ref{HARQ}(a), they introduce an end-to-end architecture called SC-HARQ with multiple isomorphic semantic coding-based encoders and decoders. When one decoder fails to interpret the correct source message, the same content will be re-transmitted at a fixed bit rate but the previously received bits will be concatenated with the latest received bits and re-used as incremental information for decoding at another decoder. In the SC-HARQ, the process repeats until the receiver successfully recovers the sentence or reaches the maximum number of re-transmissions. Notably, Sim32 is adopted to check the semantic accuracy in the recovered sentences. In this way, the whole system confirms the potential of introducing HARQ to SemCom, and can save bandwidth resources by reducing the number of re-transmission requests for some lossy channels. Different from \cite{jiang2022deep}, which employs multiple encoders and decoders for retransmission at a fixed bit rate, as shown in Fig. \ref{HARQ}(b), Zhou \etal \cite{zhou2022adaptive} exploit an incremental knowledge HARQ-based JSCC scheme, which can work even with one unified single decoder, thus further simplifying the DNN. In addition, Shao \etal \cite{shao2022task} present a selective retransmission-based communication scheme that leverages Tishby's IB principle \cite{tishby2000information} for task-relevant feature extraction and capably identify and skip redundant features to avoid compensating for the potential performance loss due to the variational approximations for IB, thus contributing to further reducing the communication load.   

\subsection{Summary of Lessons Learned}
Recalling the technical progress of SL-SemCom, end-to-end communications play a rather contributing role. In some sense, though semantics has not been taken into account, the successful replacement of conventional communication modules by end-to-end autoencoders significantly promotes the arrival of SL-SemCom. On the other hand, SL-SemCom manifests itself in many aspects. For the content semantics, benefiting from the progress of DL in many fields (\eg, NLP, CV), implicit reasoning has greatly facilitated the semantic transmission of text, images, voice and videos, and yielded fruitful results. Notably, most of the popular semantic transmission designs adopt pre-trained models, and among them, attention-based transformer combined with carefully calibrated loss functions accounts for an overwhelming percentage. Meanwhile, for non-differentiable cases where DL can not be directly applied, DRL sounds like a promising alternative. On the other hand, in order to leverage rich information that is hard to be represented and recovered at first glance, KG-based or probability theory-based explicit reasoning techniques complement the aforementioned implicit reasoning ones, by inferring the incomplete information at the receiver. For channel semantics, existing efforts primarily focus on unveiling CSI and critical surrounding environment information. Meanwhile, the existence of semantic noise and the shift towards semantic-level accuracy requires a re-thinking on noise-robustness optimization and re-transmission techniques, so as to better deal with the channel dynamics. These aforementioned results of SL-SemCom encourage to make bold endeavors towards the effectiveness level of Weaver's vision in Fig. \ref{three}.

\section{Towards Effectiveness-Level SemCom and Networked Systems}
\label{sec:el-SemCom}
As the myriad of autonomous smart devices, such as robots and UAVs, emerge and are empowered with advanced sensing, computing, and learning capabilities, the transmission of a huge amount of data (on the order of zettabytes) possibly congests the networks. For example, a swarm of mobile robots may involve the transmission of 1 GB aggregated data per second for target tracking or collaborative sensing  \cite{kountouris2021semantics}. These communication bottlenecks, if left unresolved, will severely limit the growth and utilization of networked systems \cite{uysal2022semantic}.
In this regard, adopting SemCom as a bridge between multiple agents to communicate and exchange desired information can effectively contribute to successful task execution. Notably, earlier works \cite{juba2011universal,goldreich2012theory} on goal-oriented communications address the issue of potential ``misunderstanding'' among parties, which arises from lack of initial agreement, protocols and languages used in communications due to the lack of consistent semantics \cite{strinati20216g}. Hence, SemCom can also be regarded as a special kind of goal-oriented communications.

Furthermore, for safe and successful task execution, it is crucial to take into account the timeliness and value of information. Therefore, some SemCom works focus on the semantics of specific tasks, and only the semantic information closely relevant to task execution is transmitted for decision-making at the receiver. In other words, in quest of exploring the semantic information at the effectiveness level, EL-SemCom is different from the aforementioned SL-SemCom, and is envisioned at a higher level. In particular, instead of transmitting all the semantics, EL-SemCom attempts to only transmit the essential goal-oriented semantics in a timely manner.

In this section, we first explore the latest advances towards the EL-SemCom. More specifically, we introduce point-to-point and point-to-multi-point task processing, such as MU-DeepSC \cite{xie2021task}, U-DeepSC \cite{zhang2022unified}. Meanwhile, we discuss the exploration of the age and value of semantics, as the semantics might be dynamic and some networked intelligent systems could incorporate time as part of semantics. Afterwards, we discuss some representative semantic networks and applications.

\subsection{Semantics in EL-SemCom}
\label{sec:goal-oriented}
In this part, we begin with the works of EL-SemCom frameworks on single-modal and multi-modal data transmission. Then, we generalize EL-SemCom by extending the application of semantics to effectively facilitate task execution.

For the single-modal EL-SemCom frameworks, Jankowski \etal \cite{jankowski2020deep, jankowski2020wireless} study the image re-identification task for edge devices, by remotely retrieving similar images from an edge server, such as persons or cars captured by other cameras. Correspondingly, a JSCC-based solution significantly increases the end-to-end accuracy and speeds up the encoding process. Yang \etal \cite{yang2021semantic} present a semantic communication paradigm with AI tasks (SC-AIT), in which the functionalities corresponding to the three levels of communications are clearly identified. Specifically, the effectiveness level mainly concerns the source semantics, the desired conduct of the AI tasks and the KB, while the semantic level mainly involves semantic extraction and transmission. Experiments in image classification and surface defect detection have verified the superiority over conventional schemes in terms of classification accuracy and delay. In order to be tightly coupled with downstream tasks, Kang \etal \cite{kang2022task} aim at building an aerial image transmission paradigm under limited computation resources for remote sensing, in which the high-resolution images captured by onboard cameras should be classified immediately. Furthermore, Kang \etal \cite{kang2022task} propose a DRL-based algorithm to exploit the semantic blocks most contributing to the back-end classifier under various channel conditions, thus better balancing the tradeoff between transmission latency and classification accuracy.

On the other hand, the multimodal data transmission has also been investigated under various configurations. Xie \etal \cite{xie2021task} propose a multi-user SemCom system (MU-DeepSC) for the VQA task to improve the answer accuracy, where one user transmits text-based questions about images, while the answered images are transmitted from another user. In particular, the MU-DeepSC transmitter adopts a memory, attention, and composition-based DNN to extract the essential semantic information of initial data from several transmitters, and then the receiver tries to predict the answers by directly merging different semantics, where LSTM and CNN are adopted for text and image transmitters respectively. On top of MU-DeepSC, a transformer-based framework \cite{xie2022task}, which leverages the same DNN structure for the text delivery at the transmitter side but different DNN structures for image retrieval at the receiver side, is presented to cope with various tasks (\eg, machine translation and VQA tasks). Nevertheless, though the DNN structure of the transmitter in \cite{xie2022task} can be shared across different tasks, it remains a non-unified receiver incapably adapting to various tasks. In order to cope with this drawback, Zhang \etal \cite{zhang2022unified} present a unified DL-enabled SemCom system (U-DeepSC) to serve various transmission tasks by unifying both the transmitter and the receiver, which can simultaneously handle five tasks and significantly outperform those task-oriented models designed for a specific task. In addition, to further explore the useful semantic information for executing a certain task, Farshbafan \etal \cite{farshbafan2022common} introduce the concept of beliefs and define a common language between the speaker and the listener, by which the speaker could describe the environmental observations to the listener and thus the listener becomes competent to take corresponding actions. By using a top-down curriculum learning \cite{narvekar2020curriculum} framework based on RL, \cite{farshbafan2022common} simultaneously minimizes the required time and transmission cost of task execution.

\begin{figure}[t]
    \centering
    \includegraphics[width=8.6cm]{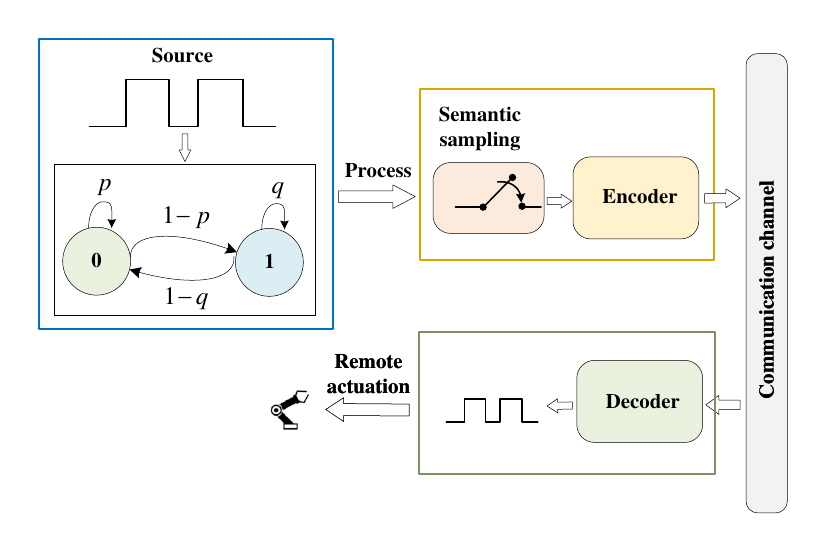}\\
    \caption{An illustration of the goal-oriented end-to-end information sampling and transmission in EL-SemCom \cite{kountouris2021semantics}.}
    \label{setup}
\end{figure}

Moreover, for edge and IoT devices with limited computation resources, some schemes also incorporate the communications between edge devices in the framework of SemCom, so as to realize an end-to-end semantics transmission and enhance communication efficiency. Kountouris \etal \cite{kountouris2021semantics} propose an end-to-end EL-SemCom framework (named E2E Semantics), in which the samples are triggered, generated and transmitted by informative data to control the remote actuator, and the smart devices have the ability to steer their traffic via semantics-aware active sampling. For example, in an end-to-end time-slotted system shown in Fig. \ref{setup}, the device monitors a two-state Markovian source with different transition probability $p$ and $q$, and transmits sampled updates of the source's status to a remote actuator. Then the real-time source reconstruction is performed at the receiver side based on the received updates, to achieve a real-time actuation goal. It is noted that different from the change-triggered sampling policy, this end-to-end semantic policy can simultaneously measure the changes at the source and track the differences between the two ends, thus avoiding some redundant transmissions. As such, the most useful semantics for goal execution is transmitted to control the remote actuator. As for performance evaluation, in the slow varying sources ($p=0.95$, $q=0.9$), the end-to-end semantic policy significantly outperforms the change-triggered policy as it can eliminate the discrepancy quickly. For the rapidly varying sources ($p=0.8$, $q=0.3$), the end-to-end semantics provides a lower actuation error without wasting resources \cite{kountouris2021semantics}.

In addition, some RL-based works \cite{chen2022performance,lotfi2022semantic} have been developed to facilitate the accomplishment of specific goals. The goal-oriented communications can be generalized to multi-agent scenarios, where the agents communicate with each other in an abstract manner over noisy channels while trying to maximize prescribed reward functions. In these problems, besides taking actions and interacting with the environment, all agents transmit and receive semantics. Here, semantics serves as a measure for the usefulness of messages with respect to the goal of data exchange, thus enabling coordination and cooperation among the agents to maximize their accumulated reward. For example, in \cite{chen2022performance}, SemCom is applied to the energy harvesting networks to reduce the transmission delay and energy consumption, where users jointly determine the transmission of partial semantic information and the number of resource blocks used. The authors in \cite{tung2021joint} formulate the multi-agent collaboration with noisy communications problem as a multi-agent MDP, where the objective of communications aims to speed up the accomplishment of a common goal, and adopt a value-decomposition-based DRL. Moreover, the authors in \cite{lotfi2022semantic} develop a semantics-aware collaborative DRL method to enable a group of untrained heterogeneous agents with semantically-linked tasks to collaborate efficiently across a resource-constrained wireless cellular network.

To further explore the relevance and importance between semantic features and corresponding tasks, Liu \etal \cite{liu2022adaptable} develop a framework for task-oriented multi-user SemCom that enables users to extract, compress, and transmit the semantics of the raw data effectively to the edge server. Specifically, \cite{liu2022adaptable} considers a scenario composed of an edge server and a set of users, where the user aims at gathering data locally and sending the extracted semantics to the edge server in a scheduled manner. Besides, the users are prioritized according to the specific service level with distinctive objectives.
Correspondingly, \cite{liu2022adaptable} proposes an adaptable semantic compression approach to compress extracted semantics. Furthermore, \cite{liu2022adaptable} introduces a compression ratio and resource allocation (CRRA) algorithm to obtain the convergent solution
and a CRRA-based dynamic user selection algorithm to handle various service levels. The simulation results \cite{liu2022adaptable} also demonstrate the superiority in reducing the size of transmitted data and successful task execution over the baselines, especially under limited communication
resources.

Similarly, as there exists a tradeoff between the compactness and informativeness (measured in the IB) of messages, the aforementioned work in \cite{yuan2021graphcomm} also applies an attention-based neighbor sampling mechanism to the graph convolutional network (GCN)-based communication (GraphComm) for multi-agent cooperation to relieve the IB. On one hand, the observations and intentions are coded compactly by variational IB (VIB) \cite{alemi2017deep} to extract the semantics as much as possible. Meanwhile, a graph IB (GIB) \cite{wu2020graph} is utilized to avoid sending similar information multiple times and thus saves the bandwidth consumption.

\subsection{Semantics-Assisted Networked Intelligent Systems}

The effectiveness of end-to-end SemCom frameworks promises potential performance improvement for intelligent networks such as industrial IoTs (IIoTs), UAVs systems, autonomous vehicles, and smart healthcare. Next, we will introduce some typical cases of these semantics-assisted networks.

\subsubsection{Industrial IoTs (IIoTs)}

IIoTs such as intelligent transportation \cite{grover2021edge}, smart grids
\cite{wang2021survey}, and industrial automation \cite{liu2019wireless,liu2019real} belong to a generic framework that exploits the abundance of available data generated by a wirelessly connected set of sensors, actuators, and a remote controller to improve the efficiency, reliability \& accuracy of an industrial manufacturing process. Essentially, the networked sensors observe the environment and report the states of interest. Based on the received report, the remote controller on the cloud generates the control signals, and then wireless actuators follow the received signals to take actions.

Since the energy consumption of sensor nodes is mainly due to data transmission, some works \cite{li2021blockchain,mohanty2020deep,shafique2020internet} concentrate on improving the energy efficiency of wireless transmission. Li \etal \cite{li2021blockchain} consider the freshness, relevance, and value of information of transmitted messages and demonstrate their superiority. Meanwhile, as for semantics-aware data processing, the optimal sampling theory for communication and control systems shall be reconsidered from the perspective of data significance. Mohanty \etal \cite{mohanty2020deep} present an RNN-LSTM-based distributed data mining model consisting of RNN and LSTM, which divides the layers of DNNs into separate parts, and deploy them to different sensors. In this sense, information can be exchanged over the network with reduced overhead at the fusion center along with a decrease in data transmission.

On the other hand, as DNN models are usually pre-trained by specific datasets under some particular channels, it becomes inevitable to fine-tune the DNN models to adapt to the changes of KB and channels. However, such a mechanism, which has been extensively discussed in Section \ref{sec:csi_inference}, becomes cumbersome for resource-limited IoT devices. Therefore, inspired by the model trading framework in collaborative edge learning \cite{lim2020federated}, Liew \etal \cite{liew2022economics} adopt a DL-based hierarchical trading system to support the joint trading of semantic model and semantic information, where model providers trade the well-trained models to support devices with limited transmission resources. In particular, a semantic model provider has more resources to train high-quality semantic models with the relevant KBs and channel models, and the edge devices can directly exploit the semantic model to extract semantics. Thus, the semantic information can be collected and traded between devices from the interested information buyers, while ensuring individual rationality and incentive compatibility. In this regard, the proposed method \cite{liew2022economics} contributes to the utility enhancement of devices by semantic model
trading.

\subsubsection{UAV Communications Networks}
UAV systems play an important role in not only working as a stand-alone system in dedicated areas \cite{wan2019toward} but also serving as complementary parts of the cellular networks \cite{mostaani2021task,azari2017ultra,seid2021multi,gu2021energy}.

Stand-alone UAV systems aim to collaboratively accomplish tasks (\eg, providing services for distributed IoT devices) under constraints like low latency, energy consumption, and high reliability. \cite{yun2021attention} introduces a graph attention exchange neural network to satisfy the stringent reliability and latency requirements for real-time air-to-ground communication networks. In particular, each UAV agent swaps attention weights with its neighbors so as to reduce the attention mismatch. Moreover, considering the public safety problems caused by illegal UAVs, by using a specifically designed tri-residual Siamese neural network (TRSN) based on the radio frequency (RF) fingerprints, \cite{liang2022few} proposes two few-shot learning UAV recognition methods (\ie, F-TRSN and SF-TRSN). Specifically, contingent on limited trading samples, F-TRSN directly extracts features of the RF signal, while SF-TRSN utilizes the semantic features of the obtained RF signal by combining the TRSN with a support vector machine (SVM). Notably, outperforming the F-TRSN in terms of UAV recognition accuracy, SF-TRSN further demonstrates that the exploitation of the semantic features provides additional benefit than that of data features only.

On the other hand, the flexibly deployed UAVs can communicate cooperatively to accomplish tasks (e.g., joint rescuing or sensing), and provide emergent ultra-reliable communications to ground users \cite{azari2017ultra} and boost the network performance \cite{seid2021multi,gu2021energy}. Especially, in the latter aspect, it becomes imperative to design appropriate path-planning methods for UAVs. \cite{wan2019toward} develops a DRL-based path-planning method to collect data generated by sensors in a manner of centralized training and decentralized execution (CTDE). In other words, a flying UAV system can be regarded as an implementation of DRL-based EL-SemCom.

\subsubsection{Autonomous Vehicles}
For autonomous vehicles, a self-driving car with no or little human intervention is becoming possible \cite{hakak2022autonomous}. The most crucial module of autonomous vehicles lies in perception, which targets to sense the surrounding environments and extract the useful information for navigation \cite{ren2022collaborative}. Though each vehicle has its individual perception sensors and is fully trained based on large-scale training data, it is still insufficient to meet the high-demand capacity to complement the limited field of view. Hence, the vehicles need to share the collective perception message (CPM) with each other in the same area from a holistic perspective \cite{ren2022collaborative}, which can be regarded as a multi-agent system.

Benefiting from the advance in V2X (Vehicle-to-Everything) communications, vehicles could reliably exchange their messages with their neighbors, thus enhancing the robustness and safety of transportation systems \cite{gyawali2020challenges}. However, sharing the raw sensory data requires huge overhead and might congest the communication network, which is impractical in most cases. Therefore, SemCom can be applied to aggregate the most useful information in an efficient manner. Coincidentally, it is natural to apply SemCom to efficiently aggregate the essential information from neighboring vehicles, which has been widely discussed in some multi-agent collaboration works \cite{liu2020who2com, liu2020when2com, jiang2018learning}. Besides, as SemCom possesses the capability of reasoning, it is also feasible to supplement the missing part of the received information and enhance its robustness. Furthermore, since autonomous vehicles require the real-time information delivery of events, locations, and time, it is significant to introduce the concept of the VoI and AoI to achieve accurate temporal and spatial alignment, such as DiscoNet \cite{li2021learning} and V2VNet \cite{wang2020v2vnet}.

\subsubsection{Smart Healthcare}
Smart healthcare promises a new technological shift toward efficient, convenient, and faster medical services based on AI and Internet of medical things (IoMT), and could significantly benefit the living quality of human beings \cite{sujith2022systematic}. For example, smart healthy monitoring (SHM) \cite{sangra2022energy} could pose many significant advantages, such as preventing some unnecessary visits to hospitals and checking the patient's condition in a real-time manner. Meanwhile, the data generated by SHM and IoMT devices can be analyzed to prevent chronic diseases and fatality of patients \cite{azimi2020data, yue2020deep}.

Considering the battery limitations of wearable sensors and healthcare devices, it belongs to one of the paramount important issues to accomplish efficient services under constrained energy. AI and edge computing techniques can be used for real-time responses and energy consumption minimization. As SemCom supports data filtering by feature extraction, it can help devices only exchange truly meaningful data between devices, thus improving energy efficiency and shortening communication latency.

\subsection{Summary of Lessons Learned}
The effectiveness of executing tasks on the basis of EL-SemCom sounds promising, as the comprehensive results from both single-modal and multi-modal data transmission provide encouraging evidence to facilitate subsequent downstream tasks. Meanwhile, together with the wide adoption of AI (\eg, RL and GCN), incorporating the EL-SemCom at the edge also brings preliminary yet appealing outcomes. On the other hand, as the process of networked intelligent systems entails a timely and effective methodology towards satisfying the urgency and value of task execution, semantics-assisted networked intelligent systems could enhance the capability to effectively accomplish tasks with minimal cost. Nevertheless, despite the research progress, fundamental performance enhancements of distributed edge learning \cite{10024766} are still at their infancy and the application of semantics-assisted networks is still worth further exploration.

\section{Challenges and Open Issues}
\label{sec:challenge}
There is a plethora of challenges that need to be tackled before SemCom can be applied to practical communication scenarios. In this section, we list several key challenges to be further explored in future investigations.
\begin{enumerate}
    \item \emph{Semantic Theory}: In the past few decades, researchers have mostly followed the classic framework of CIT and shed little light on the SIT by extending the logic probability to the scope of semantic entropy. The rationality of quantifying SemCom through semantic entropy and semantic channel capacity remains a fundamental question. In addition, the transmitted symbols in conventional communication systems are assumed to be fixed. However, in SemCom, the flexibility of semantics and the complexity of the languages causes the symbol set to change dynamically and possibly exhibit polysemy. Therefore, how to process and model these dynamic sets is still a problem worth studying. Moreover, though the source-channel separation theorem plays a vital role in reliable communications, the widely adopted JSCC in SemCom makes it interesting to study the theoretical validity of such a separation theorem.

    \item \emph{Semantic Similarity Metrics}: Several works have developed the semantic similarity metrics for different types of sources (\eg, text, image, or speech), but considering its important role in implementing the SemCom, a general semantic similarity metric suitable for various tasks needs further exploration. For instance, in order to determine the loss function and pre-train the parameters of DNNs for diversified tasks, an appropriate semantic similarity metric is essential. Meanwhile, the adopted DNN also needs to be carefully designed to avoid gradient vanishing.

    \item \emph{Real-time Requirement}: In general, the transceiver of SemCom is relatively more complicated, and it is natural to ask whether SemCom can meet the ultra-low latency requirements for future communications. In addition, although semantic reasoning can correct errors in transmission, an extra delay will be introduced in the semantic reasoning process. Meanwhile, the applicability of SemCom to low-cost IoT devices is worthy of further study. Hence, it is of significant importance by developing lightweight algorithms and improving the hardware design.

    \item \emph{Scalability}: SemCom can provide an effective and sustainable service in bandwidth saving and task processing. Besides, the physical layer and application layers are mutually independent in the widely-adopted open-system interconnection (OSI) model, and thus the upper application layer is primarily responsible for the semantic understanding of contents. Nevertheless, though some works for multimodal transmission have proven the effectiveness to process the text and image simultaneously, it is still a huge challenge to deal with more complex data types under the legacy OSI models. In other words, a general semantic-level framework for different types of sources has not been available yet.

    On the other hand, since it is a pre-requisite to share semantic KBs between the transmitter and receiver, the continual update and maintenance of a semantic KB surely involve extra storage costs and algorithm design, thus requiring intensive computational and storage costs. Consequently, how to ensure the scalability of SemCom remains to be addressed.

    \item \emph{Privacy}: As most of the implementation of SemCom depends on AI technologies, especially DL techniques, it poses significant privacy concerns to collect the possibly sensitive user data for the training data or extract the hidden user information of the trained models. In this sense, beyond the capability of ML, we need to figure out the trustworthiness \cite{strobel2022data}. In particular, privacy protection corresponds to two different levels (i.e., data level and model level). As for the former level, simply deleting sensitive features and entities might violate the data integrity and consistency, since the ``missingness'' may reveal some data properties. Hence, data obfuscation, sanitization, and synthesis are often applied to mask, scramble, or overwrite the sensitive information with a realistic fake \cite{jegorova2022survey, liu2021machine}. As for model level, the defense techniques (e.g., model augmenting, differential privacy, FL, and data encryption) can be leveraged to protect existing trained models from leaking sensitive information \cite{dong2023privacy, tian2022comprehensive}.\\
    Furthermore, considering that the correct information recovery in SemCom requires the matching of KBs at both the sender and receiver, from the perspective of privacy protection in SemCom, encrypting the KB provides a viable means to prevent semantic leakage. For example, even though a user successfully eavesdrops on bits sent to others, the user without the required KB can not accurately interpret these bits, thus enhancing privacy and security. From this regard, it is necessary to address the security of KBs and investigate the means to encrypt the semantic messages during transmission.

    \item \emph{Semantic networks}: 6G is anticipated to transform into a sizable and decentralized system, so as to better support intelligent networked systems at different communication levels \cite{meena20226g}. We boldly argue that SemCom can be applied to distributed intelligent networks and even become the dominant communication architecture of 6G with less data transmission and more knowledge exploitation. The embedded generative capability in SemCom pays the way for the orchestration of LLMs in 6G, so as to fully unleash converged communications and computing abilities \cite{chen2023netgpt}. However, until to now, there is no specific definition for the blueprint of the semantic network. Besides the aforementioned challenges for end-to-end SemCom, a semantic network also faces the difficulty to disseminate the semantic KB to distributed devices, so as adapt to the network heterogeneity.
\end{enumerate}

\section{Conclusion}
\label{sec:conclusion}
The purpose of this tutorial-cum-survey is to provide a comprehensive understanding of the state-of-the-art works for SemCom and its applications. First, we have reviewed the evolution of SemCom and provided a clear definition and explanation of the semantic-empowered mechanism in typical communication scenarios. Next, we have elaborated on the corresponding ecosystem consisting of history, theoretical guidance, metrics, toolkits, \etc. Afterwards, we have carefully reviewed two major categories of SemCom techniques, \ie, implicit reasoning and explicit reasoning, wherein the former typically relies on structured or unstructured parameterized models (e.g., DNNs), while the latter involves explicitly defined rules or entities. In particular, we have presented an overview of the recent progress on techniques to explore and exploit the content and channel semantics. On top of these end-to-end SemCom approaches, we have discussed typical semantics-assisted networked systems, where possible applications and opportunities are envisioned as well.

Through exchanging the most informative, timely, and effective information, SemCom is capable of improving resource utilization, communication efficiency and effectiveness towards accomplishing a task, and can provide a paradigm and technical basis for the next generation communication systems. Therefore, semantics-aware communication will play an important role in future intelligent systems. The objective of this article has been to provide a primer and a unified view of this semantics-aware communications.

\bibliographystyle{IEEEtran}
\bibliography{Survey}

\end{document}